\documentclass[longauth]{aa}

\usepackage{graphicx}
\usepackage{txfonts}
\begin{document}

   \title{The APEX Large CO Heterodyne Orion Legacy Survey (ALCOHOLS)
          \thanks{Based on observations collected at the European Southern Observatory under ESO programme(s) 094.C-0935(A) and
          the Swedish programme 094.F-9343.}
          }

   \subtitle{I. Survey overview}

   \author{Th. Stanke
          \inst{1}
          \and
          H. G. Arce\inst{2}
          \and
          J. Bally \inst{3}
          \and
          P. Bergman \inst{4}
          \and
          J. Carpenter\inst{5}
          \and
          C. J. Davis \inst{6}
          \and
          W. Dent \inst{5}
          \and
          J. Di Francesco \inst{7,12}
          \and
          J. Eisl\"{o}ffel \inst{8}
          \and
          D. Froebrich \inst{9}
          \and 
          A. Ginsburg \inst{10}
          \and
          M. Heyer \inst{11}
          \and
          D. Johnstone \inst{7,12}
          \and
          D. Mardones \inst{13}
          \and
          M. J. McCaughrean \inst{14}
          \and
          S. T. Megeath \inst{15}
          \and
          F. Nakamura \inst{16}
          \and
          M. D. Smith \inst{17}
          \and
          A. Stutz \inst{18}
          \and
          K. Tatematsu \inst{19}
          \and
          C. Walker \inst{20}
          \and
          J. P. Williams \inst{21}
          \and
          H. Zinnecker \inst{22}
          \and
          B. J. Swift \inst{20}
          \and
          C. Kulesa \inst{20}
          \and
          B. Peters \inst{20}
          \and
          B. Duffy \inst{20}
          \and
          J. Kloosterman \inst{23}
          \and
          U. A. Y{\i}ld{\i}z \inst{24}
          \and
          J. L. Pineda \inst{24}
          \and
          C. De Breuck \inst{1}
          \and
          Th. Klein \inst{25}
          }

   \institute{ESO,
              Karl-Schwarzschild-Stra\ss{}e 2, 85748 Garching bei M\"unchen\\
              \email{tstanke049@gmail.com}
         \and
         Department of Astronomy, Yale University, P.O. Box 208101, New Haven, CT 06520-8101, USA
        \and CASA, University of Colorado, Boulder, CO, USA
        \and Dept of Space, Earth and Environment, Chalmers Univ. of Technology, Onsala Space Observatory, 43992 Onsala, Sweden
        \and Joint ALMA Observatory, Avenida Alonso de C\'ordova 3107, Vitacura, Santiago, Chile 
        \and National Science Foundation, 2415 Eisenhower Avenue, Alexandria, VA 22314, USA
        \and NRC Herzberg Astronomy and Astrophysics, 5071 West Saanich Road, Victoria, BC, V9E 2E7, Canada
        \and Th\"{u}ringer Landessternwarte, Sternwarte 5, D-07778 Tautenburg, Germany
        \and School of Physical Sciences, University of Kent, Canterbury CT2 7NH, UK
        \and Department of Astronomy, University of Florida, PO Box 112055, USA 
        \and Department of Astronomy, University of Massachusetts, Amherst, MA, 01003, USA
        \and Department of Physics and Astronomy, University of Victoria, Victoria, BC, V8P 5C2, Canada
        \and Departamento de Astronom\'{i}a, Universidad de Chile, Casilla 36-D, Santiago, Chile
        \and European Space Agency, ESTEC, Postbus 299, 2200 AG Noordwijk, The Netherlands
        \and Department of Physics and Astronomy, University of Toledo, Toledo, OH 43606, USA 
        \and National Astronomical Observatory, 2-21-1 Osawa, Mitaka, Tokyo 181-8588, Japan
        \and Centre for Astrophysics and Planetary Science, School of Physical Sciences, University of Kent, Canterbury CT2 7NH, UK 
        \and Departmento de Astronom\'{i}a, Facultad de Ciencias F\'{i}sicas y Matem\'{a}ticas, Universidad de Concepci\'{o}n, Concepci\'{o}n, Chile 
        \and Nobeyama Radio Observatory, National Astronomical Observatory of Japan, National Institutes of Natural Sciences,
             462-2 Nobeyama, Minamimaki, Minamisaku, Nagano 384-1305, Japan\\
             \email{k.tatematsu@nao.ac.jp}
        \and Steward Observatory, University of Arizona, 933 N.\ Cherry Avenue, Tucson AZ 85721
        \and Institute for Astronomy, University of Hawaii at Manoa, Honolulu, HI 96822, USA\\ 
             \email{jw@hawaii.edu}
        \and Universidad Autonoma de Chile, Pedro de Valdivia 425, Santiago de Chile
        \and University of Southern Indiana, Evansville, IN, USA
        \and Jet Propulsion Laboratory, California Institute of Technology, 4800 Oak Grove Drive, Pasadena, CA 91109-8099, USA 
        \and European Southern Observatory, Alonso de C\'{o}rdova 3107, Vitacura, Casilla, 19001, Santiago de Chile, Chile
             }

   \date{Received \today; accepted \today}

 
  \abstract
   {The Orion molecular cloud complex harbours the nearest Giant Molecular Clouds (GMCs) and
   the nearest site of high-mass star formation. Its young star and protostar populations are thoroughly
   characterized. The region is therefore a prime target for the study of star formation.}
   {Here, we verify the performance of the SuperCAM 64 pixel heterodyne array on the Atacama Pathfinder Experiment (APEX). We give a
   descriptive overview of a set of wide-field CO(3-2) spectral line cubes obtained towards the Orion GMC complex,
   aimed at characterizing the dynamics and structure of the extended molecular gas in diverse regions of the clouds,
   ranging from very active sites of clustered star formation in Orion~B to comparatively quiet regions in southern Orion~A.
   In a future publication, we will characterize the full population of protostellar outflows and their feedback over an entire GMC.
   }
 {We present a 2.7~square degree (130~pc$^2$) mapping survey in the $^{12}$CO(3-2) transition, obtained using SuperCAM
   on APEX at an angular resolution of 19\arcsec{} (7600~AU or 0.037~pc at a distance of 400~pc),
   covering the main sites of star formation in the Orion~B cloud (L~1622, NGC~2071, NGC~2068, Ori~B9, NGC~2024, and NGC~2023),
   and a large patch in the southern part of the L~1641 cloud in Orion~A. }
 { We describe CO integrated line emission and line moment maps and position-velocity diagrams for all survey fields
   and discuss a few sub-regions in some detail.
   Evidence for expanding bubbles is seen with lines splitting into double
   components, often in areas of optical nebulosities, most prominently in the NGC~2024
   \ion{H}{II} region, where we argue that the bulk of the
   molecular gas is in the foreground of the \ion{H}{II} region. High CO(3-2)/CO(1-0) line ratios reveal warm CO along
   the western edge of the Orion~B cloud in the NGC~2023/NGC~2024 region facing the IC~434 \ion{H}{II} region.
   We see multiple, well separated radial velocity cloud components towards several fields and
   propose that L~1641-S consists of a sequence of clouds at increasingly larger distances.
   We find a small, seemingly spherical cloud, which we term 'Cow~Nebula' globule, north of NGC~2071.
   We confirm that we can trace high velocity line wings out to the 'extremely high velocity' regime
   in protostellar molecular outflows for the NGC~2071-IR outflow and the NGC~2024 CO jet, and identify
   the protostellar dust core FIR4 (rather than FIR5) as the true driving source of the NGC~2024 monopolar outflow.
   }
   {}


   \maketitle
%

\section{Introduction}

At a distance of 400-500~pc \citep[e.g.][]{mentenetal2007, kounkeletal2017, grossschedletal2018, zuckeretal2019}, the Orion
molecular cloud complex harbours the most nearby Giant Molecular Clouds (GMCs)
\citep[for an overview see, e.g.][]{genzelstutzki1989, bally2008}. The Orion~A GMC, in particular, is home to the nearest
site of high-mass star formation in the \object{BN/KL} complex in the \object{OMC-1} core behind the \object{Orion Nebula}
\ion{H}{II} region. \object{OMC-1} is at the heart of the very dense 'integral-shaped filament', featuring very active low- to
intermediate-mass star formation. The \object{L~1641} dark cloud extends over several degrees to the south-east of the
Orion Nebula and shows mostly low-mass star formation. Orion~B shows star formation mainly concentrated in a few moderate-sized
clusters with young stars up to intermediate masses. Orion~A and B thus provide the opportunity to study star formation in a
single cloud complex over a range of environments, from regions directly impacted by massive star formation and \ion{H}{II}
regions, over moderate-sized clusters, to very quiet regions with mostly isolated star formation.

The CO molecule arguably is the best tracer of the bulk of the molecular gas in clouds due to its high abundance and the low critical
densities of its lower-J rotational transitions.
Numerous studies presented wide-field observations of CO emission in Orion. Following the first detection in space of CO in the
(1-0) transition in the Orion Nebula region by \cite{wilsonetal1970}, the pioneering works of \cite{kutneretal1977} and
\cite{maddalenaetal1986} mapped out the overall extent of the Orion GMCs, while \cite{ballyetal1987} presented the first attempt
of resolving the internal spatio-kinematical structure of the Orion~A GMC in $^{13}$CO(1-0). The first attempt at wide-field mapping
in the region in the $^{12}$CO and $^{13}$CO(3-2) lines was presented by \cite{krameretal1996}, covering the \object{NGC~2023} and 
\object{NGC~2024} regions in Orion~B, including the Horsehead Nebula \object{B33}.
Further wide field lower-J CO studies in Orion include \cite{castetsetal1990}, \cite{sakamotoetal1994}, \cite{nagahamaetal1998},
\cite{aoyamaetal2001}, \cite{wilsonetal2005}, and \cite{nishimuraetal2015}, while a complete
coverage of Orion~A in the CO(4-3) line has been presented by \cite{ishiietal2016}.

Recent advancements in receiver technology (sensitivity, bandwidth, multi-pixel heterodyne array receivers) now allow
(multi)square degree sized maps to be made on larger telescopes with significant sub-arcminute spatial resolution
\citep[e.g.][]{sakamotoetal1997, shimajirietal2011, nakamuraetal2012, rippleetal2013, berneetal2014, petyetal2017, ishiietal2019,
nakamuraetal2019}. In terms of spatial
resolution combined with large area coverage, the $^{12}$CO, $^{13}$CO, and C$^{18}$O(1-0) survey by \cite{kongetal2018}, combining
CARMA interferometric observations with NRO 45~m observations using the BEARS and FOREST heterodyne receiver arrays, has set
new standards, covering 2~square degrees at 7\arcsec{} spatial resolution. Still, these surveys remain
largely limited to the lowest-J transitions, that is the (1-0) and (2-1) lines, due to the challenges posed by performing extensive
observations at sub-millimetre wavelengths.

Wide-field surveys in the CO(3-2) line, with adequate sensitivity, spatial resolution, and area coverage, have emerged only
recently \citep[e.g.][]{takahashietal2008}. The HARP 16 element receiver array at the JCMT improved significantly
on this situation, for example in the course of the JCMT Gould Belt Legacy Survey \citep{wardthompsonetal2007, davisetal2010,
gravesetal2010, buckleetal2010, buckleetal2012, curtisetal2010, whiteetal2015}, but still these studies remain mostly limited
to areas of a few hundred to a few thousand square arcminutes.


It is the latest generation of CO mapping surveys that provide the required combination of sensitivity, spatial resolution, and
area coverage to study the structure and kinematics of clouds on scales corresponding to the typical dimensions of filaments and
cores ($\sim$0.1~pc) and to reveal the full population and extents of molecular outflows driven by protostars in an unbiased way
over entire molecular clouds. The 64-pixel SuperCAM Heterodyne array, developed for use at the HHT/SMT on Mount Graham in Arizona,
opened up the opportunity to perform wide field observations in the CO(3-2) line at APEX\footnote{This publication is based on data
  acquired with the Atacama Pathfinder Experiment (APEX). APEX is a collaboration between the Max-Planck-Institut f\"{u}r
  Radioastronomie, the European Southern Observatory, and the Onsala Space Observatory. Swedish observations on APEX are supported
  through Swedish Research Council grant No 2017-00648}
during a visiting run in late 2014 and early 2015.
The 7-pixel LASMA array, now operational at APEX, will ensure efficient wide field molecular line mapping in the
future.

We have used SuperCAM at APEX to perform a survey of a 2.7~square degree area in the CO(3-2) line in the Orion GMC complex.
The purpose of the present paper is to give an overview of the survey and to present the data. The reduced data will be made
publicly available (through the ESO archive) for community use, a first example of which can be found in \cite{ballyetal2018}.
These data will allow for examining the environments of young stellar objects in the region (e.g.\ the HOPS protostars) and the
effect of environment (e.g.\ UV irradiation) on the clouds. We will explore
the population of protostellar CO outflows in the survey area in a subsequent publication.

The paper is organized as follows. In Sect.~\ref{section:observations} we describe the instrument,
the observing strategy, and the survey coverage. In Sect.~\ref{section:datareduction} we describe the data reduction procedure,
verify the calibration and basic data reduction procedure by comparing our data with previous APEX FLASH+ CO(3-2)
spectroscopic mapping observations of NGC~2023 \citep[Sect.~\ref{section:comparisonwithFLASH},][]{sandelletal2015}, and discuss
the noise properties and sensitivity of the survey
(Sect.~\ref{section:sensitivity}). In Sect.~\ref{section:results} we present the survey data, discussing each region separately;
we discuss spectra integrated over the survey fields, integrated intensity maps, position-velocity cuts, and moment maps (the latter
are presented in App.~\ref{app:momentmaps}).
We provide a summary and plans for the next steps in Sect.~\ref{section:conclusions}. We show noise maps for each survey field in
App.~\ref{app:noisemaps} and present moment maps of the CO emission in App.~\ref{app:momentmaps}. Finally, we include 870~$\mu$m dust
continuum maps\footnote{Based on observations collected at the European Southern Observatory under ESO programme(s) 086.C-0848(B), 088.C-0994(A), and 090.C-0894(B)} taken with Laboca on APEX (App.~\ref{app:laboca}), whose main purpose was to obtain sub-millimetre photometry in the
course of the HOPS survey
\citep{furlanetal2016}, but also serve in the present paper to compare the location of dense gas with the CO emission. 

\section{Observations}
\label{section:observations}

\begin{table*}
    \caption{Observing log.}
    \centering
    \begin{tabular}{l|c c c r l}
         \hline \hline
         Field & R.A.\tablefootmark{a} & Dec.\tablefootmark{a} & Size\tablefootmark{b} & P.A.\tablefootmark{c} & Dates \\
         \hline
         L~1622 & 5:54:30.0 & 1:45:10 & 27\farcm5$\times$22\farcm5 & 0 & 2014-12-23 \\
         NGC~2071\tablefootmark{d} & 5:47:25.3 & 0:29:40 & 24\farcm0$\times$32\farcm5 & 15 & 2014-12-14/17/19 \\
         NGC~2068\tablefootmark{d} & 5:46:20.0 & -0:04:30 & 25\farcm0$\times$32\farcm5 & 15 & 2014-12-11\tablefootmark{e}/12 \\
         Ori~B9 & 5:43:08.0 & -1:14:30 & 24\farcm0$\times$32\farcm5 & 35 & 2014-12-22/23 \\
         NGC~2024\tablefootmark{d} & 5:41:34.0 & -1:52:00 & 27\farcm5$\times$30\farcm0 & 3 & 2014-12-21/22 \\
         NGC~2023\tablefootmark{d} & 5:41:30.0 & -2:22:00 & 27\farcm5$\times$30\farcm0 & 3 & 2014-12-20/24 \\
         L~1641-S (1)\tablefootmark{d} & 5:40:58.0 & -7:54:30 & 24\farcm0$\times$32\farcm5 & -5 & 2014-12-14 \\
         L~1641-S (2)\tablefootmark{d} & 5:42:44.0 & -8:16:15 & 27\farcm5$\times$35\farcm0 & -5 & 2014-12-19/20 \\
         L~1641-S (3)\tablefootmark{d} & 5:42:24.0 & -8:45:05 & 40\farcm0$\times$25\farcm0 & 0 & 2014-12-25 \\
         L~1641-S (4)\tablefootmark{d} & 5:40:58.0 & -8:21:20 & 27\farcm5$\times$20\farcm0 & 0 & 2014-12-23/24/25 \\
         \hline
    \end{tabular}
    \tablefoot{
    \tablefoottext{a}{Coordinates of the adopted fields centre (epoch J2000, as for all coordinates in this paper)}
    \tablefoottext{b}{Field size in arcminutes}
    \tablefoottext{c}{Field position angle (in degrees east of north)}
    \tablefoottext{d}{Fields \object{NGC~2071} and \object{NGC~2068}, fields \object{NGC~2023} and \object{NGC~2024},
      and fields \object{L~1641-S} (1)--(4) were combined into larger fields during data reduction.}
    \tablefoottext{e}{first coverages taken with a position angle of 5$^\circ$}
    }
    \label{tab:observinglog}
\end{table*}

\begin{figure}
   \centering
   \includegraphics[width=\linewidth]{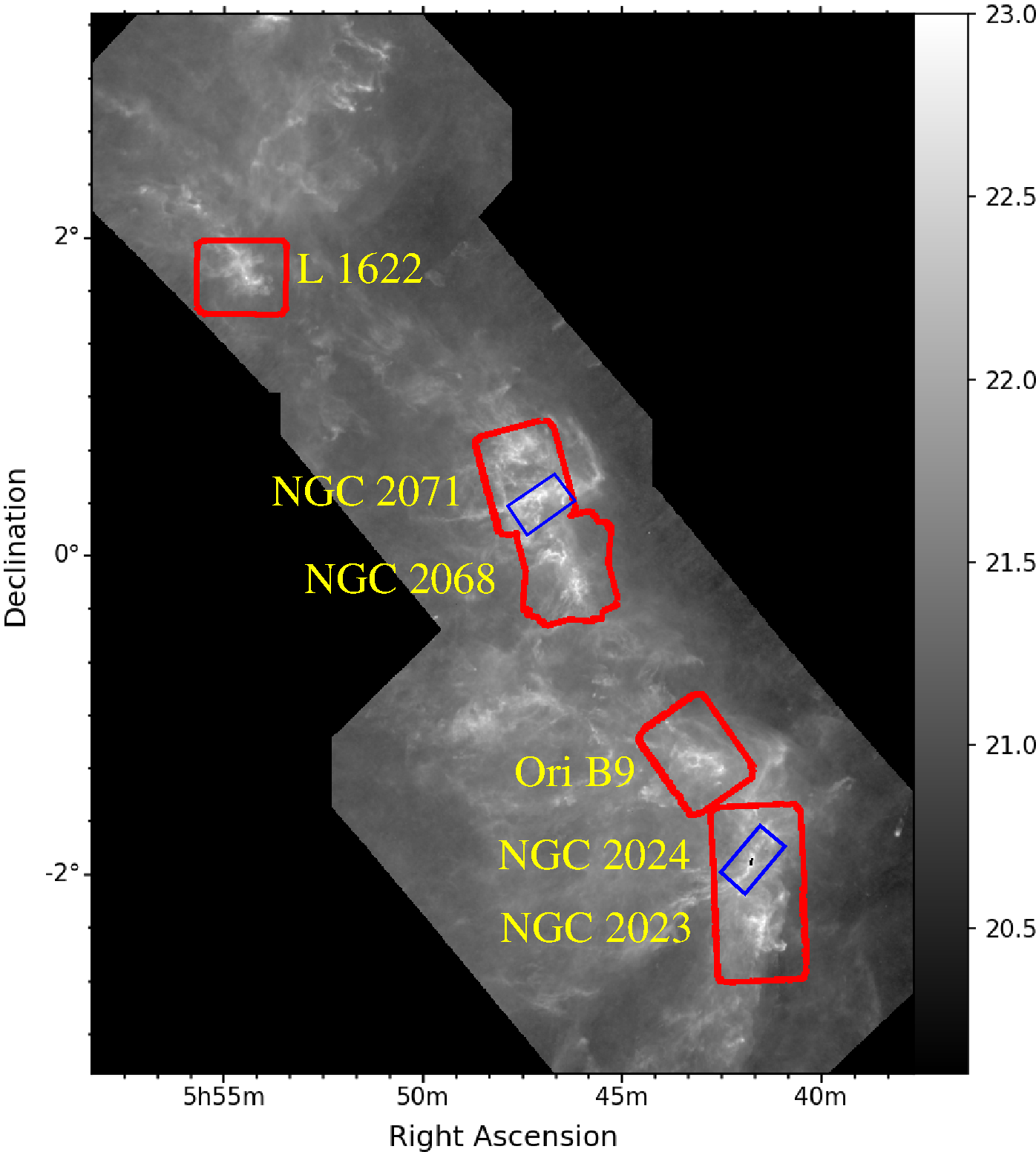}
   \caption{Column density map ($\log{N_{\rm H}}$ [cm$^{-2}$]) towards the Orion~B GMC
     \citep[greyscale; based on Herschel dust continuum mapping:][]{stutzkainulainen2015, stutzgould2016, stutz2018}
     with the coverage of the ALCOHOLS survey in Orion~B
     overplotted in red. The blue rectangles mark the coverage of the JCMT Legacy Survey of the Gould Belt
     in the NGC~2024 and NGC~2071 regions \citep{buckleetal2010}.
              }
   \label{fig:coverage_OrionB}
\end{figure}

\begin{figure}
   \centering
   \includegraphics[width=\linewidth]{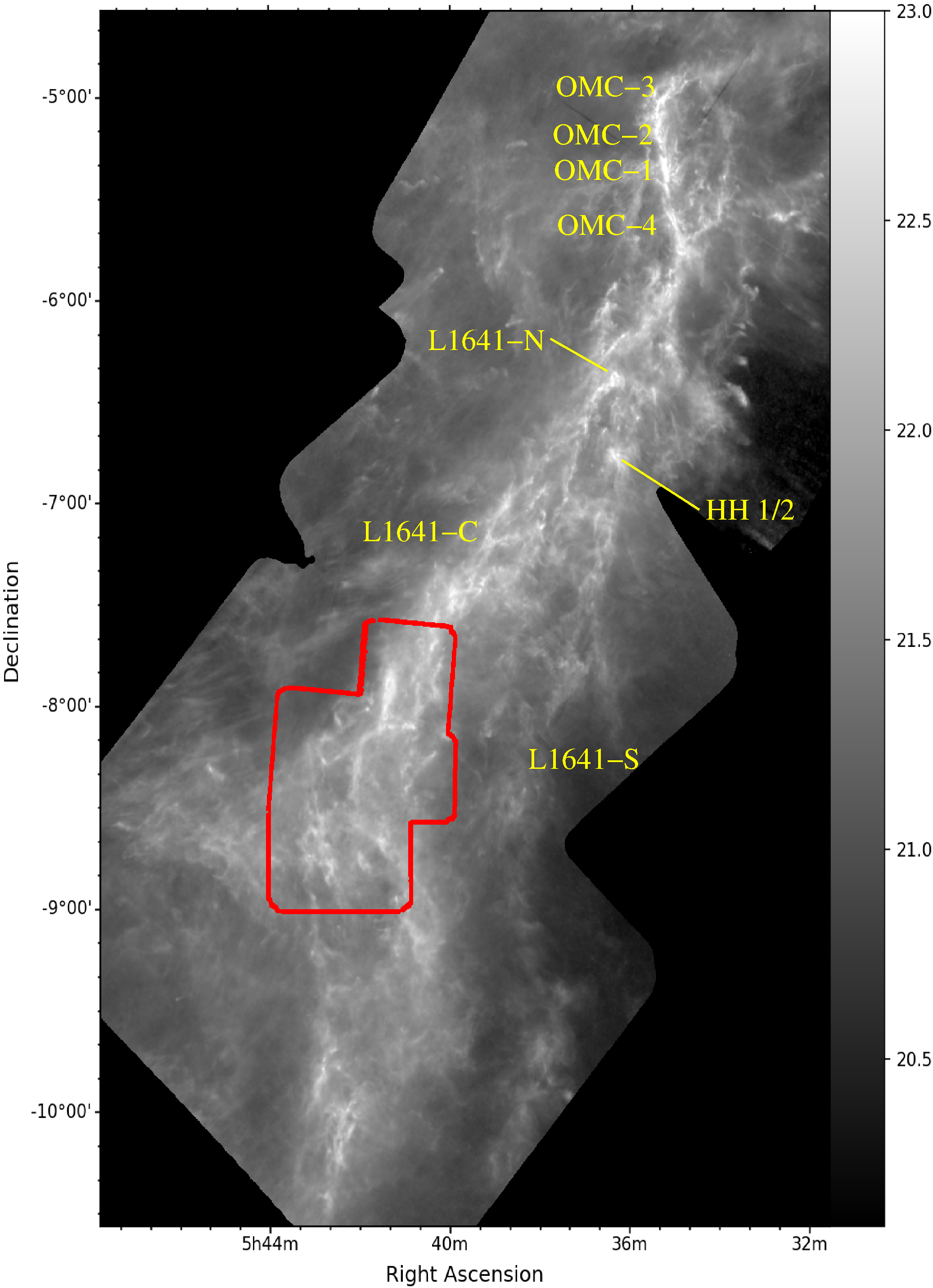}
   \caption{As Fig.~\ref{fig:coverage_OrionB} for the Orion~A GMC, but with the coverage of the ALCOHOLS survey
            in Orion~A overplotted in red.
              }
   \label{fig:coverage_OrionA}
\end{figure}

We used the SuperCAM heterodyne receiver array \citep{kloostermanetal2012} mounted at the APEX 12~m
telescope \citep{guestenetal2006} located on Chajnantor at an elevation of 5100~m to map a total
area of $\sim$2.7 square degrees in the Orion~A and Orion~B GMCs
in the $^{12}$CO~(3-2) transition at 345.8~GHz.
The observations were obtained between December 9 and 25, 2014 (see Table~\ref{tab:observinglog}), and took a total of 51~hours of
telescope time.

SuperCAM consists of a total of 64 heterodyne receivers (individual receivers will
be referred to as 'pixels' in the following), of which 49 were operational and connected at the time
of the observations. The receivers were tuned to the frequency of the $^{12}$CO(3-2) transition
at 345.7~GHz. The spectra were recorded with a 64$\times$250~MHz Fast Fourier Transform spectrometer, giving
900 channels with a native channel width of $\sim$0.278~MHz/0.24~km/s.

The maps were done in the 'On--The--Fly' (OTF) mode, where a rectangular (or square) area is scanned
in a sequence of parallel rows. We chose a spacing between subsequent rows of 150\arcsec{} (about half the size of the SuperCAM
field of view at APEX), and alternated the scanning direction in a 'zigzag' pattern. Along the rows, spectra were read
every 0.5~seconds, spaced by 6\arcsec{} ($\sim$1/3 of the main beam HPBW of 19\arcsec{}), corresponding to a
scanning speed of 12\arcsec{}/second. Several coverages were done on each field, with the orientation of the scanning
direction alternating between E--W and N--S. Hot/Sky-calibration observations were taken every 3 to 4 rows (depending on the
geometry of the map).

Pointing was checked during the observations about once per hour, and the focus was checked at the beginning of each observing session.
An observation of a standard Eccosorb (AN-72) blackbody load at the ambient
temperature ($\sim$285~K) of the APEX C-cabin and a blank-sky observation were taken every few OTF rows
to provide hot and sky measurements to compensate for atmospheric
extinction by the standard 'chopper wheel' calibration method of \cite{ulichhaas1976}.
These measurements were passed through the APEX online
calibrator pipeline \citep{mudershafok2019} to provide a facility measure
of T$_{\rm cal}$ for a central 'fiducial' pixel and subsequently the system
noise temperature T$_{\rm sys}$ through per-pixel sky/(hot-sky) 'gain' arrays.

The precipitable water vapour at the APEX telescope ranged from 0.5~mm to
2.2~mm during observations of these Orion maps, corresponding to an
zenith atmospheric opacity at 345.6~GHz from 0.1 to 0.35.  Receiver
noise temperatures across the (44) selected pixels ranged from 75~K to
120~K DSB with a median of 90~K DSB.  The heterodyne focal plane array is
coupled to the APEX telescope via a compact refractive optical relay
comprised of two lenses machined from high density polyethylene (HDPE),
which allowed SuperCAM to be mounted into the APEX C-cabin with minimal
disturbance to existing instruments.  The warm losses of the two
relay lenses used to couple the SuperCAM feed horns to the APEX
telescope were, however, unfortunately significant: $1.5 \pm 0.1$~dB.  The impact of
these losses raised the noise temperature of the combination of receiver
plus coupling optics to a range of 200-250~K DSB.  The resulting
single-sideband T$_{\rm sys}$ varied with atmospheric water vapour content, from
700~K to about 1000~K.

Because T$_{\rm cal}$ was ultimately derived from a single central pixel, a
per-pixel scaling factor was required to correct the antenna
temperatures of each array pixel. This mean correction factor, derived
from OTF maps of Jupiter and Saturn, was a factor of 1.3. This per-beam
correction, relative to the T$_{\rm cal}$ that was calculated for the 'fiducial'
pixel, results in a uniform T$_{\rm a}^*$ over the array.

\subsection{Coverage}
\label{section:coverage}

The survey fields were chosen to complement corresponding CO(3-2) observations done
at the JCMT (as of the time when the proposals were written, i.e. in March 2014), in the course of the
JCMT Goulds Belt survey \citep{wardthompsonetal2007, buckleetal2010, buckleetal2012}, and by the HARP instrument
team (unpublished). In \object{Orion~A} (Fig.~\ref{fig:coverage_OrionA}), we cover 1.1~square degrees in \object{L~1641-S} \citep{allendavis2008} south of a declination
of $-$7:30. In \object{Orion~B} (Fig.~\ref{fig:coverage_OrionB}), 1.6 square degrees were mapped, covering the major star forming clumps around \object{NGC~2023} \&
\object{NGC~2024} \citep[e.g.][]{meyeretal2008}, \object{NGC~2068} \& \object{NGC~2071} \citep{gibb2008}, the \object{Ori~B9} region, and the \object{L~1622} cloud \citep{reipurthetal2008}.

\section{Data reduction}
\label{section:datareduction}

For data reduction we used the GILDAS CLASS software\footnote{http://www.iram.fr/IRAMFR/GILDAS} and,
for the flatfielding, the ESO MIDAS software.
As a first step, bad receiver pixels were identified, and, for each scan, pixels with an
average rms higher than two times the median rms of all pixels during the scan were rejected. This typically
left $\sim$44 (of the original 64) good pixels. Linear baselines were then subtracted from all spectra. The fitting
range was restricted to 10--20~km/s beyond the maximum velocity extent of the CO emission in
each survey field (and excluding the full velocity range where CO emission was detected). The
full CO velocity range was identified from spectral cubes created from the raw data, and refined later--on,
where necessary.

The spectral cubes initially showed clear striping patterns on regions of bright emission. These patterns
were greatly reduced by applying a 'flat--field' correction for the individual receivers. We derived, for
each receiver pixel, a rescaling factor, as the flux ratio of the cube using all pixels and a cube including
only the respective pixel (using only spectral and spatial regions with emission above a certain brightness
threshold). We then compared the total fluxes of the cubes (including all receivers) before and after the
flatfielding, rescaling the flatfielded cubes to reproduce the original total flux, where necessary
(typically for fields with only fairly faint emission). A separate set of pixel scaling values was derived
for each survey field (as the signal--to--noise ratio was generally too low to do it on a per--scan basis).
While the cube for any given survey field may generally include scans taken on different days, the scaling
factors are seen to be consistent between the fields, implying that they are also stable with
time. 

The spectra frequently showed `spikes' (unrealistically high or low values, in individual velocity channels).
The spikes show up in individual spectra and frequently in larger groups of consecutive spectra. Often a set of 
`typical' velocity channels consistently is affected for several receivers and in several scans. While the
off--line calibration removed a significant fraction of the spikes, visual inspection of all spectra was
necessary to identify the remaining, obvious groups of spikes, which were then corrected by interpolating
between the adjacent, valid velocity
channels (CLASS commands DRAW KILL and DRAW FILL). Visual inspection also showed a number of clearly bad
spectra, which were removed. In addition, the preliminary spectral cubes showed a number of
spectra which clearly were outside the intended map coverage. We identified the corresponding
integration time stamps and removed all spectra belonging to these.

We then applied an iterative procedure to remove any remaining spikes (mostly individual spikes in
individual spectra) and to subtract higher order polynomial baselines from the spectra.
At this stage, we subtracted from each spectrum a `model spectrum'
of the CO emission. The model spectrum was created using the resulting spectral cubes of the previous
iteration, averaging spectra extracted from the cube over a small aperture (of a radius of 8--10~arcsec,
smaller than or comparable to the beamsize of the observations) around the position of the respective spectrum.
Adjacent and overlapping survey
fields were processed together to maximize the quality of the `model spectra' in the overlap
regions (fields NGC~2023 and NGC2024, fields NGC~2068 and NGC~2071, fields L1641S-1 and L1641S-2 and
L1641S-3 and L1641S-4). The model spectrum was added back to the spectra after de-spiking/de-baselining.

As spikes, we considered outliers beyond certain multiples of the rms of each (model--subtracted) spectrum,
starting at large values (10-15~$\sigma$) and reducing to 4.8~$\sigma$ in the final iterations.
At this stage, spikes were replaced by the brightness value of the corresponding velocity channel
of the `model spectrum'. 

Similarly, for baseline
removal, we increased the baseline polynomial order, starting with linear baselines, up to
polynomial orders of 3 in the final iterations. We also decreased the spectral window to be excluded from the
baseline fits, from almost the full velocity extent of the emission in the beginning for the linear baselines,
to using essentially the full spectrum for baseline fitting in the final iterations.

At the end of the automated de-spiking/de-baselining (after typically 4 iterations) the resulting spectra were again
checked for pixels and integration ranges requiring even higher order baseline subtraction, resulting from
receiver instability or unstable weather conditions. Higher order baselines were fitted and subtracted where
necessary, again after subtracting a model spectrum. Here the highest order used was 18, but it has to be noted that very
high order baselines only were used on a very minor fraction of the spectra (e.g.\ order 10 and higher were used for
only $\sim$0.2\% of the spectra). A final round of de-baselining with a baseline order of 3 was then done on all
spectra (again removing a model spectrum before fitting the baselines).

Finally, we divided the resulting spectra by a mean main beam efficiency
of 48\% as derived from OTF maps of Jupiter and Saturn, assuming a intrinsic beam-convolved brightness temperature of
150~K for Jupiter (Butler-JPL-Horizons 2012 model, ALMA Memo 594\footnote{http://library.nrao.edu/public/memos/alma/main/memo594.pdf})
and 124~K for Saturn given the ring inclination at the time of observation \citep{weilandetal2011}. This somewhat
low beam efficiency is due to a small but measurable amount of field
distortion introduced by the aforementioned refractive relay that
couples the SuperCAM focal plane to the APEX telescope.  The principal
impact is a measurable beam elongation that increases for pixels further
from the central optical axis.  This elongation reaches a maximum aspect
ratio of 1.5:1 for pixels at the corners of the 8x8 square focal plane
array. We note that these edge pixels were not used in the analysis.

The error beam of SuperCAM is thus dominated by a very narrow component, only a few arcseconds
wider than the 18\arcsec{} diffraction limited PSF core. The mean coupling efficiency rises to $>$0.7
in a slightly larger beam convolved to 22\arcsec{} (corresponding well to main beam efficiencies measured
in that frequency range with the APEX facility receivers before the upgrade of the telescope in 2018
\citep[e.g.][]{guestenetal2006}). For further details on the derivation of the SuperCAM
beam efficiencies (and their pixel-to-pixel variations) and the error beam see \cite{kloosterman2014}.
As will be demonstrated in Section \ref{section:comparisonwithFLASH}, high imaging fidelity is achieved with SuperCAM in this
configuration at APEX with a small amount of spatial smoothing.
Regardless, a general warning about the interpretation of both spectral
line imaging and spectral line profiles at the diffraction limit must be
made. It is important to emphasize that the results of this first survey
paper do not depend upon data analysis at these very highest spatial
frequencies.

We then created cubes at a velocity resolution of 0.25~km/s and with a pixel size of
6~arcseconds, over a velocity range wide enough to include the full CO emission and 10-20~km/s beyond on
each side. As some residual baseline instabilities were still apparent in the cubes, we performed another
round of baseline subtraction on the cube, where the spectral range to exclude from the fit was set to the
full spectral extent of the CO emission; special care was given to areas with very high velocity emission,
making sure that the exclusion range was set large enough (and the baseline order low enough) to avoid subtracting
out or artificially introducing any high velocity line wing emission.

\subsection{Comparison with previous APEX CO(3-2) map of NGC~2023}
\label{section:comparisonwithFLASH}

\begin{figure}
   \centering
   \includegraphics[width=\linewidth]{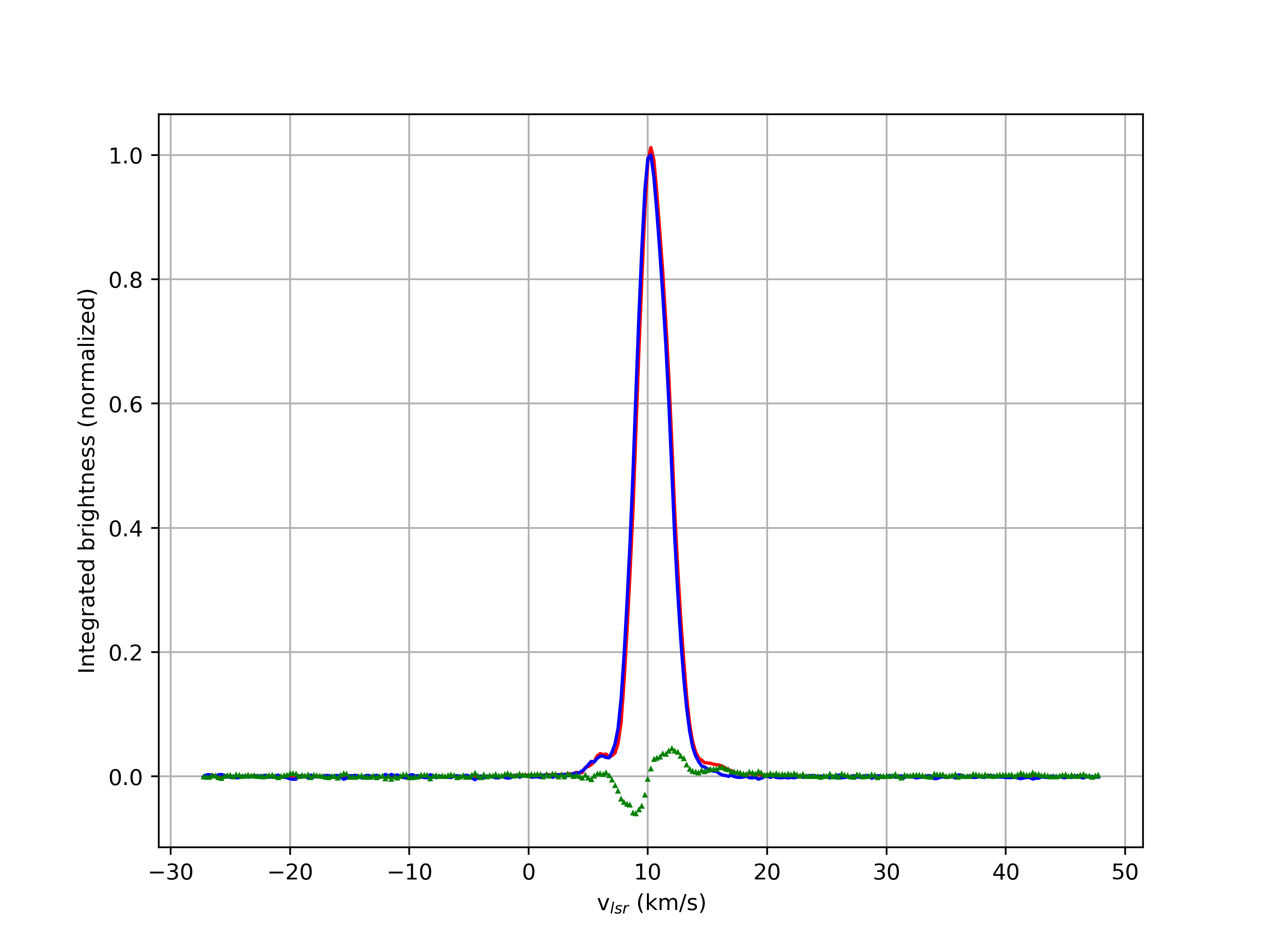}
   \caption{Comparison between the integrated (per channel) brightness of the FLASH+ CO(3-2) observations of
     NGC~2023 by \cite{sandelletal2015} (red line) and the ALCOHOLS dataset (blue line). The green
     triangles show the difference between the two datasets. The data were normalized to give a peak integrated
     brightness of 1.0 for the ALCOHOLS dataset.
              }
   \label{fig:GoeranVSALC_fluxperchannel}
\end{figure}

\begin{figure}
   \centering
   \includegraphics[width=\linewidth]{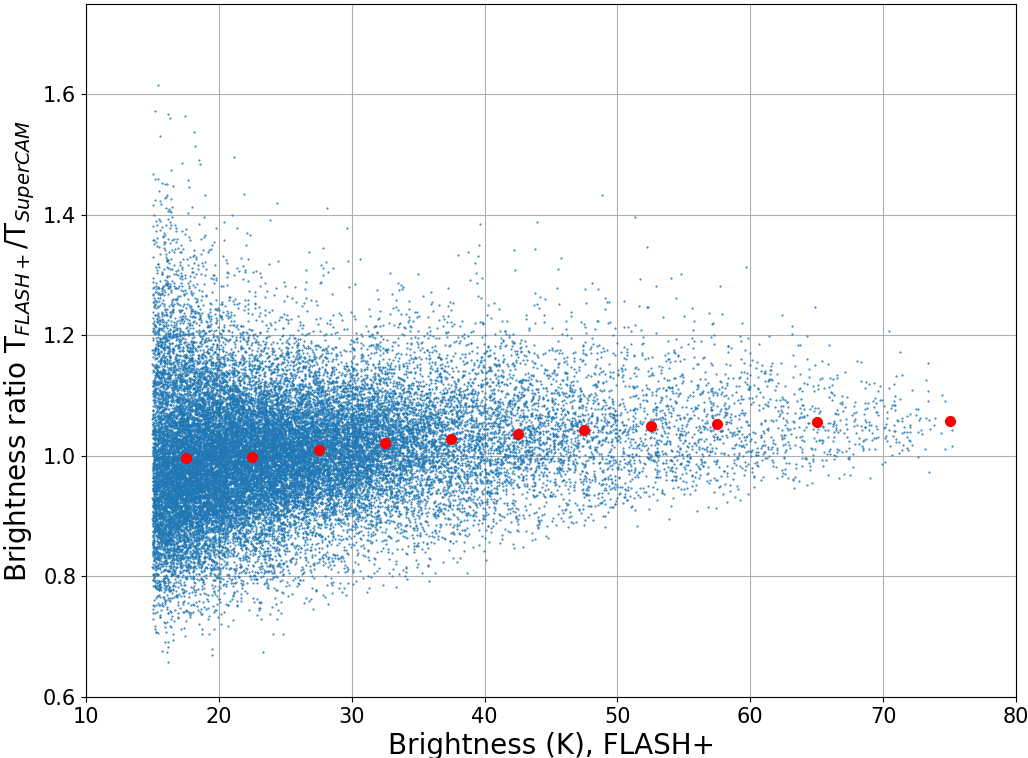}
   \includegraphics[width=\linewidth]{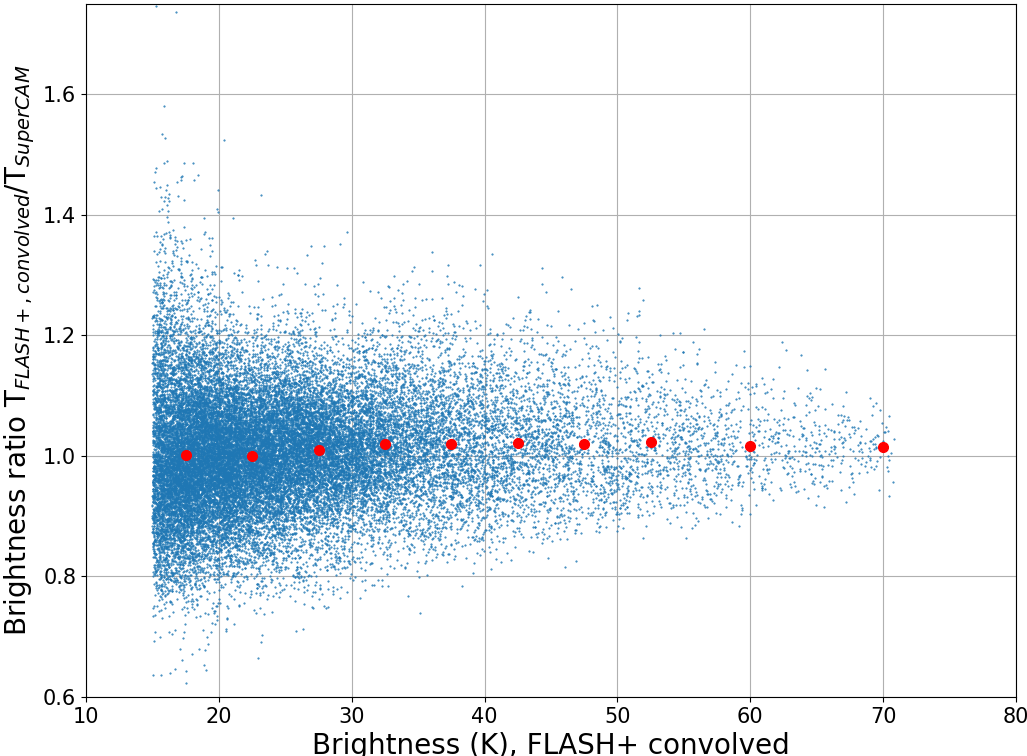}
   \caption{Per pixel flux ratios (blue dots) of the \cite{sandelletal2015} FLASH+ CO(3-2) NGC~2023 data to the
           ALCOHOLS SuperCAM data, as a function of the brightness of the FLASH+ data, for pixels with a FLASH+ 
           brightness greater than 15~K. Red bullets represent flux ratios averaged over 5~K temperature bins 
           (10~K bins for the two rightmost points). Top: direct ratio of the FLASH+ and SuperCAM data;
           bottom: ratio after convolving the FLASH+ data with a 9\arcsec{} Gaussian.}
   \label{fig:ratioGoeranALC}
\end{figure}

A subset of our survey region in NGC~2023 was previously mapped in the CO(3-2) line with the FLASH+ receiver
\citep{kleinetal2014} on APEX,
that is, with the same telescope and on the same site \citep{sandelletal2015}. We use this dataset to check the
performance of SuperCAM and to verify the calibration and the robustness of the data reduction against
artificially introducing or subtracting out high velocity line wing emission.

To do so, we created a subcube of the ALCOHOLS data at the same spatial and spectral gridding as the
FLASH+ data cube (which also used a pixel size of 6\arcsec{} and a velocity channel width of 0.25~kms/s).
After correcting for a small pointing offset between the two datasets (3\arcsec{} in declination),
we used two approaches
to compare the flux scaling. First, we summed up the brightness temperature
values of all pixels within a velocity channel, and compared the resulting integrated spectra
(Fig.~\ref{fig:GoeranVSALC_fluxperchannel}).
We find an excellent agreement of the peak values (as derived from Gaussian fits) of the two spectra, which
differ only by 0.7\% (where the FLASH+ map is slightly brighter). We note a small velocity shift between the
two datasets ($\sim$0.11~km/s) producing the negative and positive residual immediately to the blue and red side, respectively,
of the line in the difference spectrum (green symbols in Fig.~\ref{fig:GoeranVSALC_fluxperchannel}. The
magnitude of this shift is very close to half the channelwidth and might be due to a different definition of the
channel velocity between the two datasets. The ALCOHOLS data also reproduce the shape
of the blue-shifted line wing emission very well. In the redshifted wing, at velocities between $\sim 14$~km/s to
  $\sim 18$~km/s the FLASH+ data show a mild excess (of the order of 1-1.5\% ($\sim$0.3~K) of the flux at the line centre).
The excess cannot be attributed to compact features as may be expected from outflows, but is apparently due
to a diffuse, very low surface brightness component, which might not be real but due to, for example, slight differences
in the spectral baseline fitting and subtraction procedure.

For each pixel of all channel maps that were brighter than a threshold of 15~K, we also derived
the brightness ratio between the two datasets. The cutoff at 15~K was chosen as a compromise between having
  a significant number of flux values at good signal-to-noise in both datasets, as the value around which the flux ratios started
to show increasingly significant scatter greater than $\sim$20\%. Again, we find an overall excellent agreement, with a mean and
median flux ratio of 1.011 and 1.008, respectively, with the FLASH+ map being slightly brighter. We have also
plotted the pixel-to-pixel brightness ratios as a function of brightness (in the FLASH+ map, 
Fig.~\ref{fig:ratioGoeranALC}, top panel). Here, we notice
a trend of increasing flux ratio with brightness by about 5\% when going from 15~K to the maximum of $\sim$75~K.
We attribute this trend to a small difference between the
main core of the point spread functions (psf)
of the two observations. The brightest emission
is in very compact knots, where the peak flux will be much more affected by blurring by a broader psf
than more extended structures. To test this assumption, we convolved the FLASH+ data cube by Gaussians 
with FWHMs of 6\arcsec{} and 9\arcsec and found that for a convolution with a 9\arcsec{} Gaussian
the flux ratios as a function of (convolved) FLASH+ brightness become essentially flat 
(Fig.~\ref{fig:ratioGoeranALC}, bottom panel). Convolving a
18\arcsec{} or 19\arcsec{} Gaussian beam with a 9\arcsec{} Gaussian would result in a 20\farcs{}1 and
21\arcsec{} FWHM beam, respectively, that is, would correspond to an only marginally broader psf.
There remains some residual structure in the flux ratio vs.\ brightness plot, which might point
to a more complicated, possibly non-axisymmetric difference between the psfs of the FLASH+ and SuperCAM
observations, due to the substantial differences in the optics of the two systems.

\subsection{Sensitivity and noise properties}
\label{section:sensitivity}

\begin{figure*}
   \centering
   \includegraphics[width=\textwidth]{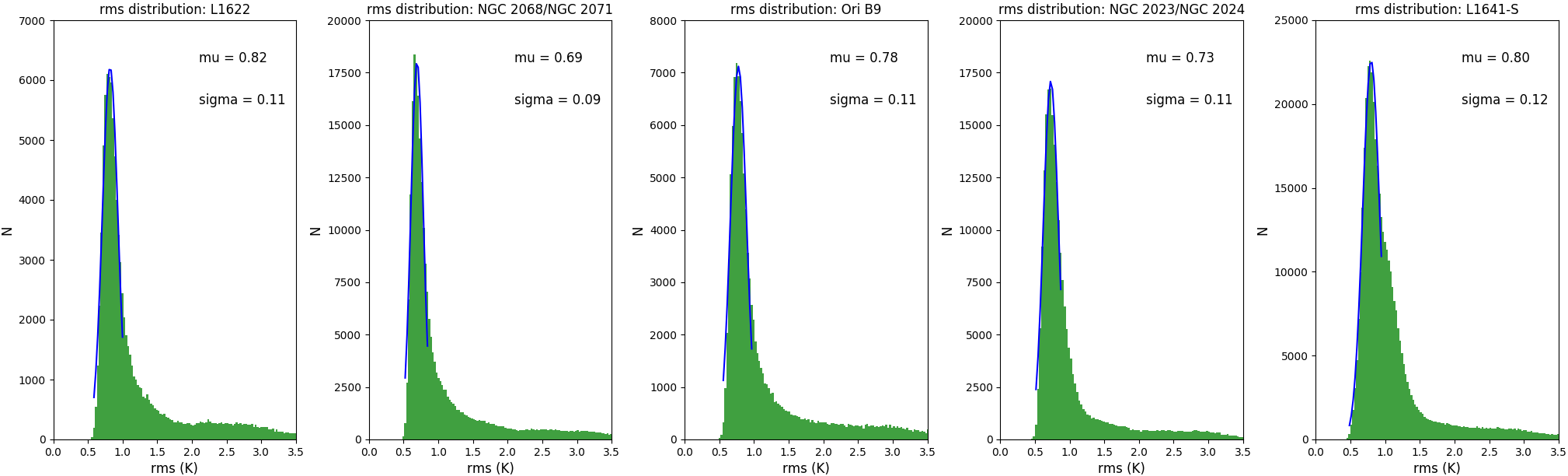}
   \caption{Distribution of rms values measured for each spatial pixel (excluding the spectral range showing CO emission)
     for each survey field (histograms, restricted to rms$<$3.5). The blue curves
     indicate Gaussian fits to the peak of the histograms (with the resulting mean ($\mu$) and
     standard deviation ($\sigma$) indicated).}
              \label{fig:rmsdist}%
\end{figure*}

\begin{table*}
\caption{Noise properties (rms in K), by survey field.}            
\label{table:rms}      
\centering                          
\begin{tabular}{l | r c c c c c c c}        
\hline\hline                 
Field & N$_{\rm pix}$\tablefootmark{a} & $\mu$, $\sigma$\tablefootmark{b} & rms min, max\tablefootmark{c} & median\tablefootmark{d} & mean\tablefootmark{d} & stddev\tablefootmark{d} \\    
\hline                        
L1622       &  96355 & 0.82, 0.11 & 0.518, 6.831 & 0.91 & 1.21 &   0.69 \\
N2068/N2071 & 239639 & 0.69, 0.09 & 0.456, 8.484 & 0.78 & 1.10 &   0.74 \\
Ori B9      & 112406 & 0.78, 0.11 & 0.490, 5.134 & 0.87 & 1.19 &   0.74 \\
N2023/N2024 & 221477 & 0.73, 0.11 & 0.442, 6.381 & 0.80 & 1.03 &   0.63 \\
L1641-S     & 413827 & 0.80, 0.12 & 0.416, 9.889 & 0.91 & 1.14 &   0.69 \\
\hline                                   
\end{tabular}
\tablefoot{
    \tablefoottext{a}{N$_{\rm pix}$: total number of valid spatial pixels in the field.}
    \tablefoottext{b}{$\mu$, $\sigma$: mean value and standard deviation from Gaussian fit to the peak of
         the rms histogram, respectively.}
    \tablefoottext{c}{rms min, max: minimum and maximum rms, respectively.}
    \tablefoottext{d}{median, mean, stddev: median, mean and standard deviation including all rms values, respectively.}
    }
\end{table*}

The scanning pattern used during the observations was not designed to provide fully sampled maps for each
individual receiver pixel (which would naturally result in near uniform coverage overall), but assumed
that the differences in the sensitivity between receivers and lower effective coverage due to dead or poor
pixels would average out, as still any position in the map was observed in a single scan by several pixels,
and several coverages were taken for each field, with the array orientation changing from scan to scan
due to the rotation of the target area on the sky. To check the validity of this approach we
derived noise maps for each field by calculating the rms of the spectra at any spatial pixel in the cube (over
the velocity range beyond the maximum velocity extent of CO emission in the respective cube). 

We show histograms of the rms distribution for each survey field in Fig.~\ref{fig:rmsdist}, while maps
of the rms are shown in Appendix~\ref{app:noisemaps}
(which also shows full frequency distributions of the brightness values for all fields, including the full
spectral range, in Fig.~\ref{fig:rms_hist}).
The histograms are characterized by a narrow peak
centred around 0.7--0.8~K and a tail extending to higher rms values. As can be seen from the rms maps,
the high--rms tails stem from the edges of the maps, where coverage is lower. The histogram for L1641-S
shows a shoulder
next to the main peak, which is due to the southernmost sub--field having lesser coverage and correspondingly
higher rms (see Fig.~\ref{fig:rmsmap_L1641S}). While the maps show a clearly distinguishable striping pattern, indicating some degree of
non--uniform rms and pixel--to--pixel variations, the rms values within the central parts of the survey
fields are fairly uniform, as indicated by the narrow main peaks of the histograms. To quantify further
the typical rms in the central parts of the survey fields we fitted the peak of the rms histograms with
a Gaussian curve. The resulting mean values and standard deviations are indicated in the plots and listed
in Table~\ref{table:rms}, indicating typical rms values of 0.7--0.8~K and pixel--to--pixel variations
of the order of 10--20$\%$. We take the mean of the Gaussian fits as the final sensitivity of the survey
(at a pixel size of 6$\arcsec$ and a velocity resolution of 0.25~km/s). This level is comparable to the target sensitivity
of the JCMT Gould Belt Legacy survey of 0.3~K per pixel at a velocity resolution of 1~km/s \citep{wardthompsonetal2007},
but significantly poorer than achieved in the JCMT HARP maps in Orion~B \citep[0.1~K per pixel at 1~km/s][]{buckleetal2010}. 

\section{Results}
\label{section:results}

\begin{figure*}
   \centering
   \includegraphics[width=3.6cm]{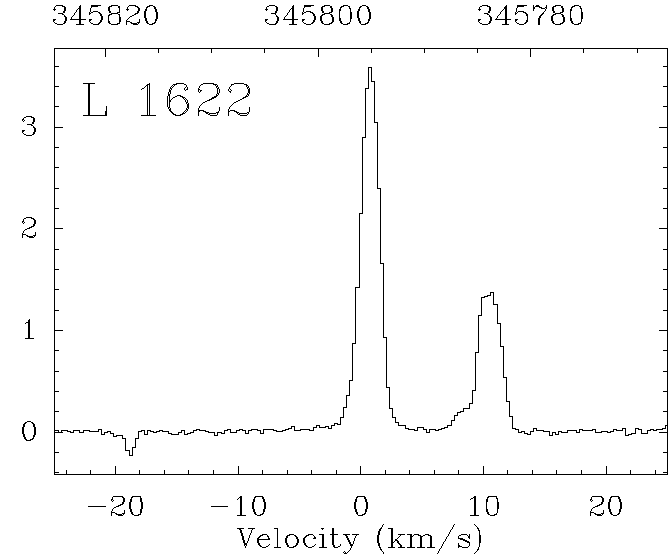}
   \includegraphics[width=3.6cm]{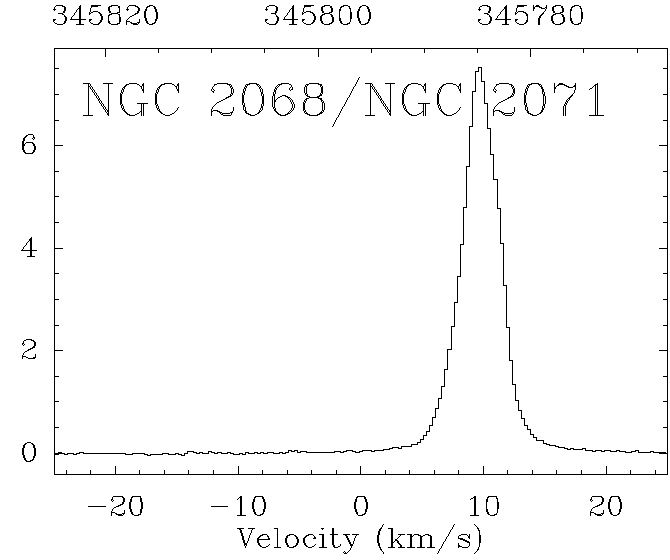}
   \includegraphics[width=3.6cm]{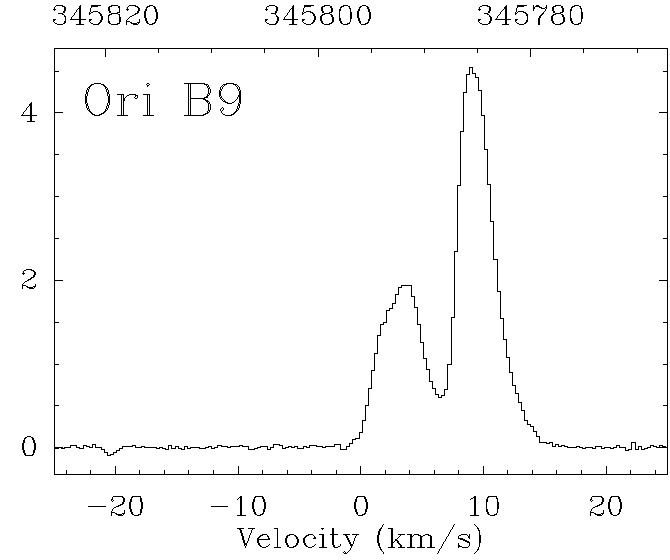}
   \includegraphics[width=3.6cm]{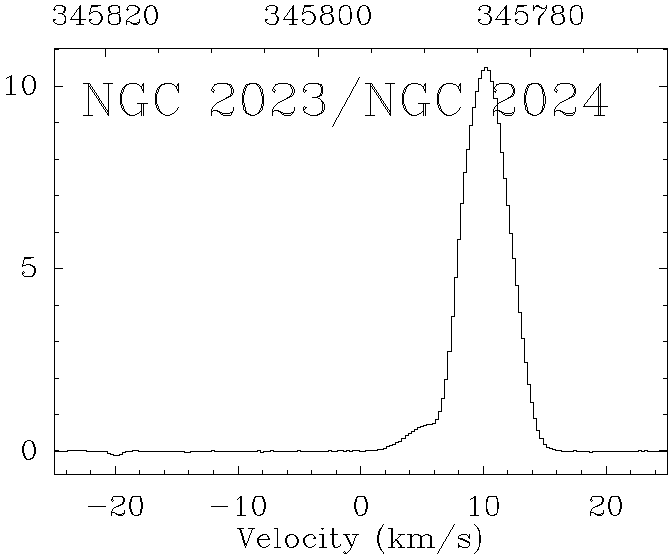}
   \includegraphics[width=3.6cm]{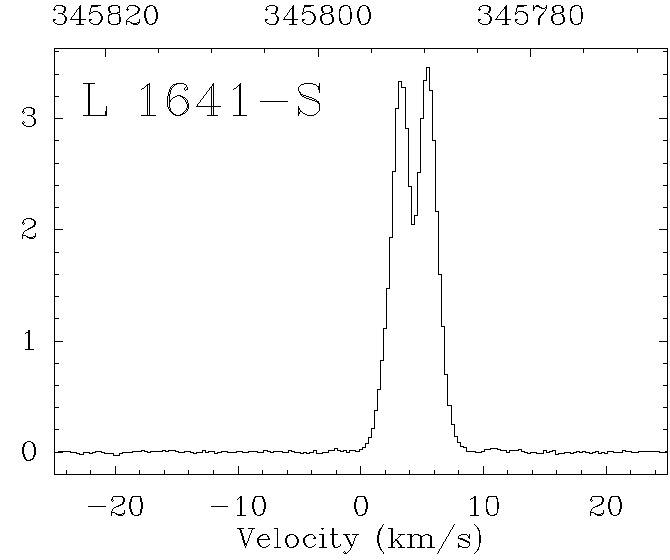}
   \caption{CO spectra averaged over the full survey fields. L~1622 and Ori~B9 each show two distinct
            velocity components, L1641-S also features a double peaked spectrum, while the NGC~2023,
            NGC~2024, NGC~2068, and NGC~2068 areas show single peaked spectra, though a hint of a
            secondary component is seen as a blue shoulder in the NGC~2023/NGC~2024 spectra.}
              \label{fig:meanspectra}%
\end{figure*}

\begin{table}
\caption{Velocities and line widths of the main CO components identified from Fig.~\ref{fig:meanspectra}  derived
      from Gaussian fits}            
\label{table:integratedlines}      
\centering                          
\begin{tabular}{l | c c}        
\hline\hline                 
Field & v (km/s) & FWHM (km/s)  \\    
\hline                        
L1622 (1~km/s)   &  0.8 & 1.8 \\
L1622 (10~km/s)  & 10.5 & 2.1 \\
N2068/N2071      &  9.8 & 3.7 \\
Ori B9 (3~km/s)  &  3.2 & 3.8 \\
Ori B9 (10~km/s) &  9.4 & 3.3 \\
N2023/N2024      & 10.2 & 4.5 \\
L1641-S (3~km/s) &  3.2 & 1.9 \\
L1641-S (6~km/s) &  5.6 & 1.8 \\
\hline                                   
\end{tabular}
\end{table}

\begin{figure}
   \centering
   \includegraphics[width=\columnwidth]{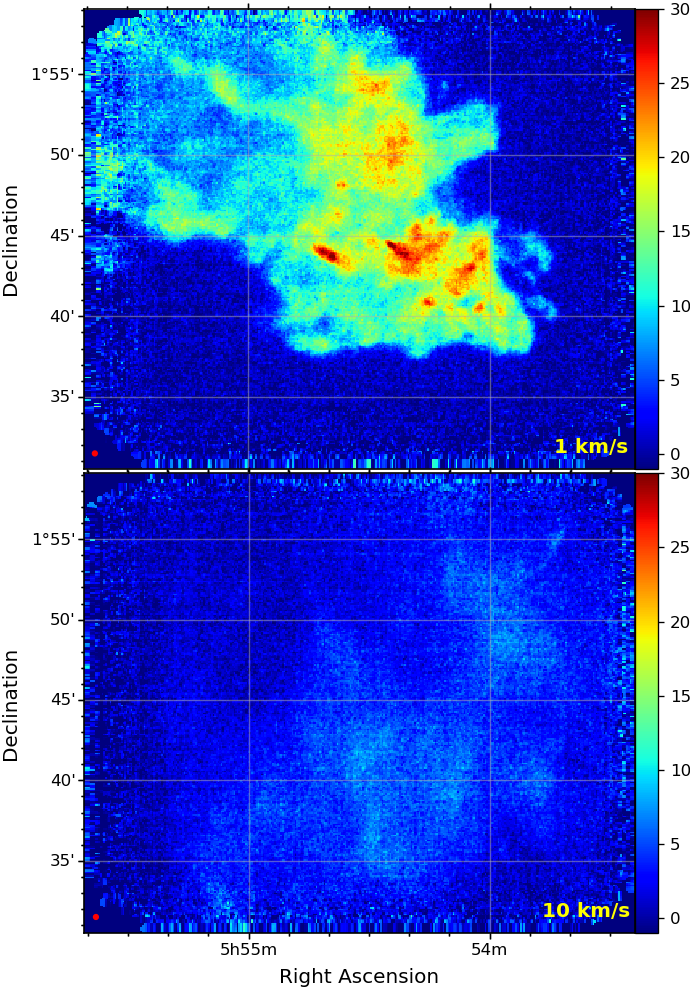}
   \caption{CO emission in the L~1622 field, integrated from -1~km/s to 2.25~km/s (top, in K~km/s)
            and from 9.25~km/s to 12~km/s (bottom). The red dots in the bottom left corners indicate the beam size.}
              \label{fig:intmap_L1622}%
\end{figure}

\begin{figure}
   \centering
   \includegraphics[width=\columnwidth]{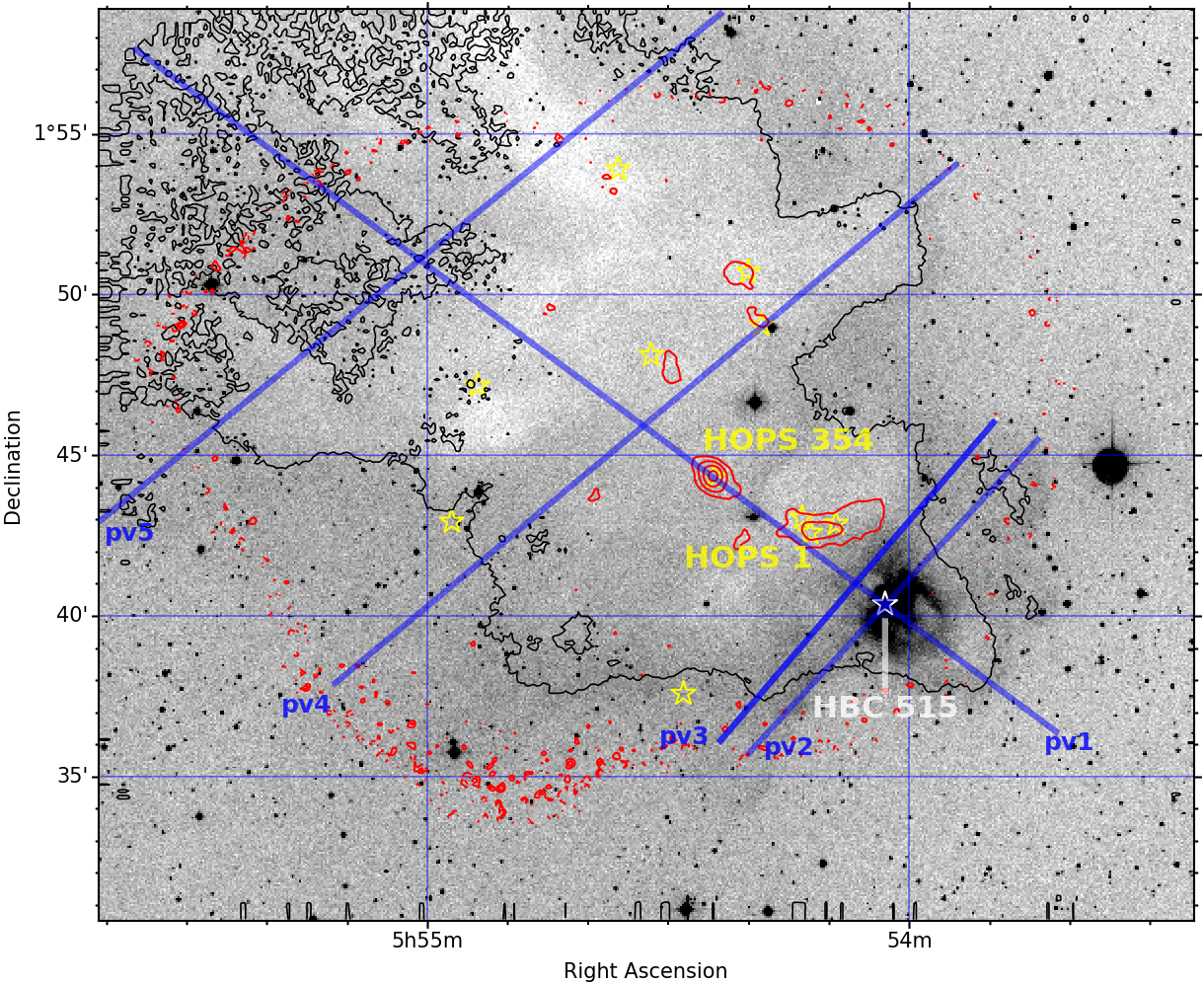}
   \caption{Overview of the L 1622 field. The greyscale shows an optical (DSS2 blue) image. Black and red contour overlays mark the
   extent of the CO(3-2) emission of the 1~km/s component and APEX/Laboca 350~$\mu$m dust continuum emission, respectively. Yellow stars
   mark the location of HOPS protostars, while the white star marks the position of the \object{HBC~515} pre-main-sequence multiple
   system. Blue lines mark the location of the position-velocity cuts shown in Fig.~\ref{fig:L1622_pv}.
   }
              \label{fig:L1622_overview}%
\end{figure}

\begin{figure}
   \centering
   \includegraphics[width=\columnwidth]{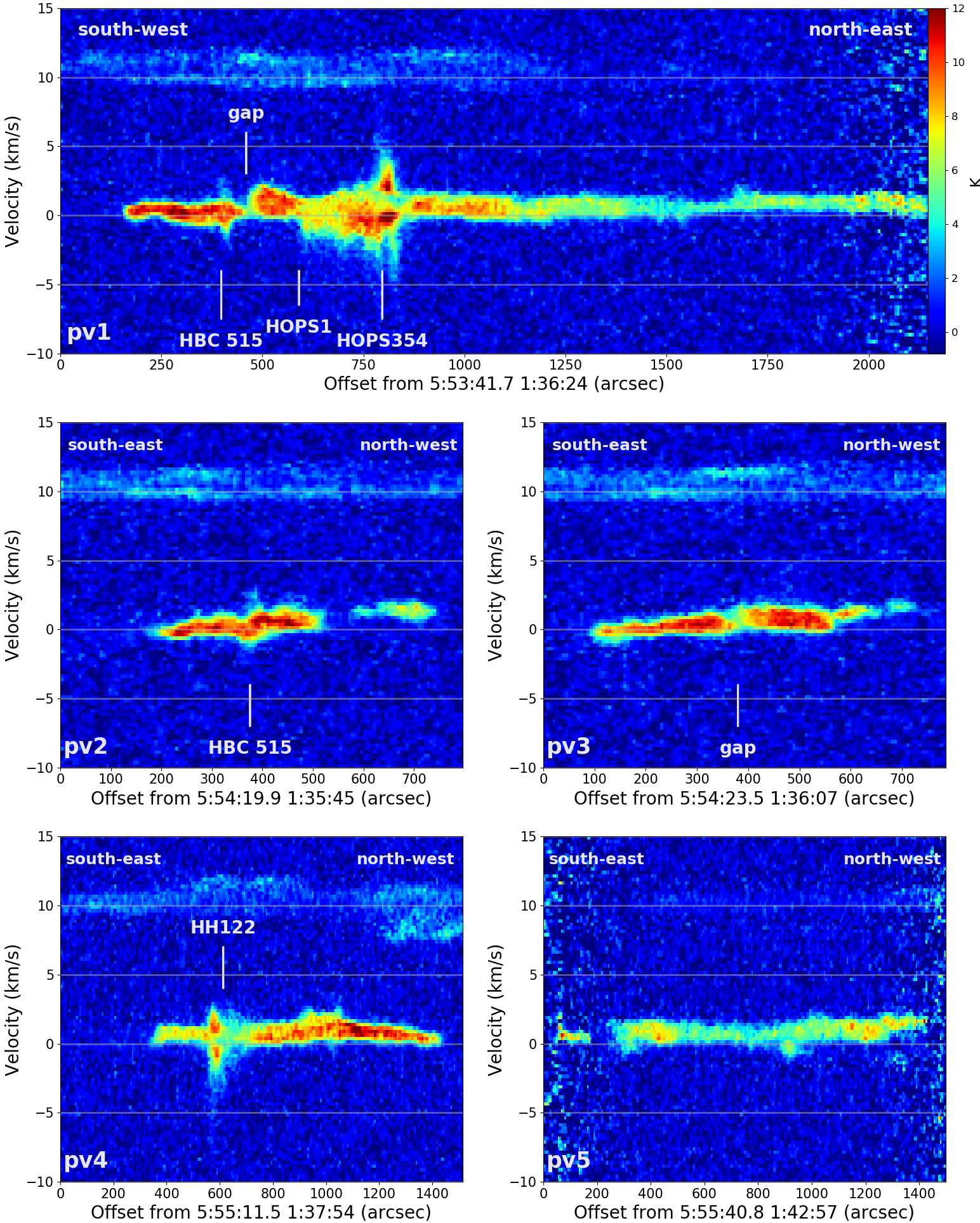}
   \caption{Position-velocity cuts across L~1622, as marked in Fig.~\ref{fig:L1622_overview}. Top: cut along the clouds major axis,
        starting from 5$^h$53$^m$41.$\!\!^s$7,
        1$^\circ$36\arcmin24\arcsec{} to the south-west of the cloud, ending at 5$^{h}$55$^m$36.$\!\!^s$3, 1$^\circ$57\arcmin37\arcsec{}
        in the north-eastern corner of the mapped area. Middle row: cuts along
        the minor axis of the cloud; left panel (pv2): starting at 5$^h$54$^m$19.$\!\!^s$9, 1$^\circ$35\arcmin45\arcsec{} (south-west),
        ending at 5$^h$53$^m$44$^s$, 1$^\circ$45\arcmin29\arcsec{} (north-east);
        right panel (pv3): starting at 5$^h$54$^m$23.$\!\!^s$5, 1$^\circ$36\arcmin07\arcsec{} (south-west), ending at 5$^h$54$^m$23.$\!\!^s$5,
        1$^\circ$36\arcmin07\arcsec{} (north-east). Bottom row, cuts along the minor axis of the cloud, further east in the tail of
        the cloud; left panel (pv4): starting at 5$^h$55$^m$11.$\!\!^s$5, 1$^\circ$37\arcmin54\arcsec{} (south-west), ending at 5$^h$53$^m$54.$\!\!^s$2, 1$^\circ$54\arcmin02\arcsec{} (north-east); right panel (pv5):
        starting at 5$^h$53$^m$54.$\!\!^s$2 1$^\circ$54\arcmin02\arcsec{} (south-west), ending at 5$^h$53$^m$54.$\!\!^s$2 1$^\circ$54\arcmin02\arcsec{} (north-east).}
              \label{fig:L1622_pv}%
\end{figure}

\begin{figure}
   \centering
   \includegraphics[width=\columnwidth]{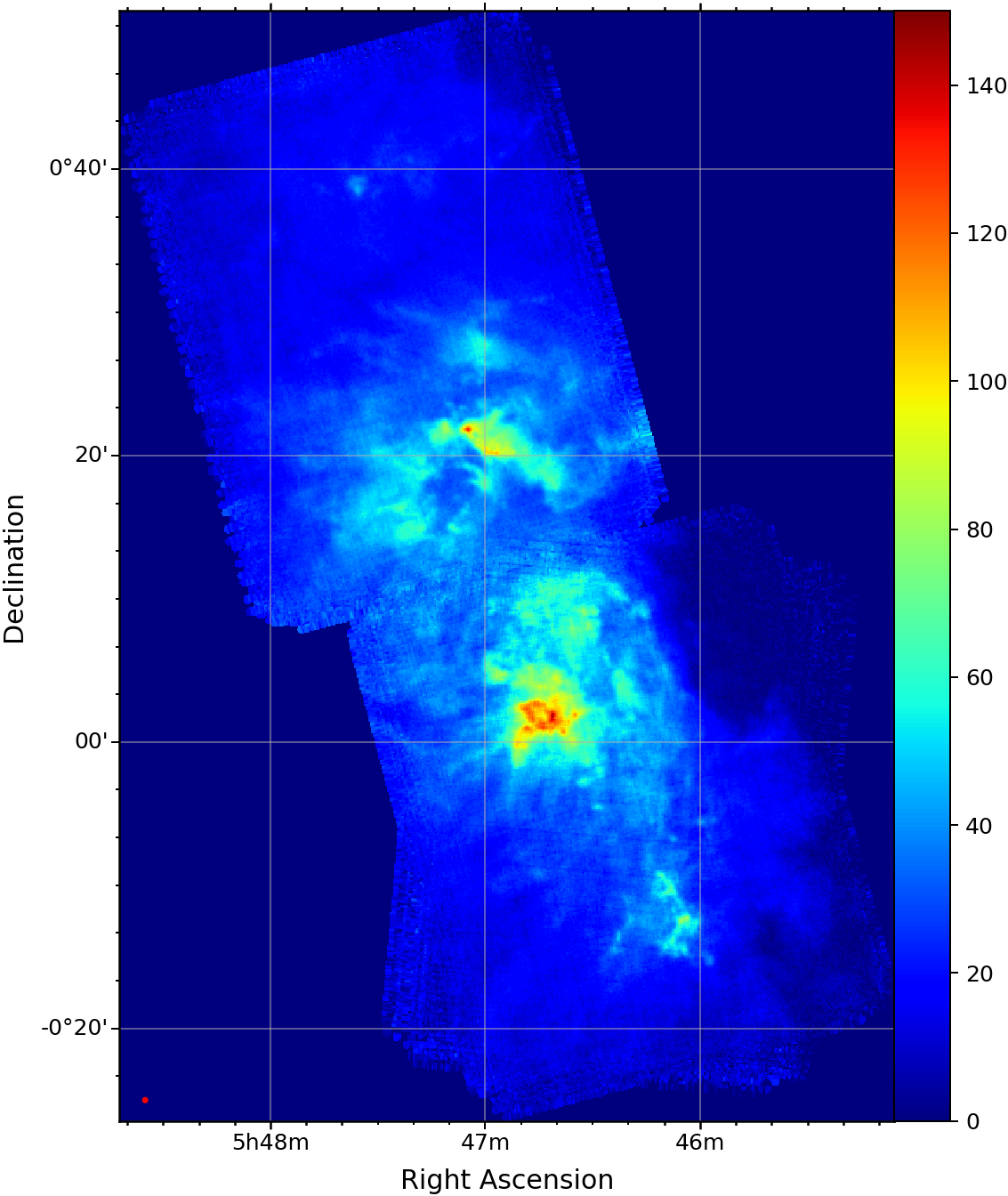}
   \caption{CO emission in the NGC~2068 and NGC~2071 field, integrated from 7.5~km/s to 13~km/s (in K~km/s).
        The red dot in the bottom left corner indicates the beam size.}
              \label{fig:intmap_N2068N2071}%
\end{figure}

\begin{figure}
   \centering
   \includegraphics[width=\columnwidth]{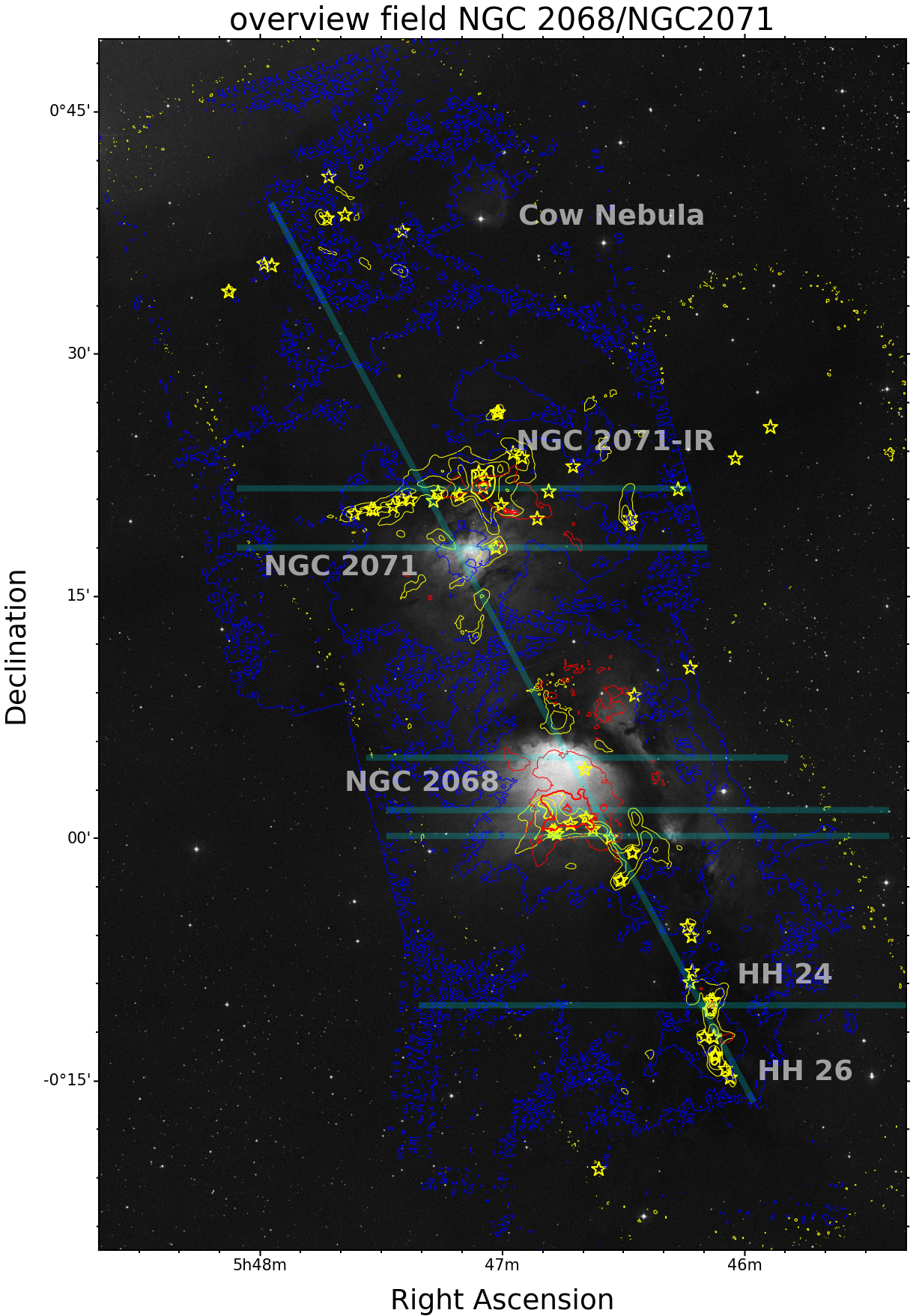}
   \caption{Overview of the NGC~2068/NGC~2071 region, showing a DSS2-blue optical image of the region, with the integrated CO emission
   from Fig.~\ref{fig:intmap_N2068N2071} overplotted as blue and red (for the brightest regions) contours, and APEX Laboca 870~$\mu$m
   dust continuum as yellow contours. Yellow stars mark the position of HOPS protostars, and cyan lines mark the location of the 
   position-velocity cuts shown in Fig.~\ref{fig:N2068N2071_pv}.
   }
              \label{fig:N2068N2071_overview}%
\end{figure}

\begin{figure}
   \centering
   \includegraphics[width=\columnwidth]{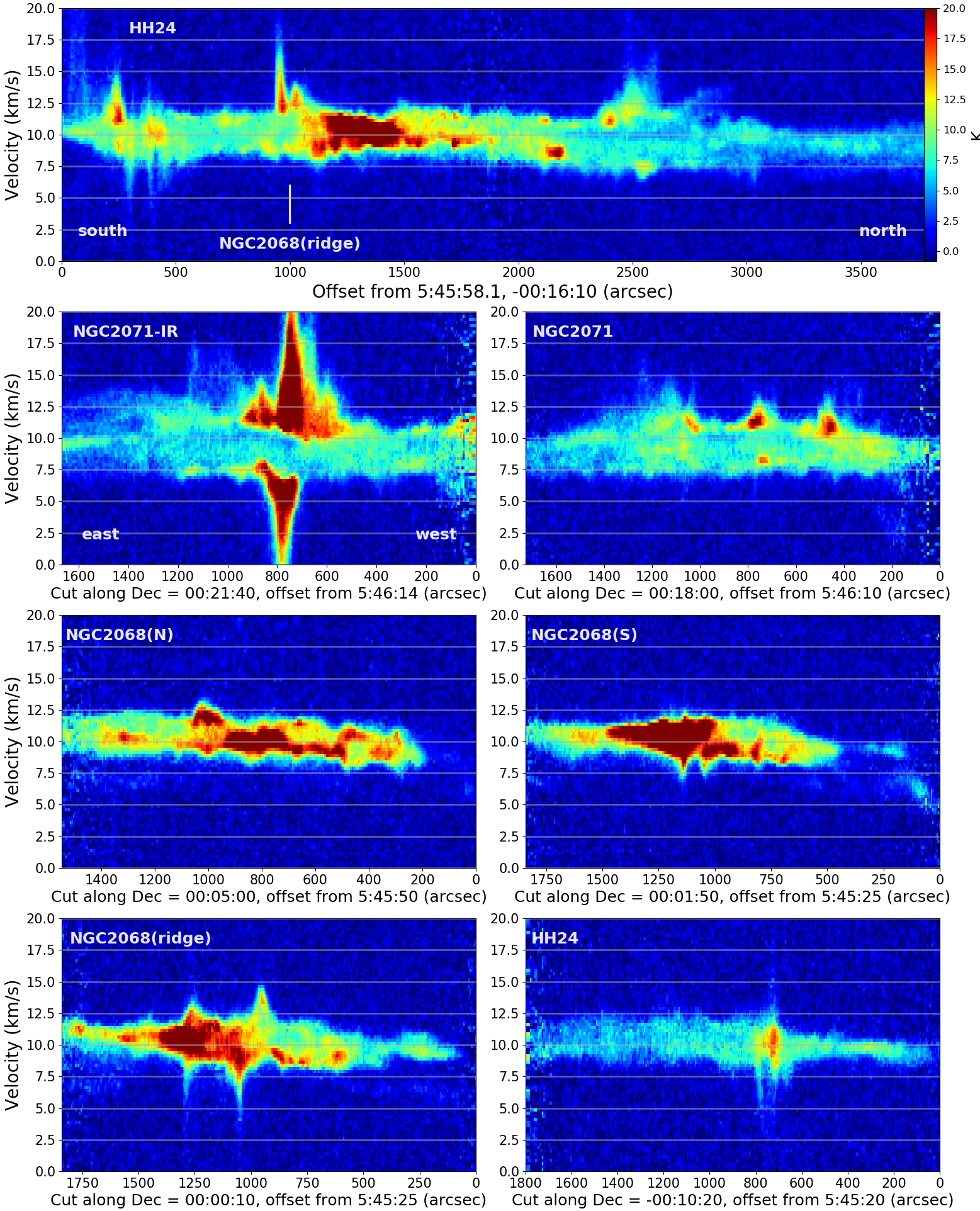}
   \caption{Position velocity cuts across the NGC~2068/NGC~2071 survey field. Top: cut starting from 5:45:58.1, -0:16:10 (south-west),
   ending at 5$^h$47$^m$57.$\!\!^s$0, $+$00$^\circ$39\arcmin10\arcsec{} (north-east).
   }
              \label{fig:N2068N2071_pv}%
\end{figure}

\begin{figure*}
   \centering
   \includegraphics[width=13cm]{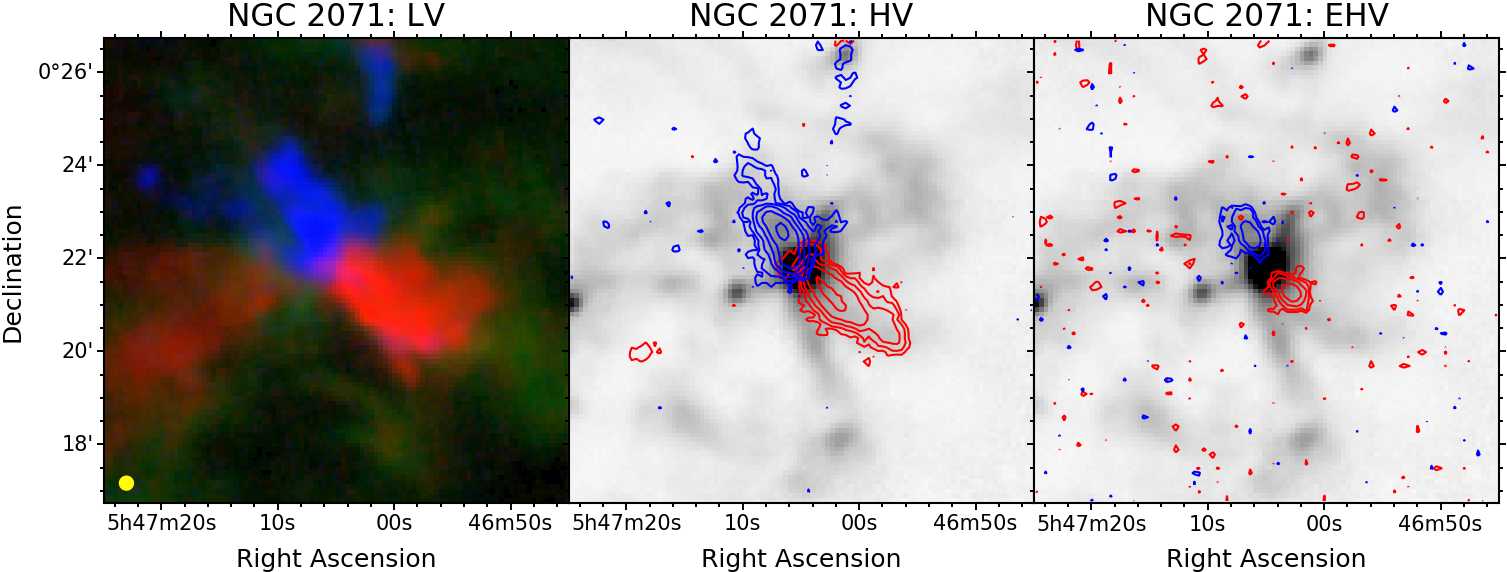}
   \includegraphics[width=5cm]{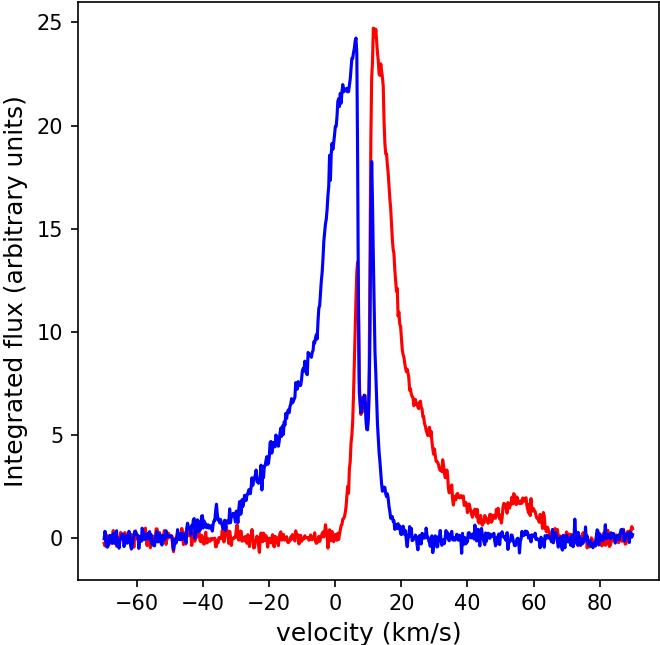}
   \caption{Maps and spectra of the NGC~2071 molecular outflow. The leftmost panel shows the low velocity component of the
           outflow (red: $v = +$13 to $+$20~km/s, blue: $v = -$5 to $+$5~km/s) along with CO emission around the clouds rest velocity
           (green: $v = +$6 to $+$12~km/s). The yellow dot indicates the beam size. The two central panels show contour plots of the high-velocity (HV) emission (red: $v = +$21 to $+$44~km/s, blue: $v = -$25 to $-$6~km/s) and extremely high-velocity (EHV) emission (red: $v = +$45 to $+$67~km/s, blue: $v = -$41 to $-$26~km/s); the greyscale shows APEX Laboca 870~$\mu$m dust continuum emission. The panel to the right shows
           spectra extracted over a 12\arcsec{} radius on the peak position of the EHV emission component, showing a well separated
           EHV feature particularly in the red lobe.}
              \label{fig:N2071flow}%
\end{figure*}

\begin{figure}
   \centering
   \includegraphics[width=\columnwidth]{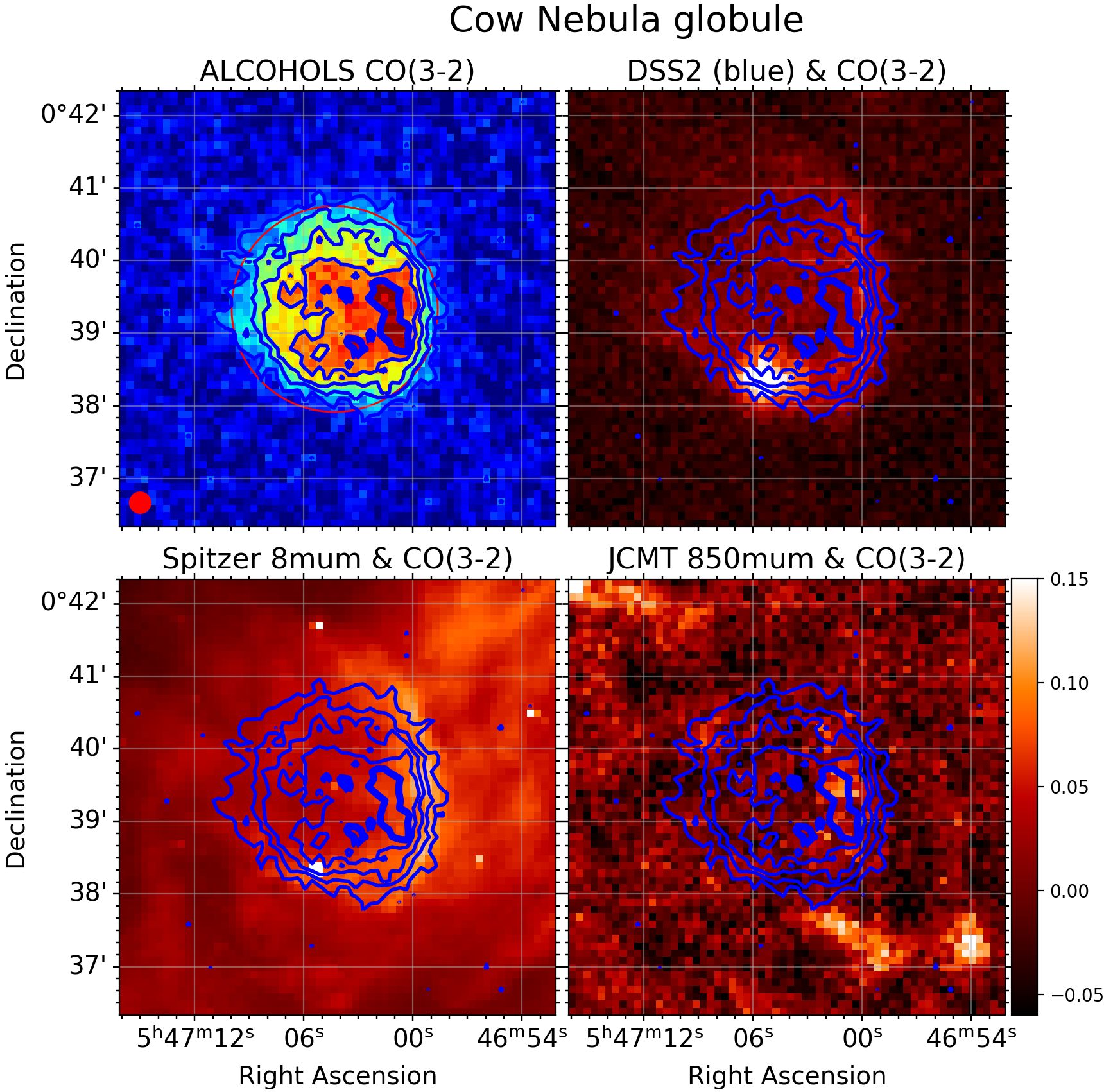}
   \caption{CO emission from the Cow~Nebula globule. Top left: emission integrated over the velocity interval $v = -6.25$~km/s to $-4.5$~km/s.
     The circle is centred at 5$^h$47$^m$04.$\!\!^s$3,  $+$00$^\circ$39\arcmin20\arcsec{} and serves to illustrate the circular
     morphology of the globule.
     Top right, bottom left, bottom right: integrated CO (top left panel) overlaid as contours over an optical image of the
   region (DSS2, blue), a Spitzer 8~$\mu$m image \citep{megeathetal2012}, and a JCMT 850~$\mu$m map \citep{kirketal2016}, respectively.}
              \label{fig:comcloud_maps}%
\end{figure}

\begin{figure}
   \centering
   \includegraphics[width=\columnwidth]{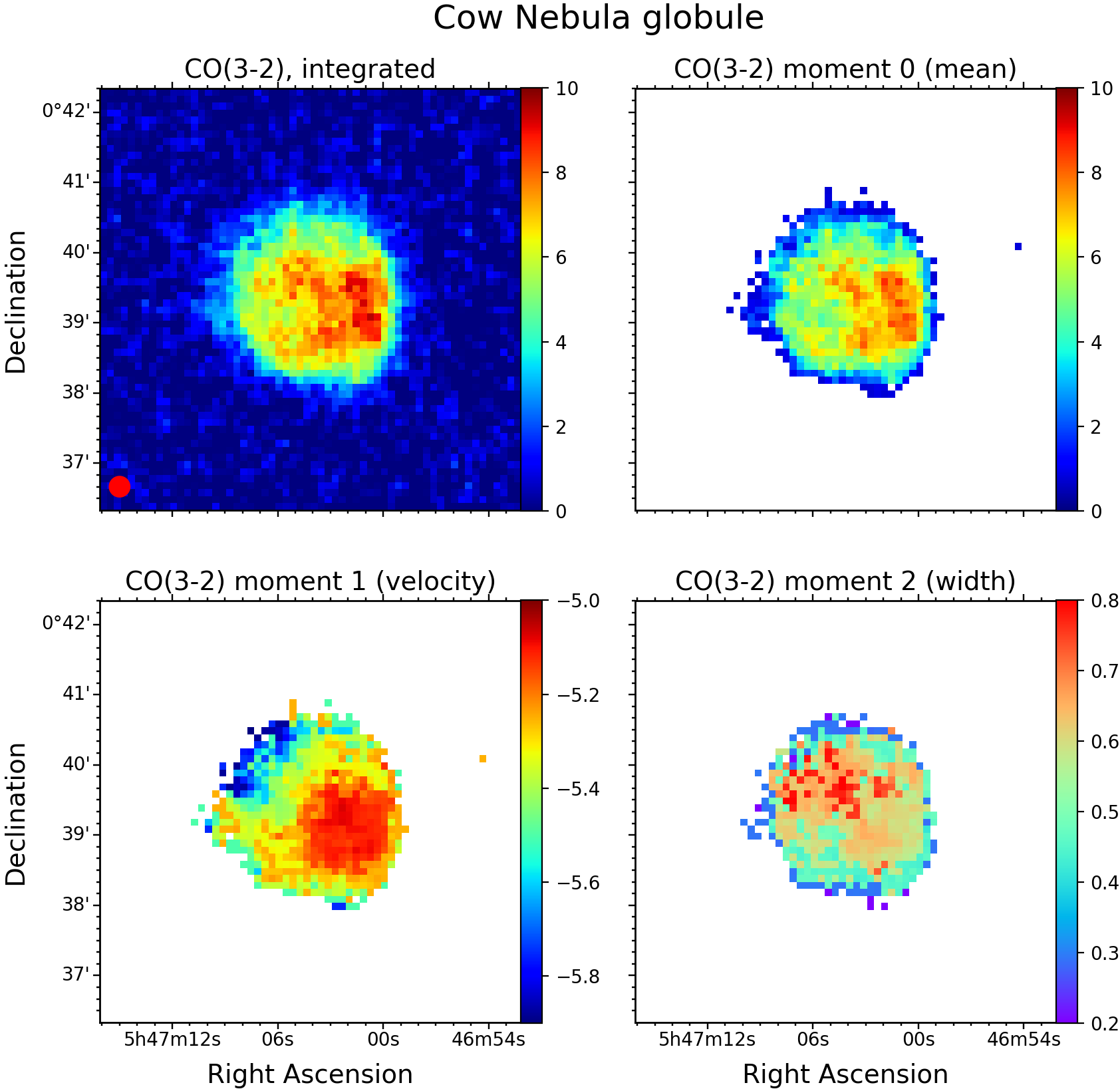}
   \caption{CO moment maps for the Cow~Nebula globule. Top left: emission integrated over the velocity interval $v = -6.25$~km/s to $-4.5$~km/s. The red dot indicates the beam size. Top right, bottom left, bottom right: moment~0, moment~1 (velocity), and moment~2 (line-width), respectively.}
              \label{fig:comcloud_moments}%
\end{figure}

In this section, we provide a general description of the properties of the $^{12}$CO(3--2) emission
in the survey area. The main scientific goal of the survey, a characterization of the population of
protostellar molecular outflows, is deferred to a companion paper. We here present the legacy data for community
use. To illustrate the potential use, we show
examples of outflow maps, to be compared with previous results.

In the following subsections, we discuss integrated intensity maps, position velocity cuts, and moment maps
for each survey field separately, and a few remarkable sub-regions.
In order to obtain a first impression on the properties of the CO emission, we show for each survey
field spectra integrated over the entire area in Fig.~\ref{fig:meanspectra}, and
Table~\ref{table:integratedlines} gives the line parameters as derived from Gaussian fits to those integrated spectra.
For L~1622 \citep[cf.][]{kunetal2008} we notice the presence of two well separated velocity components, which we refer
to as the 1~km/s component and the 10~km/s component. In Ori~B9 we also see two distinct
spectral components, which we refer to as the 3~km/s and 10~km/s components. The NGC~2068/NGC~2071 and NGC~2023/NGC~2024 fields
are dominated by a single, broad component, with a faint secondary component seen at lower (blue-shifted) velocities in the
NGC~2023/NGC~2024 fields. The L~1641-S spectrum shows two closely spaced, narrow components, which we refer
to as the 3~km/s component and the 6~km/s component.

\subsection{L~1622}
\label{section:L1622}

L~1622 has been described as a cometary cloud \citep[e.g.][]{reipurthetal2008}. Its overall shape is suggestive of being subject to radiation or wind
from the south-west. We see two well separated velocity components in its integrated spectrum in Fig.~\ref{fig:meanspectra}.
Consistent with previous observations, we see a clear cometary shaped cloud at velocities around $\sim$1~km/s
(Fig.~\ref{fig:intmap_L1622}, top), which corresponds to the actual L~1622 cloud. The emission around 10~km/s is much more
diffuse (Fig.~\ref{fig:intmap_L1622}, bottom),
and may trace the outskirts of the L~1617 complex of cometary clouds, located about 1~degree to the north-west
\citep[e.g.][]{reipurthetal2008}. The velocity maps in Fig.~\ref{fig:momentmaps_L1622} indicate an overall rather smooth velocity
field, with indications for a velocity gradient along the clouds minor axis. Together with the relatively small line width
(less than 1.5~km/s over the bulk of the cloud, larger line widths can mostly be associated with outflows),
the agent responsible for the cometary appearance is likely not an effective trigger of turbulence within the cloud.

Figure~\ref{fig:L1622_pv} shows position-velocity cuts through the cloud; the top panel shows a cut along the cloud's
major axis. It confirms
the overall fairly quiet nature of L~1622, as the emission is seen with very little variation in velocity from head to tail,
except for offsets between 400\arcsec{} and 850\arcsec{}. Between $\sim$450 and 850\arcsec{},
the cut intersects two molecular outflows, driven by \object{HOPS~1} and \object{HOPS~354} \citep[the latter outflow is seen at optical
and infrared wavelengths as \object{HH~962},][]{ballyetal2009}. At an offset around 400\arcsec{}, there is faint emission extending over
1-2~km/s beyond the ambient clouds velocity, which we attribute to a faint outflow originating from the \object{HBC~515} multiple
pre-main sequence system \citep{reipurthetal2010}. We note an emission gap, with only faint emission, between the position of
\object{HBC~515} and the onset of the bright, redshifted emission from the \object{HOPS~1} outflow.

Figure~\ref{fig:L1622_pv} also shows four position-velocity cuts through L~1622 along the cloud's minor axis in its middle and bottom rows.
The cuts in the middle row are in the head of the cloud, while the cuts in the bottom row are taken further west, in the tail.
The cuts through the head indeed seem to show an overall velocity gradient, going from more blue-shifted velocities in the south-east
to more red-shifted in the north-west. The cut in the middle-left panel intersects the position of \object{HBC~515} (marked with
the dashed line) and again shows faint emission extending beyond the ambient cloud velocity. The cut in the middle-right panel
intersects the emission gap seen in the cut along the cloud's major axis, and shows up as small gap in this direction as well.
Together, this gap suggests the presence of a small cavity to the north-east of \object{HBC~515}, possibly created by outflow from
this system and corresponding to the cavity seen extending to the south-west of \object{HBC~515} in optical images
\citep[e.g.][]{reipurthetal2008, ballyetal2009}.

The cut in the bottom-left panel of Fig.~\ref{fig:L1622_pv} intersects a bright CO emission feature, which stands out in all three
moment maps as bright (moment 0), blueshifted (moment 1) and broad (moment 2), and shows up as high-velocity emission in the
position-velocity cut. It is located near the position of the bright, optical shock system of HH~122
\citep{reipurthmadsen1989, ballyetal2009} and might indicate the terminating shock of the molecular outflow driven by \object{HOPS~1}.
In contrast to the cuts through the head of the cloud, the cuts through the tail do not show a pronounced, systematic velocity
gradient, possibly indicating that the agent creating the gradient is more efficient at the head of the cloud (e.g. uneven
exposure to wind or radiation, between the south-eastern and north-western extents of the cloud).
 
\subsection{The NGC~2068/NGC~2071 area}
\label{section:N2068N2071}

Figure~\ref{fig:intmap_N2068N2071} shows the CO emission in the NGC~2068/NGC~2071 fields integrated over the velocity interval from
7.5~km/s to 13~km/s, and Fig.~\ref{fig:N2068N2071_overview} shows an overview of the region, comparing optical, CO, and dust emission.
The brightest CO emission broadly corresponds to the \object{NGC~2068} nebula, with the very brightest region being seen at the
interface between the optical nebulosity and the dense filament seen in dust emission in the southern part of NGC~2068
\citep[e.g.][see Fig.~\ref{fig:N2068N2071_overview}]{motteetal2001, kirketal2016}. Further bright
emission is seen towards the NGC~2071-IR region (while the area of the brightest optical emission in \object{NGC~2071} corresponds
to a local minimum in CO emission) and in the \object{HH~24}-\object{HH~26} region. 

In all three moment maps
(Fig.~\ref{fig:momentmaps_N2068N2071}) the NGC~2071-IR outflow (see Sect.~\ref{section:N2071flow} below) is the most outstanding
feature, but a few additional outflows in the HH~26 and NGC~2068 region are also readily visible. There is an overall north-south velocity
gradient, with the northernmost parts of the area being more blueshifted. The area around the NGC~2071 nebula stands out as
having consistently large ($> 3$~km/s) line widths, but also the HH~26 and NGC~2068 region features line widths greater than
2~km/s over wide areas.

Figure~\ref{fig:N2068N2071_pv} shows position-velocity cuts through  the NGC~2068/NGC2071 area. The top panel shows a cut going largely from
south to north, while the remaining panels show east-west cuts (at fixed declination). The south-to-north cut shows the overall
velocity gradient already noted in the moment~1 map. At offsets around 400\arcsec{} and 1000\arcsec{} the cut intersects the
HH~24 to HH~26 region and the dense ridge of cores in the southern part of NGC~2068,
respectively, which both display strong molecular outflow activity. The brightest CO emission is seen in NGC~2068 as a single
velocity component. Starting at an offset of $\sim$2200\arcsec{} the CO emission splits into two components, roughly coinciding
with the appearance of the NGC~2071 optical nebulosity. At around 2500\arcsec{}, the cut intersects a dust ridge extending
east from the NGC~2071-IR core that is associated with several protostars, where again high-velocity CO due to protostellar outflow
activity is present. Further north, at offsets greater than 3000\arcsec{}, only one velocity component is seen.

The east-west cuts in Fig.~\ref{fig:N2068N2071_pv} are ordered from north to south. 
The panels in the second row show cuts going through NGC~2071-IR (left panel) and the brightest part (at optical wavelengths)
of the NGC~2071 nebula. Both cuts show CO to be split into two components over a large fraction of the cut. The cut going through
the brightest part of the optical nebula in particular shows a dominant, single component at both ends, splitting up when going
further towards the middle of the cut. The northernmost cut through NGC~2071-IR shows very bright emission extending to high
velocities due to the NGC~2071-IR molecular outflow (Sect.~\ref{section:N2071flow}).
The cuts in the third row and fourth row, left, go through the NGC~2068 optical nebulosity. The northernmost (third row, left)
intersects the brightest optical emission, the second (third row, right) goes through the brightest CO emission (just north of the
bright dust continuum ridge), while the third cuts through the bright dust ridge. In all three cuts the brightest emission is seen
as a single component, while the fainter emission (particularly to the west) tends to show double-peaked lines, typically at offsets
that correspond to the location of the dark (in the optical) ridge marking the northern and western boundary of the NGC~2068
nebula and the fainter optical emission nebulosities further west. While we can't exclude CO self-absorption as the cause of the
double-peaked line profiles seen in several areas in NGC~2071 and NGC~2068, its association with optical nebulosities suggests that
the double-peaked lines are due to expanding bubbles in the molecular gas, driven by the early-type stars illuminating the
NGC~2068 and NGC~2071 nebulae.


Finally, the southernmost cut (Fig.~\ref{fig:N2068N2071_pv}, bottom row, right panel) intersects the \object{HH~24} region,
known to be an active site of an outflow, and indeed shows clear high velocity emission in the middle of the cut. Perhaps more
remarkable is the sharp transition from a narrow line profile to the west to a broad profile to the east of the HH~24/HH~26 dust
ridge. Indeed, examining the moment~2 map (line width) more closely, the entire HH~24-HH~26 ridge is located in a transition
zone between linewidths hardly exceeding 1~km/s to its south and west, and linewidths mostly greater than 2~km/s to its north
and east. We speculate that this transition in velocity dispersion might be related to the peculiar overabundance
of particularly young protostars (PBRs) in this area \citep{stutzetal2013}.

\subsubsection{The \object{NGC~2071-IR} outflow}
\label{section:N2071flow}

The outflow driven by the intermediate-mass protostar NGC~2071-IR \citep{walthergeballe2019} was among the first to be detected
in high-velocity CO emission \citep{bally1981, lichten1982, snelletal1984}. Subsequent, deeper observations showed that the
outflow extends to much higher velocities than originally seen \citep{margulissnell1989, choietal1993}, with a full velocity extent of
over 100~km/s ($-$50 to $+$65). \cite{cherninmasson1992} show that the extremely high velocity (EHV) component of the flow,
particularly in the red lobe, clearly stands out as a separate bump in the spectrum and is in the form of compact blobs of emission.
\cite{cherninwelch1995} interpreted the red EHV feature as originating from a jet interacting with a dense clump in its way.

Our ALCOHOLS survey data for this outflow are presented in Fig.~\ref{fig:N2071flow}. We show red- and blue-shifted CO emission
at low outflow velocities, together with emission at rest velocities around 10~km/s, in the leftmost panel, reproducing
the overall shape and extent of the outflow as seen before \citep{snelletal1984, moriartyschievenetal1989, cherninmasson1992},
revealing additional outflowing CO gas at fainter intensity levels \citep[see also][]{buckleetal2010}. At higher velocities
we recover the well collimated outflow shape seen also in \cite{moriartyschievenetal1989} and \cite{cherninmasson1992}. Finally,
at extreme velocities (Fig.~\ref{fig:N2071flow}, rightmost two panels) we see the emission only from small, compact regions
relatively close to the protostar, with particularly the red EHV component also clearly standing out in the spectrum.
Consistent with earlier observations \citep{cherninmasson1992}, the red component is more compact, while the blue component
extends over a larger range along the outflow axis. The velocity extent we see in our data, from about $-$50~kms in the blue
lobe, to about $+$65~km/s in the red lobe, also agrees well with previous, targeted observations. Overall, this agreement confirms the quality
of our data reduction procedure, particularly with respect to spectral baseline removal.

\subsubsection{The Cow~Nebula globule}
\label{section:comcloud}

Figure~\ref{fig:comcloud_maps} shows the CO(3-2) emission of a remarkable cloud in the northern part of the NGC~2071 survey field
(see Fig.~\ref{fig:N2068N2071_overview}) with an almost perfectly circular appearance.
Its morphology is highly suggestive of a spherical cloud, and might serve well as an observational testbed for numerical and
theoretical considerations. Due to its simplicity, we refer to it as the "Cow~Nebula" globule.

The globule is well separated in velocity from the Orion~B cloud, being centred at $v = -5.25$~km/s (while the bulk of the
cloud in this area is centred at around $+8$~km/s.
The top-left panel in Fig.~\ref{fig:comcloud_maps} shows the CO(3-2) emission towards the Cow~Nebula globule, integrated over its
full velocity extent ($v = -6.25$~km/s to $-4.5$~km/s). The circle serves to illustrate the circular shape of the globule. It is
centred at 5$^h$47$^m$04.$\!\!^s$3,  $+$00$^\circ$39\arcmin20\arcsec{}, which we take as the position of the globule, and has
a radius of 85\arcsec{}. Assuming that the globule is at the same distance as the Orion~B clouds ($\sim 400$~pc), this extent corresponds
to a radius of 34\,000~AU or 0.16~pc.

The top-right panel shows the CO integrated emission overlaid as contours on a DSS2-blue optical image. A small, cometary
feature is clearly associated with the CO emission. At optical wavelengths, the 'head' of the feature points towards the south-west
(towards the Orion Belt stars), and its rim aligns perfectly with the CO emission. Apart from being mentioned in the
compilations of (reflection) nebulae
by \cite{dorschnerguertler1963} (\object{DG 82}) and \cite{bernes1977} (\object{$[$B77$]$ 100}) \citep[see also][]{magakian2003} this
object has apparently not yet been studied in any detail. A diffuse tail extends towards the north-east, and is seen to extend further
in the optical images than in CO. The star seen in projection against the clouds southern edge (\object{HD~290857})
seems to be an unrelated foreground object.

The panel to the bottom-left shows the CO map overlaid on a Spitzer IRAC 8~$\mu$m map \citep{megeathetal2012}. Here, a sharp bright rim
is seen outlining the western edge of the CO globule. Finally, the panel to the bottom-right shows the CO map overlaid on a JCMT 850~$\mu$m
(dust continuum) map \citep{kirketal2016}. We see a faint, diffuse ridge of emission at the western edge of the globule, coinciding
broadly with
the area of the brightest CO emission. We do not see any compact continuum source that could indicate the presence of an embedded
protostar in the globule or even a dense, pre-stellar core. We also do not see any HOPS protostars within or close to the boundary of
the CO emission.

In Fig.~\ref{fig:comcloud_moments} we show moment maps for the Cow~Nebula globule. The radial velocity is fairly constant, around
$-$5.2~km/s, over the south-western part of the cloud, that is, the head. In the tail towards the north-east, we observe a shift to more
blue-shifted velocities. We only see a marginal trend in the line widths, which are of the order of 0.6--0.65~km/s in the south-western
half of the globule, and slightly broader (0.7--0.75~km/s) in the north-eastern half of the globule.

Taken together the data suggest that the Cow~Nebula Globule, despite showing a cometary structure in the optical, is intrinsically
a round (possibly spherical, in 3D) starless cloud. We interpret the steeper rise in CO emission and the enhanced brightness seen around
the western edge of the globule as being due to heating from the west (as indicated by the bright rim seen in the
Spitzer image outlining the western globule edge) rather than being due to larger column density in the western half of the globule.
Clearly, subsequent multiline observations are needed to determine better the column density structure of the globule, and
higher spectral resolution will be required to determine its dynamical state.

\subsection{Ori~B9}
\label{section:ORIB9}

\begin{figure}
   \centering
   \includegraphics[width=\columnwidth]{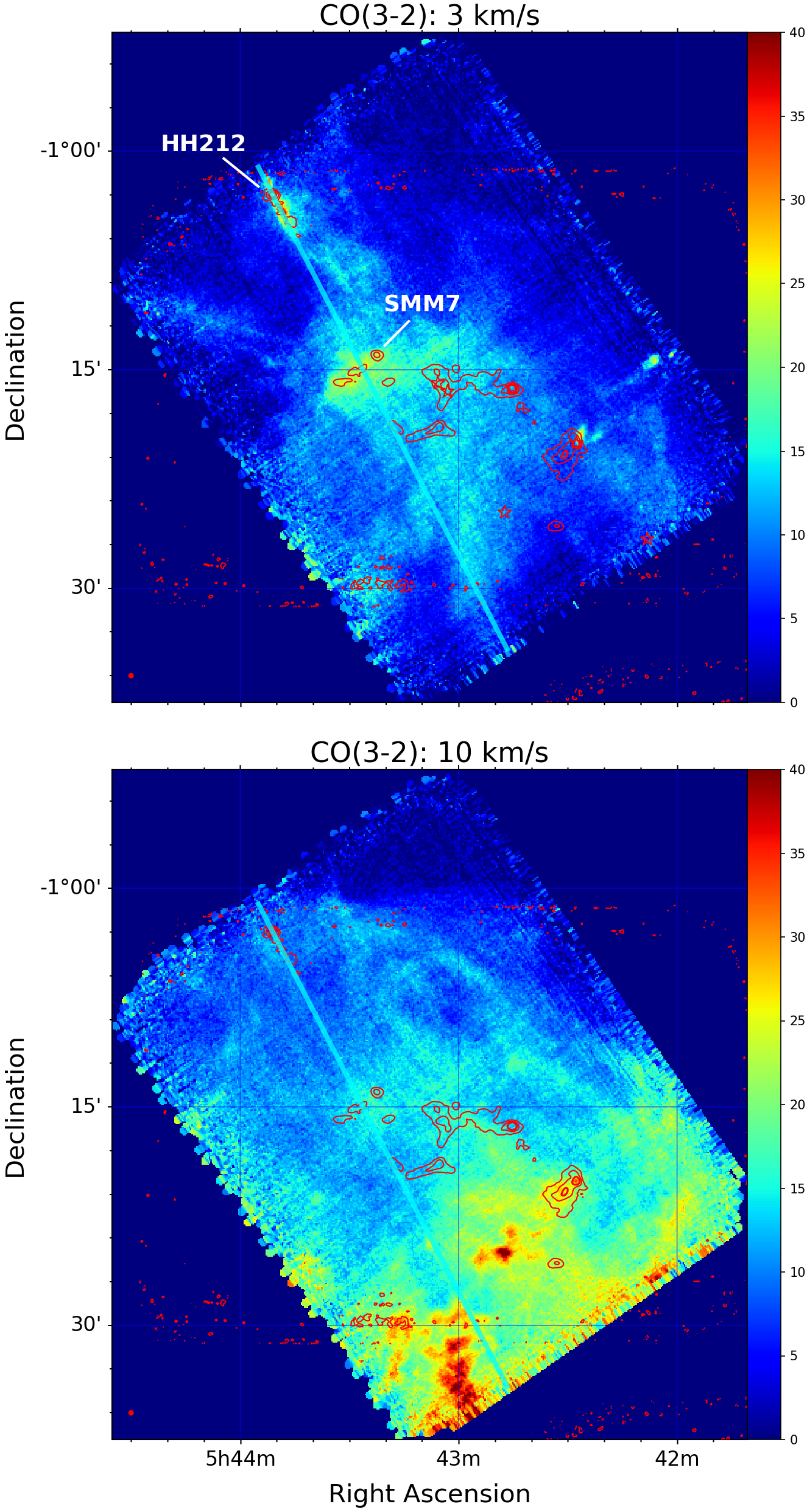}
   \caption{CO emission in the Ori~B9 field, integrated from 0.5~km/s to 5.5~km/s (top, in K~km/s) and from
            8~km/s to 12~km/s (bottom). Contours mark APEX/Laboca 870~$\mu$m dust continuum. The blue line marks the location of the
            position-velocity cut shown in Fig.~\ref{fig:ORIB9_pv}.  The red dots in the bottom left corners indicate the beam size
            (which is the same for the ALCOHOLS and Laboca data).
            }
              \label{fig:intmap_ORIB9}%
\end{figure}

\begin{figure}
   \centering
   \includegraphics[width=\columnwidth]{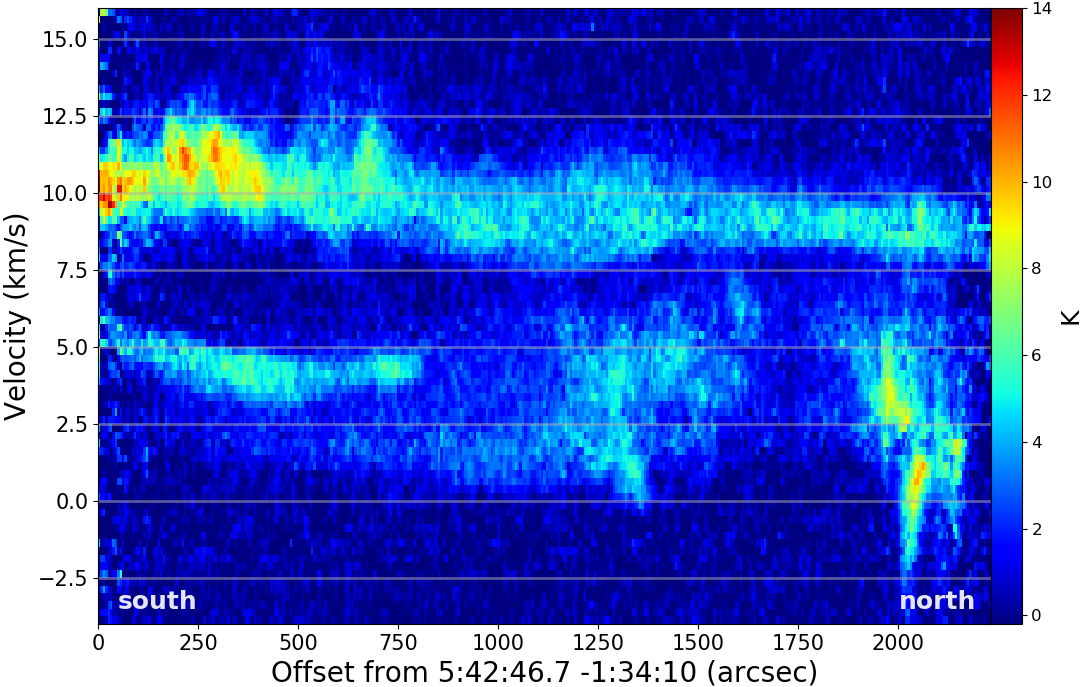}
   \caption{Position-velocity cut through the Ori~B9 field, starting at 5$^h$42$^m$46.$\!\!^s$7, $-$1$^\circ$34\arcmin10\arcsec{}
   (south) and ending at 5$^h$43$^m$55.$\!\!^s$0, $-$1$^\circ$01\arcmin05\arcsec{} (north).
   }
              \label{fig:ORIB9_pv}%
\end{figure}

The Ori~B9 region shows two distinct velocity components, centred at around 3~km/s and 10~km/s. We show integrated intensity maps
for both components in Fig.~\ref{fig:intmap_ORIB9}, and also moment maps separately for the two components in 
Figs.~\ref{fig:momentmaps_ORIB9_2kms} and \ref{fig:momentmaps_ORIB9_10kms}. A largely south-to-north position-velocity cut for the field
is shown in Fig.~\ref{fig:ORIB9_pv}, with the velocity range covering both components.

The component at velocities around 3~km/s has a patchy, incoherent appearance in the moment~1 map (Fig.~\ref{fig:momentmaps_ORIB9_2kms}),
with an overall filamentary spatial structure. This behaviour is also reflected in the position-velocity cut in Fig.~\ref{fig:ORIB9_pv}, where
the 3~km/s component is visible between $\sim$1~km/s and $\sim$7~km/s, with some regions of brighter, narrower emission embedded in
a broad, faint, diffuse background. The presence of CO emission at velocities lower than (i.e. blue-shifted)
the main cloud emission at around 10~km/s has been noted by \cite{wilsonetal2005}, and \cite{miettinenetal2009} noted the presence
of N$_2$H$^+$ emission from this component as well. The cloud core harbouring the protostellar driving source of the \object{HH~212}
jet seems to be associated with the 3~km/s component (i.e. the NH$_3$ core associated with \object{IRAS~05413-0104}: \cite{harjuetal1993}),
with mainly blue-shifted high-velocity CO due to this outflow seen at the northern end of the position-velocity cut. We also note
some patches of faint emission at around 5~km/s in the position-velocity cuts in the NGC~2023/NGC~2024 region
(Fig.~\ref{fig:N2023N2024_pv}) to the south-west of Ori~B9. Together with the presence of the blue shoulder in the integrated
spectrum of the NGC~2023/NGC~2024 field, this emission indicates that the 3~km/s component extends further the south-west, into
the NGC~2023/NGC~2024 region.

The CO component at velocities around 10~km/s is seen at similar velocities as the bulk CO emission in the NGC~2023/NGC~2024 region
located to the south-west of Ori~B9. It seems to trace the north-eastern outskirts of that region, albeit it appears separated in
$^{13}$CO from the NGC~2024 cloud, see \cite{casellimyers1995}. We see a north-south velocity gradient, with more blue-shifted
CO found towards the north. Along with the shift in radial velocity we see a trend for decreasing linewidth further north.

The population of dust continuum cores in the Ori~B9 region has been studied by \cite{miettinenetal2009}, implicitly assuming that
all dust sources are associated with the main cloud component at velocities around 10~km/s. While no detection of CS emission is
reported in the survey by \cite{ladaetal1991}, the presence of dense, cold gas in the 3~km/s cloud is indicated by the detection
of N$_2$H$^+$ at these velocities \citep{miettinenetal2009}. Fig.~\ref{fig:intmap_ORIB9} shows 870~$\mu$m dust continuum emission
as observed with
APEX \citep[merging our own observations with those presented by][]{miettinenetal2009} overlaid as contours on the integrated CO 
intensity maps. We note that the compact source \object{SMM~7} of \cite{miettinenetal2009} and the ridge of fainter cores extending
to its south-east correspond well to the region of brightest integrated CO emission seen in the 3~km/s cloud component. We suggest
that these cores actually reside within the 3~km/s cloud, providing the second evidence of ongoing (or imminent) star
formation in that cloud in addition to the HH~212 system.

\begin{figure}
   \centering
   \includegraphics[width=\columnwidth]{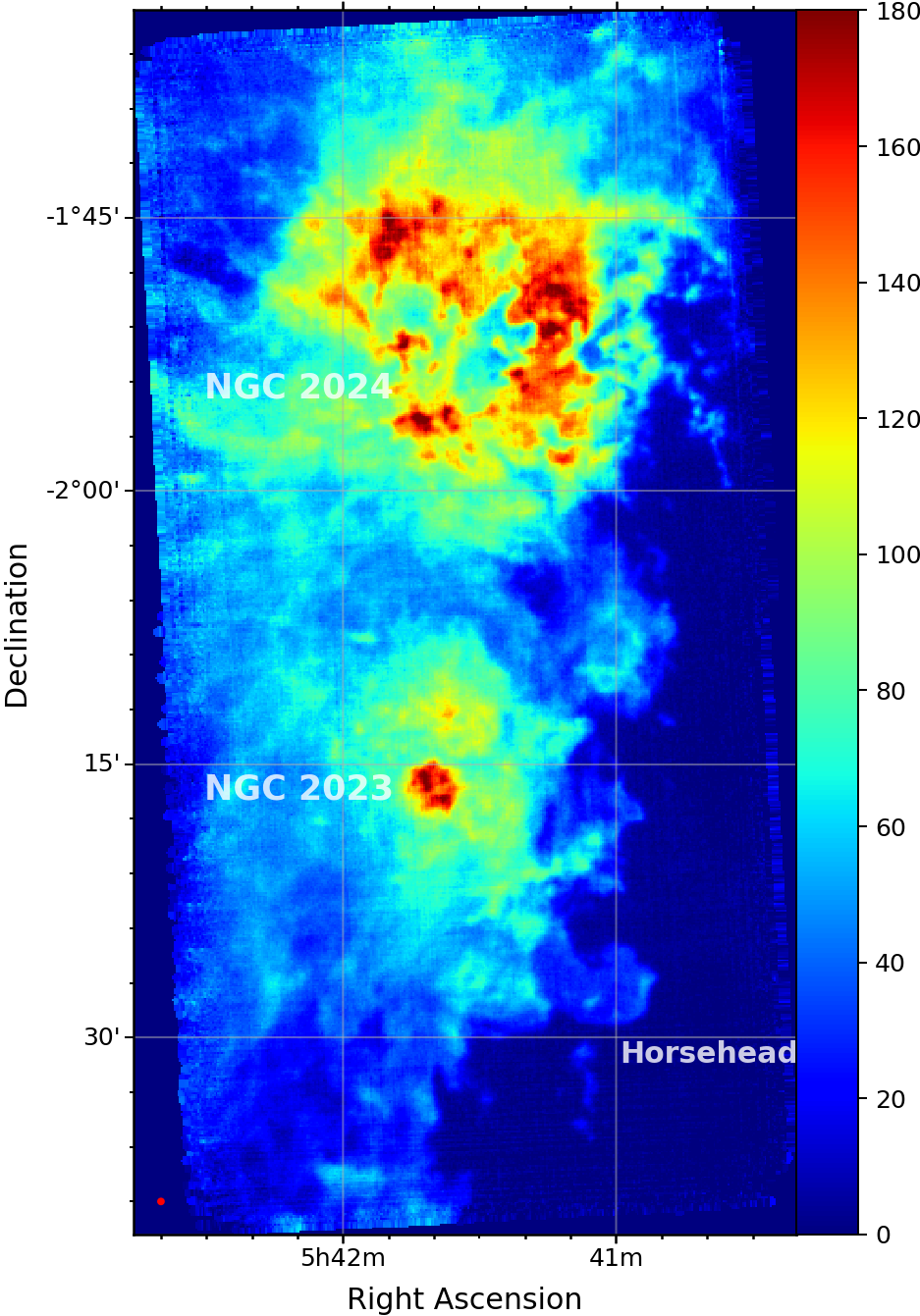}
   \caption{CO emission in the NGC~2023 and NGC~2024 field, integrated from 6.5~km/s to 14~km/s (in K~km/s)
        The red dot in the bottom left corner indicates the beam size.}
              \label{fig:intmap_N2023N2024}%
\end{figure}

\begin{figure}
   \centering
   \includegraphics[width=\columnwidth]{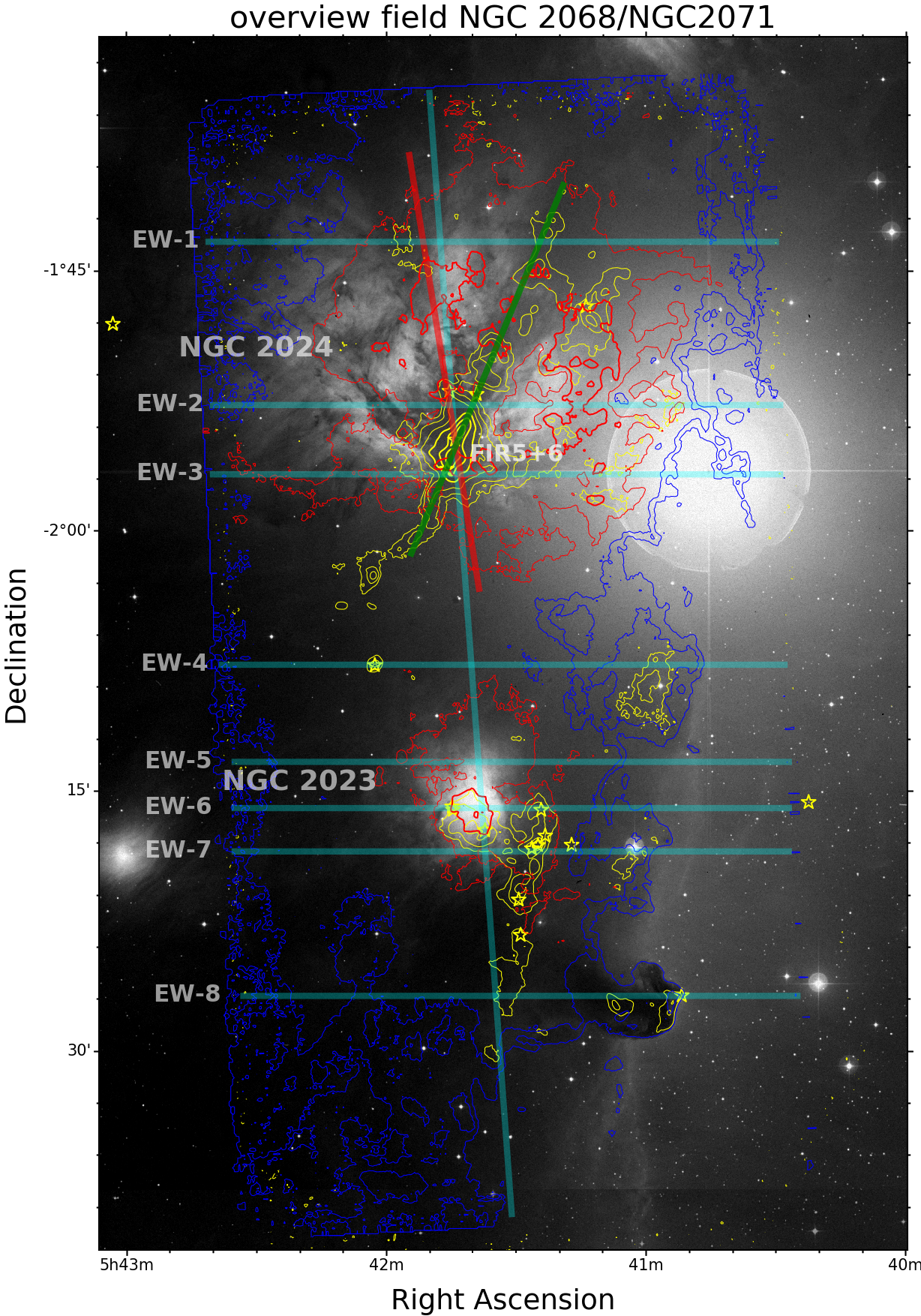}
   \caption{Overview of the NGC~2023/NGC~2024 region, showing a DSS2-blue optical image of the region, with the integrated CO emission
   from Fig.~\ref{fig:intmap_N2023N2024} overplotted as blue and red (for the brightest regions) contours, and APEX Laboca 870~$\mu$m
   dust continuum as yellow contours. Yellow stars mark the position of HOPS protostars, and cyan lines mark the location of the 
   position-velocity cuts shown in Fig.~\ref{fig:N2023N2024_pv}. The red line marks the cut along the NGC~2024 monopolar outflow axis
   (shown in Fig.~\ref{fig:N2024_flow_pv}), the green line marks the cut along the dark lane (shown in Fig.~\ref{fig:N2024_pv_darklane}).
   }
              \label{fig:N2023N2024_overview}%
\end{figure}

\begin{figure}
   \centering
   \includegraphics[width=\columnwidth]{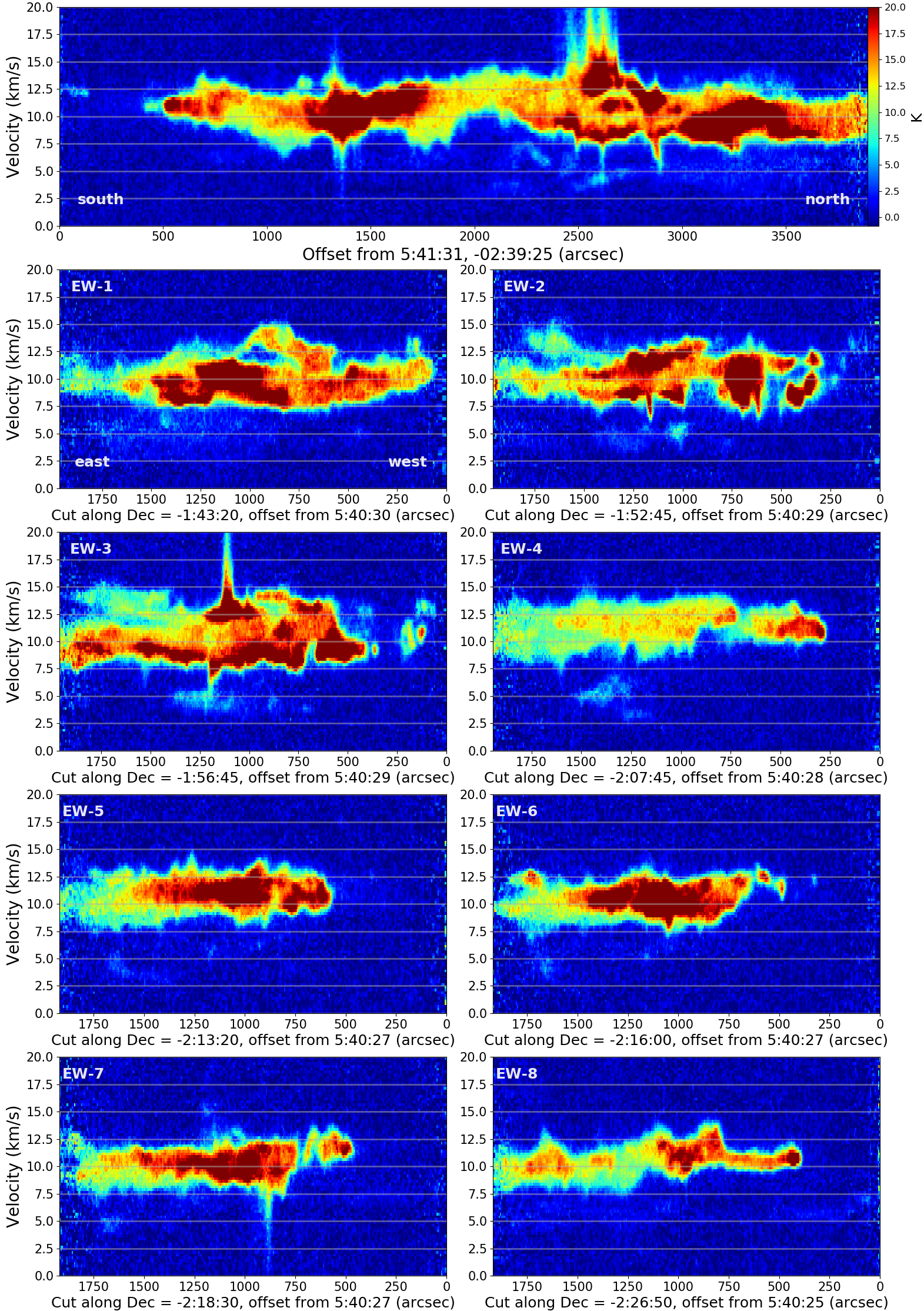}
   \caption{Position velocity cuts across the NGC~2023/NGC~2024 survey field. Top: cut starting from 5$^h$41$^m$31$^s$,
     $-$2$^\circ$39\arcmin25\arcsec{} (south), ending at 5$^h$41$^m$50$^s$, $-$01$^\circ$34\arcmin45\arcsec{} (north).
     The remaining cuts go from east to west.
   }
              \label{fig:N2023N2024_pv}%
\end{figure}

\subsection{NGC~2023, NGC~2024, and Horsehead Nebula}
\label{section:N2024}

Figure~\ref{fig:intmap_N2023N2024} shows the CO emission in the NGC~2023/NGC~2024 fields integrated over the velocity interval from
6.5~km/s to 14~km/s, and Fig,~\ref{fig:N2023N2024_overview} shows an overview of the region, comparing optical, CO, and dust emission.
Our map reproduces the main features seen in the CO(3-2) map of \cite{krameretal1996}, albeit at better angular resolution.
The area around the \object{NGC~2024} \ion{H}{II} region is seen prominently in CO emission. The CO emission in the northern and
north-western part of the nebula is
suggestive of being shaped by the early-type stars in the NGC~2024 cluster, showing cometary features with their
heads pointing towards the centre of the nebula. Overall, the nebula also stands out in the moment~2 map
(Fig.~\ref{fig:momentmaps_N2023N2024}) as a region of large line widths.
\object{NGC~2023} shows bright CO over the same extent as the bright optical reflection nebula, but is much smaller than NGC~2024,
and hardly stands out in the moment~1 and 2 maps.

As for the overall velocity field, the moment~1 maps show a prevalence for more blue-shifted lines in and around the two optical
nebulae, with more red-shifted emission between them. We also notice a trend for red-shifted emission to prevail at the western
edge of the cloud, where it is exposed to the intense radiation from \object{$\sigma$~Ori}. 

Figure~\ref{fig:N2023N2024_pv} shows position-velocity cuts through the NGC~2023/NGC~2024 region. The top panel shows a cut going largely from
south to north, intersecting the brightest part of the NGC~2023 nebula, NGC~2024~FIR5, and the NGC~2024 \ion{H}{II} region. The
remaining panels all show cuts going east-to-west through the cloud.

The south-to-north cut shows the large scale pattern noticed
also from the moment~1 map, that is, more blue-shifted emission in the NGC~2023/NGC~2024 region, with more red-shifted emission
in the very south of the region and between the two nebulae. Only one bright emission component is seen towards NGC~2023, while
two or even three velocity components are seen towards NGC~2024.

The eight east-west cuts shown in Fig.~\ref{fig:N2023N2024_pv}, EW-1 through EW-8, are ordered from north to south. The first three, EW-1 to EW-3, go through the NGC~2024 \ion{H}{II}
region. They all show a very wide overall velocity extent, with two components (locally even three) seen  over much of the
second and third cut. The southernmost cut of the three intersects NGC~2024~FIR5, showing strong high-velocity emission. In all three cuts,
the emission at the western edge appears systematically shifted towards lower (red-shifted) velocities. 

The fourth east-west cut, EW-4, is located between the NGC~2024 and NGC~2023 nebulae and shows fainter emission mostly in a single
line component. The final three cuts intersect the NGC~2023 reflection nebula. All three show bright CO emission in the
area of the optical nebula, but only in one velocity component. The southernmost cut shows high velocity emission associated with the
outflow from \object{NGC~2023~MM1} \citep{sandelletal1999}. The emission closest to the western cloud edge in EW-6 and EW-7 again is
seen to be slightly more red-shifted than the CO further east. This trend is also seen in EW-8, going through the
Horsehead Nebula (\object{B33}), showing the entire 'head' and 'neck' to be redshifted compared to the emission in the western
part of the cloud (for a more detailed discussion of the CO and $[$\ion{C}{I}$]$ kinematics of the Horsehead region see \cite{ballyetal2018}).

Taken together, the ALCOHOLS data show NGC~2023/NGC~2024 to be the most dynamic region studied in the survey.
The dynamic nature of the emission can also clearly be seen when scrolling through channel maps of the line data cube.
We clearly see the strong impact of the NGC~2024 \ion{H}{II} region shaping the morphology of the surrounding gas and boosting the
line width in the area. The prevalent double peaked line there suggests the presence of an expanding cavity (see Sect.~\ref{section:N2024HII}
below). NGC~2023 has a much smaller
impact on the cloud. While still creating bright line emission (presumably by heating the gas) it is neither associated with
double-peaked line profiles indicative of an expanding bubble, nor does it significantly enhance the velocity dispersion in its
environment. The western edge of the cloud is systematically more red-shifted compared to the bulk of the cloud, presumably due to
the influence of $\sigma$~Ori to its west, which creates the IC~434 ionization front at the western surface of the cloud.  

\begin{figure}
   \centering
   \includegraphics[width=\columnwidth]{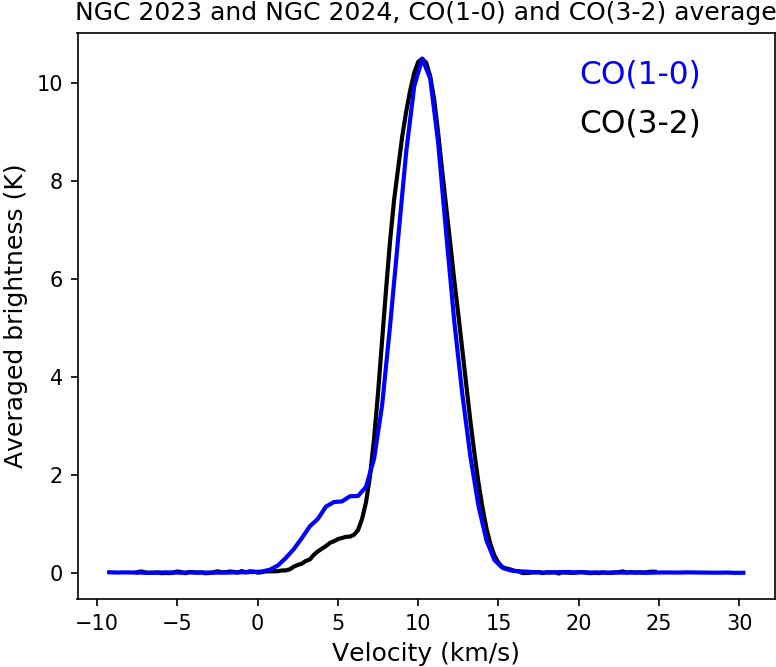}
   \caption{CO(1-0) spectrum (blue), averaged over the survey field covered by \cite{petyetal2017}
     (see their Fig.~4), and CO(3-2) spectrum,
     averaged over our NGC~2023/NGC~2024 survey field (black), with the CO(1-0) spectrum renormalized to the maximum of the CO(3-2) spectrum.
   }
              \label{fig:N2024_avespectrum}%
\end{figure}

\begin{figure}
   \centering
   \includegraphics[width=\columnwidth]{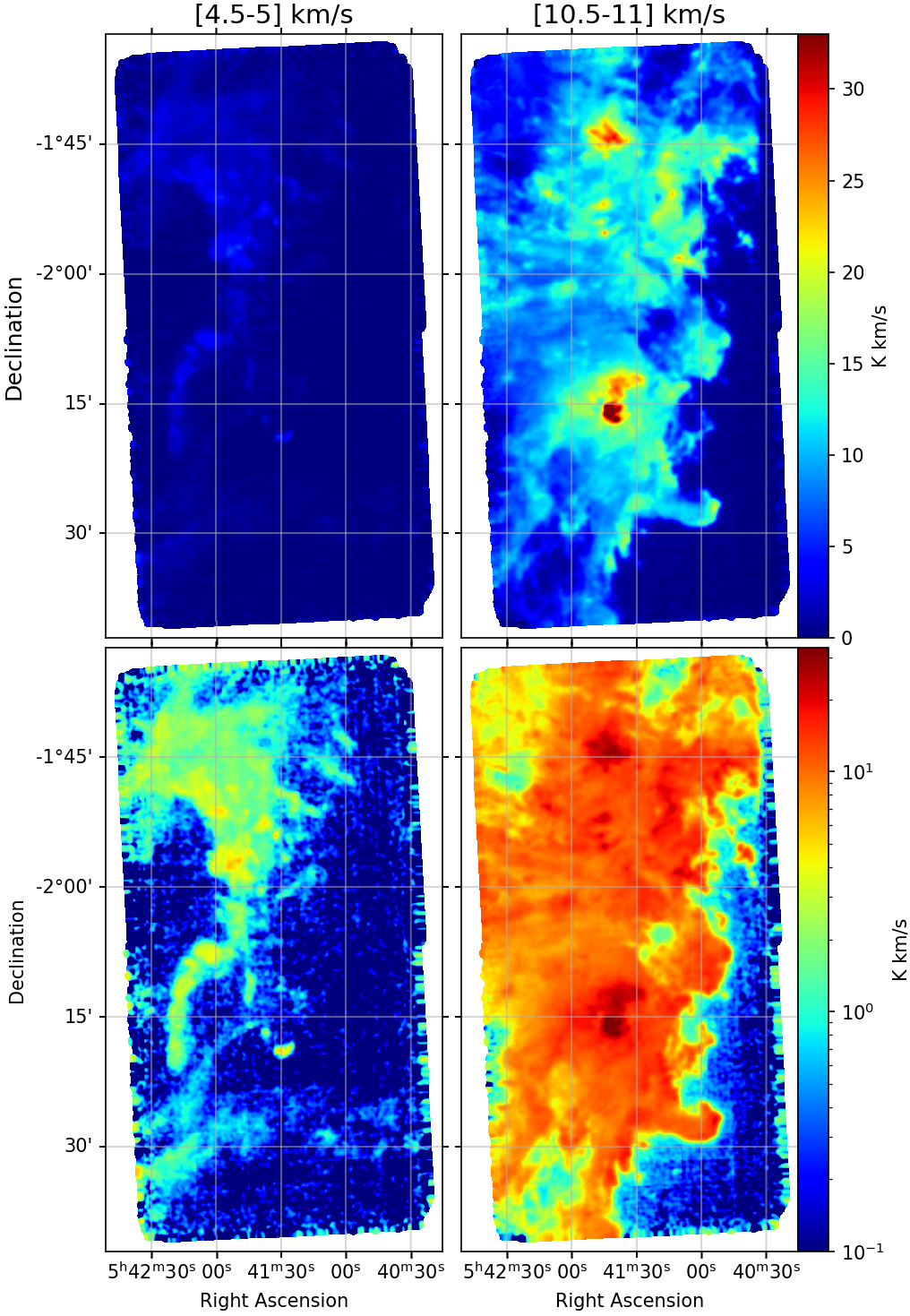}
   \caption{CO(3-2) maps over the NGC~2023/NGC~2024 survey field, integrated over the same velocity ranges (covering 0.5~km/s around the
       peaks of the two separate velocity components) and convolved to the same
     31\arcsec{} beam as in Fig.~5 of
     \cite{petyetal2017}, in linear (top) and logarithmic (bottom) intensity scaling.
   }
              \label{fig:N2024_5kms_11kms}%
\end{figure}

\begin{figure*}
   \centering
   \includegraphics[width=14cm]{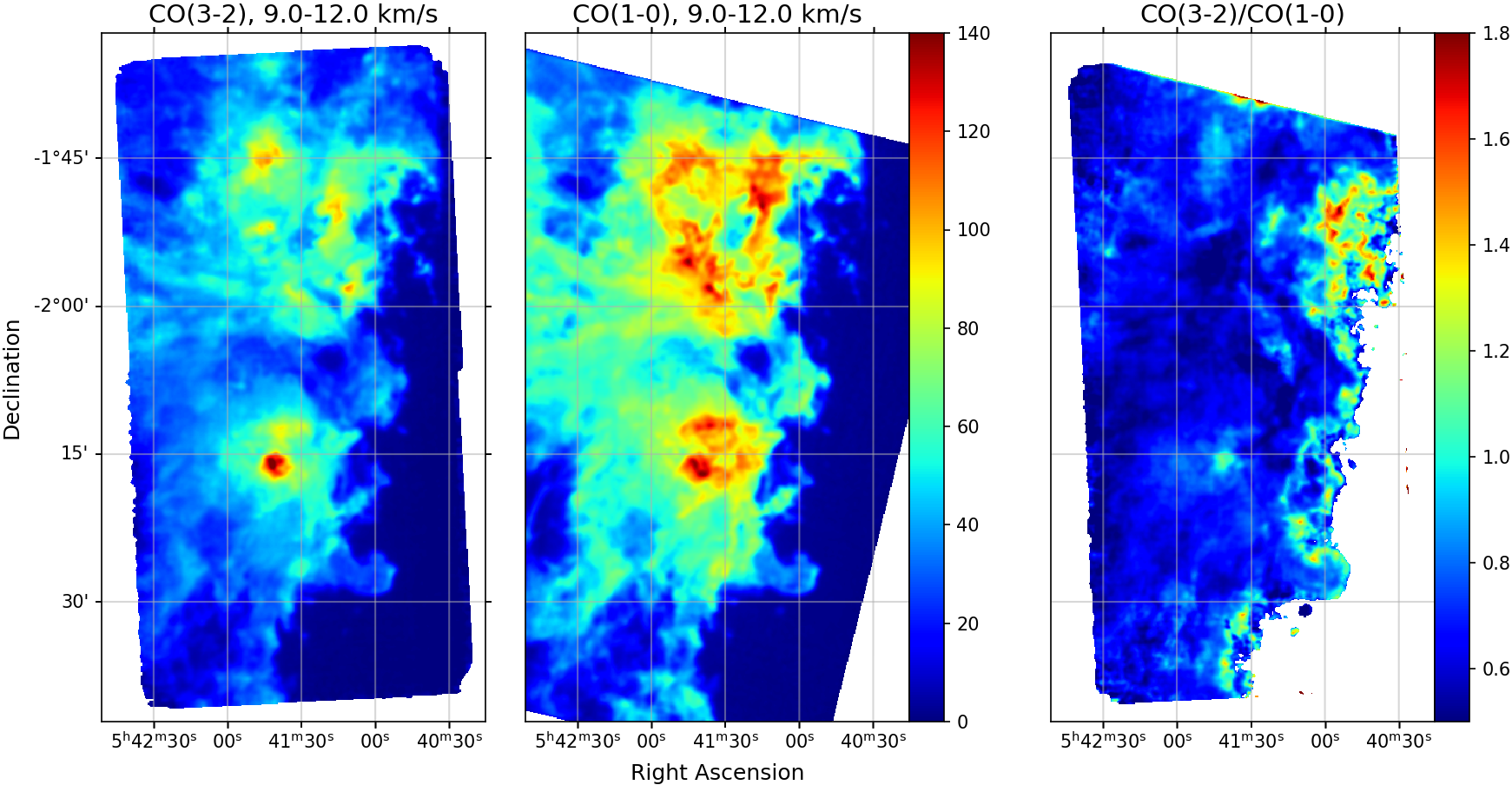}
   \caption{CO(3-2) map over the NGC~2023/NGC~2024 survey field, integrated over the same velocity range (9-12~km/s) and convolved to
     the same 31\arcsec{} beam as in Fig.~2 of
     \cite{petyetal2017}, along with their CO(1-0) velocity integrated map, and the ratio of the CO(3-2)/CO(1-0) maps (where the ratio
     was computed for all pixels with a CO(3-2) line brightness temperature greater than 2~K, i.e. excluding pixels with noisy
     ratios due to low signal-to-noise ratios).
   }
              \label{fig:N2024_co32_co10}%
\end{figure*}

\begin{figure}
   \centering
   \includegraphics[width=0.9\columnwidth]{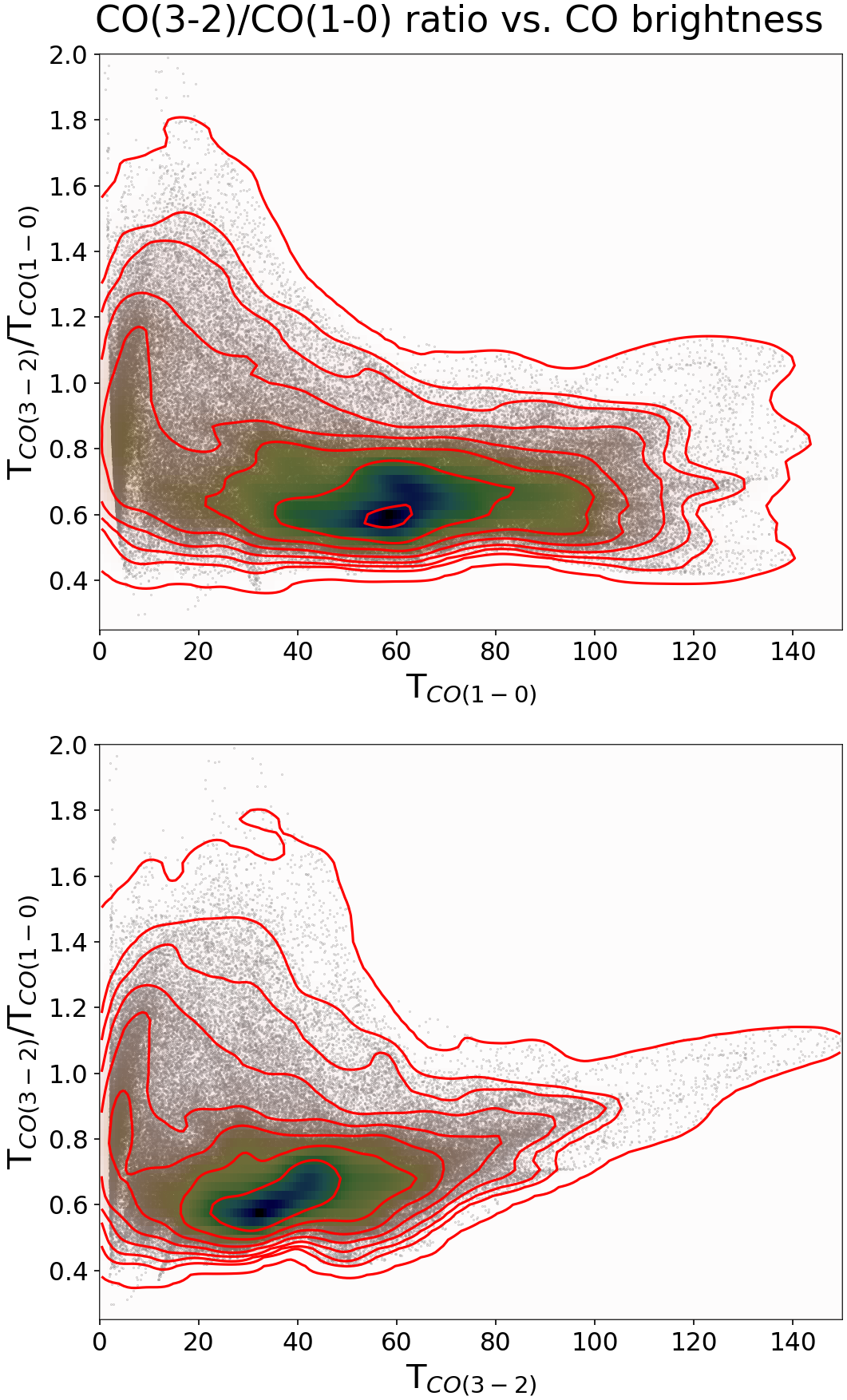}
   \caption{CO(3-2)/CO(1-0) line ratios in the  NGC~2023/NGC~2024 survey field as a function of CO line brightness.
   }
              \label{fig:N2024_ratioCO_TCO}%
\end{figure}

\subsection{CO(1-0) versus CO(3-2) in NGC~2023/NGC~2024}
\label{section:N2024_CO10_CO32}

The fields covered by our CO(3-2) survey overlap significantly with the area covered by the IRAM~30m 100~GHz spectral mapping survey.
We provide here a comparison between our data and its CO(1-0) data as presented in \cite{petyetal2017}.

Figure~\ref{fig:N2024_avespectrum} compares the CO(1-0) and CO(3-2) spectra averaged over the respective survey
areas\footnote{Only the spectrum integrated over the full field of view covered by \cite{petyetal2017} is publicly available
(https://www.iram.fr/~pety/ORION-B/data.html)}. With the caveat that
the areas over which the spectra have been obtained are not the same (in particular, the area covered by \cite{petyetal2017} extends
further to the east, that is, away from the western, UV illuminated cloud edge), the shapes of the main component of the lines compare
fairly well, with some excess emission seen on the blue-shifted side of the CO(3-2) line. This secondary component at velocities
around 4~km/s is
comparatively much fainter than the main component in the CO(3-2) transition, by about a factor of two. We attempted to fit
a double Gaussian to the CO(3-2) profile for a more direct comparison with the ratio between the two components in the CO(1-0) line derived
in \cite{petyetal2017}, but the fit did not converge due to the more non-Gaussian shape of the main component, and the secondary
component being significantly fainter than the main component.

Figure~\ref{fig:N2024_5kms_11kms} shows CO(3-2) maps integrated over the same velocity ranges and convolved to the same 31\arcsec{} beam
as in Fig.~5 of \cite{petyetal2017}, to allow
for a direct comparison of the CO(3-2) and CO(1-0) distributions. The map at 4.5-5~km/s, representative of the secondary line component,
shows that our CO(3-2) survey covers the majority of the secondary component as seen in the CO(1-0) map, thus the low ratio with
respect to the primary component (as compared to the ratio in the CO(1-0) line) is unlikely due to the CO(3-2) survey missing a
significant part of the secondary component. Similarly, no significant part of emission in the main line component is missed by the
CO(1-0) map, thus the difference in the ratios between main and secondary components between the (3-2) and (1-0) lines is most likely real.
This difference may indicate that the secondary component traces a layer of molecular gas that is colder and more optically thin than the main
component.

Overall the CO(3-2) map of the secondary component corresponds well to the CO(1-0) maps in terms of the structures detected. The
streak of emission running east-west around the declination of the Horsehead Nebula seems comparably fainter in the CO(3-2) maps, which
in part may be due to the fact that the bulk of this feature is at slightly lower velocities (as can be seen from the southern-most
E-W oriented position-velocity cut at the bottom right of Fig.~\ref{fig:N2023N2024_pv}). Perhaps the most noteable difference is the much
more prominent appearance of the blue-shifted lobe of the outflow driven by \object{NGC~2023~MM1} \citep{sandelletal1999} in the CO(3-2)
map, confirming that CO(3-2) is well suited for the main purpose of the survey, that is an unbiased search for molecular outflows.

In Fig.~\ref{fig:N2024_co32_co10} we present a detailed comparison of the main line component, integrated over a velocity interval of
9-12~km/s, between
our CO(3-2) data and the CO(1-0) data of \cite{petyetal2017}, where the CO(3-2) data are convolved to the same 31\arcsec{} beam as
used in \cite{petyetal2017}. The panel to the right shows the ratio of the two lines. While the absolute value of the ratio may be
skewed by how exactly the integrated line maps were created, some spatial trends in the distribution of the line ratio can be seen.
Over most of the area covered by both surveys, the ratio is fairly similar, at values between $\sim$0.5 and 0.7. Interestingly, this
includes the NGC~2024 \ion{H}{II} region, which features dominantly in the line brightness maps. Assuming that generally the CO emission
is optically thick, this near-constant ratio would imply relatively small variations of the gas temperature over the field, even in the gas associated
with the NGC~2024 \ion{H}{II} region. As can be seen in the following Section~\ref{section:N2024HII}, however, the brightest emission
in the NGC~2024 \ion{H}{II} region actually falls just outside the velocity range used for the integrated intensity, particularly on the
blue-shifted side, so velocity-resolved line-ratio maps would probably show a different picture.

Higher line ratios are found in the NGC~2023 nebula, over a fairly small area, and
all along the western edge of the cloud. In NGC~2023, the area of high line-ratio coincides with bright line emission, consistent
with the presence of warm, optically thick molecular gas. In contrast, the areas
of high line-ratio along the western cloud edge predominantly arise in comparatively faint CO emitting regions. This behaviour is further
illustrated in Fig.~\ref{fig:N2024_ratioCO_TCO}, plotting the distribution of line ratios against brightness in the CO(1-0) and CO(3-2)
lines. Fig.~\ref{fig:N2024_ratioCO_TCO} also shows that the line ratios over most of the cloud are at fairly uniform values, with high
line ratios found associated mainly with fainter line emission. Apparently the regions with high line ratios trace the warm molecular
gas, presumably with low optical depth, associated with the PDR at the
western edge of the Orion~B cloud. NGC~2023 and the western cloud edge are in fact also seen as regions of elevated dust temperature
\citep[e.g.][]{lombardietal2014}, whereas the high dust temperatures in NGC~2024 are not reflected in correspondingly higher line
ratios (in the velocity interval used for the line integration used in \cite{petyetal2017} and here).

\begin{table}
\caption{RADEX model inputs and resulting parameter ranges for the three diverse CO emission regimes in the NGC~2023/NGC~2024 region}             
\label{table:RADEX}      
\centering                          
\begin{tabular}{c c c c}        
\hline\hline                 
 & typical & 4\,km/s & high excitation \\    
\hline                        
  CO(3-2)/(1-0)          & 0.5-0.75 & 0.25-0.4 & 0.9-1.8 \\      
  T$_{mb}$(3-2) (K)       & 8-16     & $< 10$    & 2-13 \\
  T$_{mb}$(1-0) (K)       & 15-23    &      & $< 10$ \\
\hline
  $N_{\rm CO}$ (cm$^{-2}$) & 10$^{17}$-10$^{19}$ & 2$\cdot$10$^{16}$-5$\cdot$10$^{18}$    & 5$\cdot$10$^{15}$-5$\cdot$10$^{16}$ \\
   $T$ (K)                & 10-60 & 10-50    & 30-200 \\ 
   $n$ (cm$^{-3}$)        & 10$^3$-10$^6$ & 10$^2$-10$^6$    &  2$\cdot$10$^2$-10$^6$\\ 
\hline                                   
\end{tabular}
\end{table}

  In Appendix~\ref{app:radex} we present exploratory RADEX \citep{vandertaketal2007} radiative transfer modelling of the CO emission
  in the CO(3-2) and CO(1-0) lines, in order to constrain the differences in physical properties of three characteristic emission regimes:
  the multi-colour scale in Fig.~\ref{fig:N2024_Radex} represents models reproducing the 'typical' emission, which according to
  Fig.~\ref{fig:N2024_ratioCO_TCO} is characterized by CO(3-2)/CO(1-0) line ratios around 0.6, a
  CO(1-0) brightness temperature around 20~K (i.e. 60~K~km/s if integrated over 3~km/s), and a CO(3-2) brightness temperature
  around $\sim$12~K ($\sim$35~K~km/s if integrated over 3~km/s). The red colour scale in Fig.~\ref{fig:N2024_Radex} corresponds to models
  resembling the 'low excitation' emission, corresponding to the $\sim 2$~times lower CO(3-2)/CO(1-0) line ratio
  inferred for the velocity component around 4-5~km/s, which we assume to be characterized by CO(3-2)/CO(1-0) line ratios around 0.3 and
  a CO(3-2) brightness temperature less than 10~K (the maximum brightness seen in our line cubes for this velocity component). The blue
  colour scale in Fig.~\ref{fig:N2024_Radex} shows the parameter range reproducing the 'high excitation' emission reflecting the conditions
  at the western cloud edge, with CO(3-2)/CO(1-0) line ratios greater than 0.9 and brightness temperatures less than 13~K and 10~K for the
  CO(3-2) and CO(1-0) lines, respectively.

  Table~\ref{table:RADEX} summarizes the restrictions in terms of line ratios and line brightnesses imposed on the RADEX models representing
  the three distinct emission regimes, as well as the ranges of CO column densities, temperatures, and densities that yield models
  reproducing the respective observed line properties. For all three regimes we find that, in principle, wide ranges in CO column
  densities, temperatures, and H$_2$ densities may yield CO emission consistent with the respective input restrictions. Nevertheless,
  we can identify some trends. Regardless of the emission regime and column density considered, there is a general degeneracy between
  density and temperature, and valid models reproducing a certain set of constraints at a given column density typically fall into a narrow
  stripe in the density-temperature plane, with lower temperatures requiring overall higher densities. Varying the column density
  typically leads to a shift of that stripe, with lower column densities usually requiring a combination of higher temperatures and
  densities, that is, shifting the stripe in the density-temperature plane upwards and to the right. Additional observations would be needed
  to further limit the physical conditions of the gas, in particular, rarer CO isotopologues ($^{13}$CO) should help to constrain the
  CO column density, while additional transitions should help to further constrain the gas temperature. As an additional caveat, it has
  to be kept in mind that the CO(3-2) and (1-0) lines may have a different beam filling factor, for example warmer gas emitting stronger in
  the (3-2) transition might be limited to smaller volumes.

  Despite the wide range in parameters allowed for each of the three emission regimes considered, we can identify some physical differences
  between them. In terms of column density, the 'typical' emission requires the highest CO column densities (no models meet the
  restrictions imposed for column densities
  $N < 10^{17}$cm$^{-2}$), the 'high excitation' component requires the lowest CO column densities ($< 10^{17}$cm$^{-2}$), while the
  'low excitation' component of the 4-5~km/s cloud requires CO column densities around $10^{17}$cm$^{-2}$, intermediate between the
  'typical' and the 'high excitation' component. This indicates that the 'typical' component indeed traces the high column density bulk
  molecular gas of the GMC, while the 4-5~km/s 'low-excitation' emission traces a significantly more tenuous cloud. In terms of temperature
  and density the 4-5~km/s 'low-excitation' might not be too different from the bulk cloud, with temperatures less than $\sim$50~K. In
  particular if the lowest densities considered are excluded, as they would imply unrealistically large cloud extents along the line of
  sight, temperatures for both components should be well below 40~K.

  In contrast, we find that the 'high excitation' component most
  probably requires significantly higher temperatures $>$40~K (with temperatures as low as
  $\sim$30~K possible only for very high densities). This is similar to the findings of \cite{zhangetal2019} in a study of
  $^{12}$CO(3-2)/(2-1)/(1-0) line ratios in the \object{N131} dust bubble, who see high line ratios outlining the inner rim of the bubble.
  They attribute the high line ratios to the high temperatures and pressure generated by the impact of the massive hot stars creating the
  bubble. While our RADEX models do not provide unambiguous evidence for higher pressure in the high excitation component, it seems clear
  that the temperature is significantly higher than in the bulk of the molecular cloud. We conclude that the
  'high excitation' component seen along the western cloud edge indeed corresponds to warm gas heated and possibly compressed by the
  incident UV radiation field. The 4-5~km/s 'low-excitation' emission instead indicates a fairly cool cloud not exposed to significant
  irradiation and might trace a tenuous cloud in the fore- or background of the Orion complex, at a significantly larger physical
  distance from the OB stars creating the IC~434 ionization front at the western edge of the bulk cloud in the NGC~2023/NGC~2024 region.
  
\begin{figure}
   \centering
   \includegraphics[width=\columnwidth]{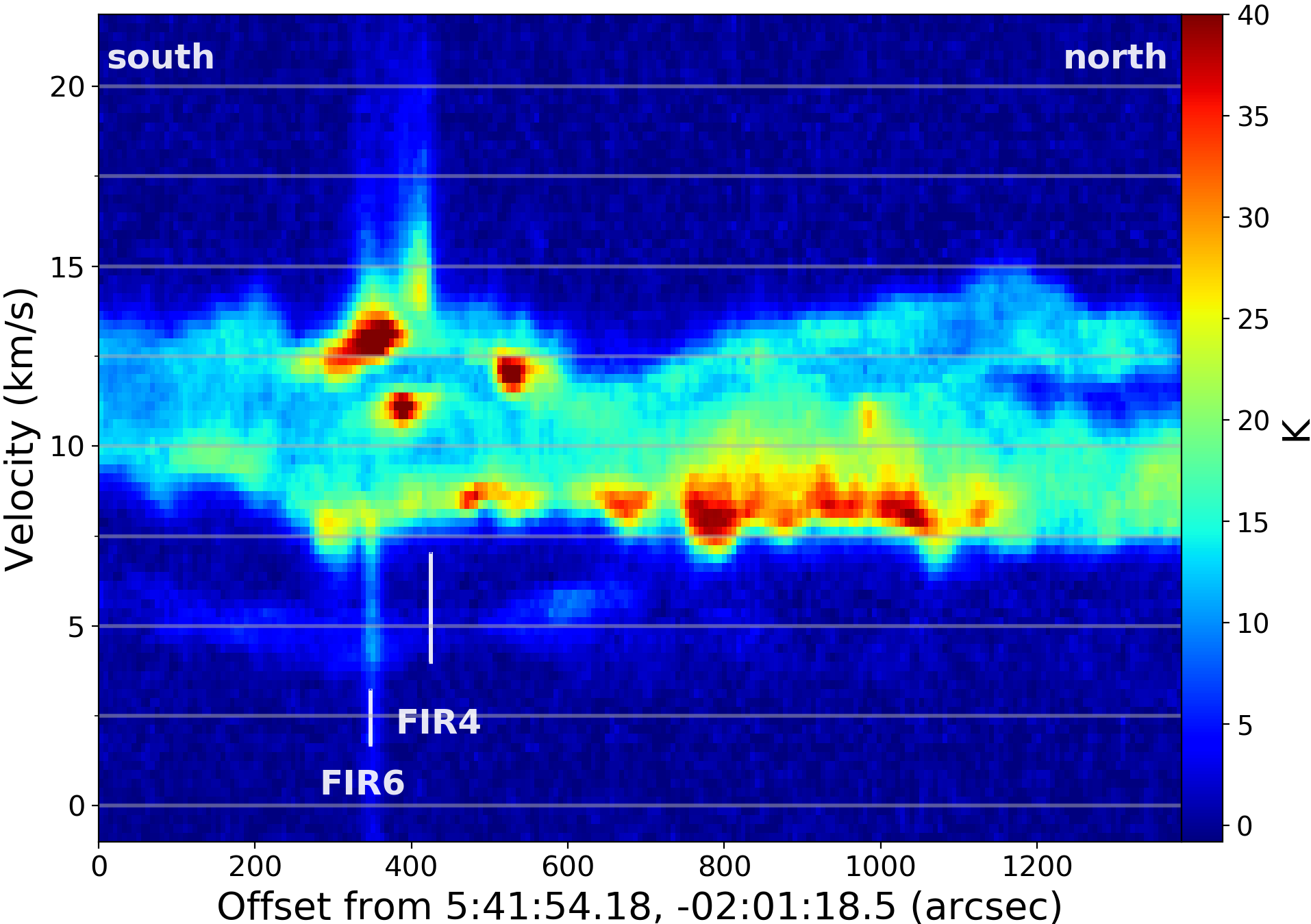}
   \caption{Position-velocity cut along the NGC~2024 dark lane along the green line shown in Figs.~\ref{fig:N2023N2024_overview},
     \ref{fig:N2024_PDRs}, and \ref{fig:N2024_opt_8kms}.
   }
              \label{fig:N2024_pv_darklane}%
\end{figure}

\begin{figure}
   \centering
   \includegraphics[width=\columnwidth]{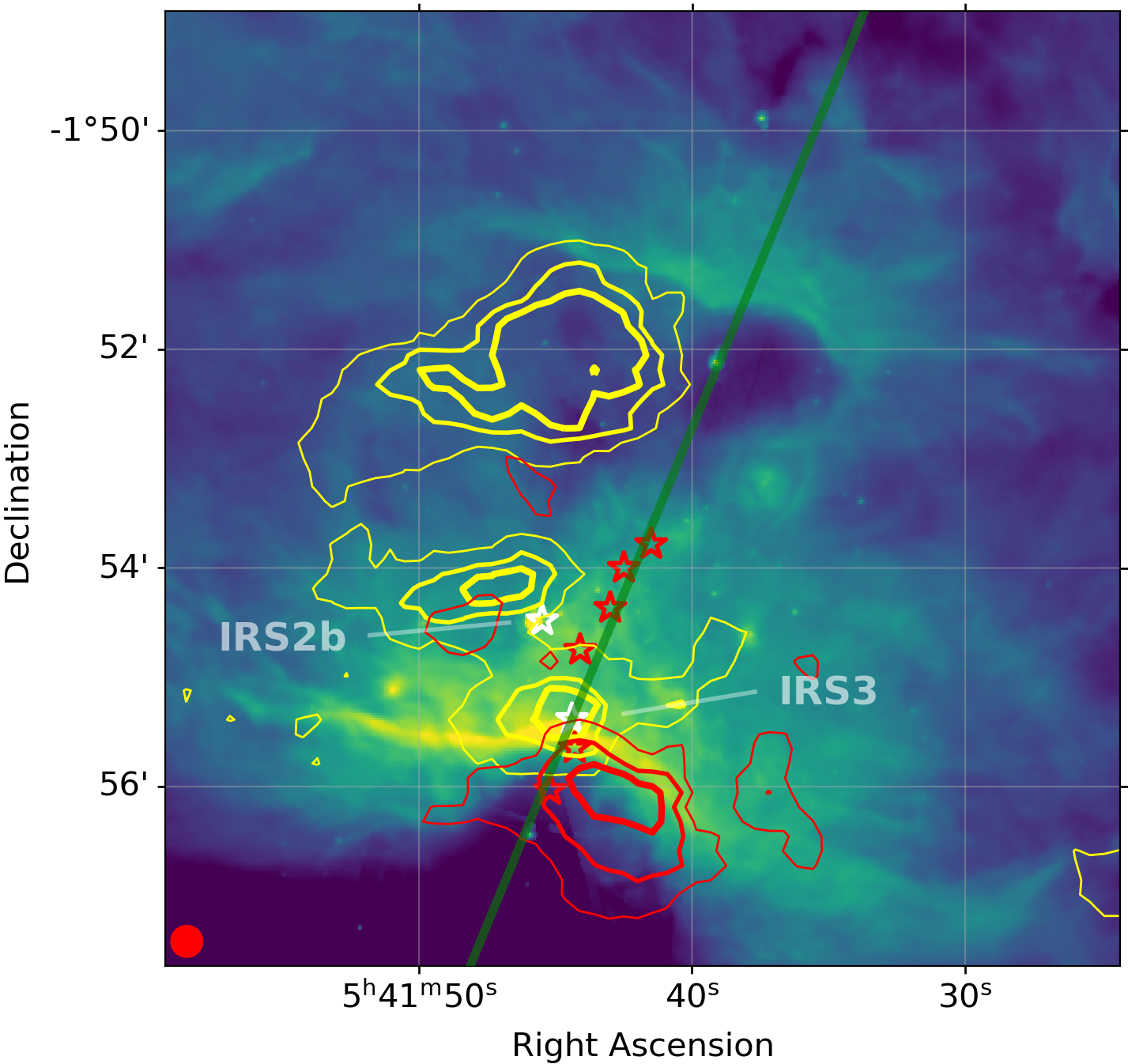}
   \caption{Spitzer IRAC4 (8.0~$\mu$m) image of the central NGC~2024 region (logarithmic colour scale) overlayed with contours of CO(3-2)
     emission at 10.75 to 11.25~km/s (yellow) and 12.75 to 13.25~km/s (red); red stars mark the positions of FIR1 to 6 from \cite{mezgeretal1988},
     the green line marks the location of the position-velocity cut shown in Fig.~\ref{fig:N2024_pv_darklane}. The red circle in the lower left
     marks the beam size of the CO(3-2) maps.
   }
              \label{fig:N2024_PDRs}%
\end{figure}

\subsubsection{The NGC~2024 \ion{H}{II} region}
\label{section:N2024HII}

The basic picture of the structure of the NGC~2024 \ion{H}{II} region is summarized, for example, in Fig.~6 of \cite{matthewsetal2002}.
It is a blister-type \ion{H}{II} region opening towards the north, east, and west, but bounded to the south \citep{barnesetal1989}.
A dense ridge of molecular cores \citep{mezgeretal1988,mezgeretal1992} is located in the background cloud at the southern end of the
\ion{H}{II} region. A layer of foreground material protruding from the main cloud at the southern end of the \ion{H}{II} region
causes the prominent dark optical lane running through the middle of the nebula, while the bulk of the cloud is thought to be
located behind the \ion{H}{II} region. A cluster of young stars is found in infrared images of the region
\citep[e.g.][]{barnesetal1989,biketal2003}, with the source \object{IRS2b}, having a late O-type spectrum, identified as the most
likely exciting source of the \ion{H}{II} region \citep{biketal2003}.

The structure of the gas towards the dense ridge has been the subject of previous investigations that attempted to explain the
profiles of CO lines in the direction of FIR5 with a wide range of upper energy levels and including various isotopologues
\citep[e.g.][]{grafetal1993, emprechtingeretal2009}. The resulting structure along the line of sight corresponds to the picture
sketched above, with the bulk of the molecular gas behind the \ion{H}{II} region at velocities of $\sim$11~km/s and a thinner
foreground layer with a velocity of $\sim$9.3~km/s. Both of these layers are separated from the \ion{H}{II} region by thin layers
of warm molecular gas representing the interfacing PDRs. An additional cold foreground layer is postulated to explain the
depression of the CO low-J main isotopologue lines at 12~km/s, where the location and extent of that foreground layer remains
unclear.

\begin{figure*}
   \centering
   \includegraphics[width=12cm]{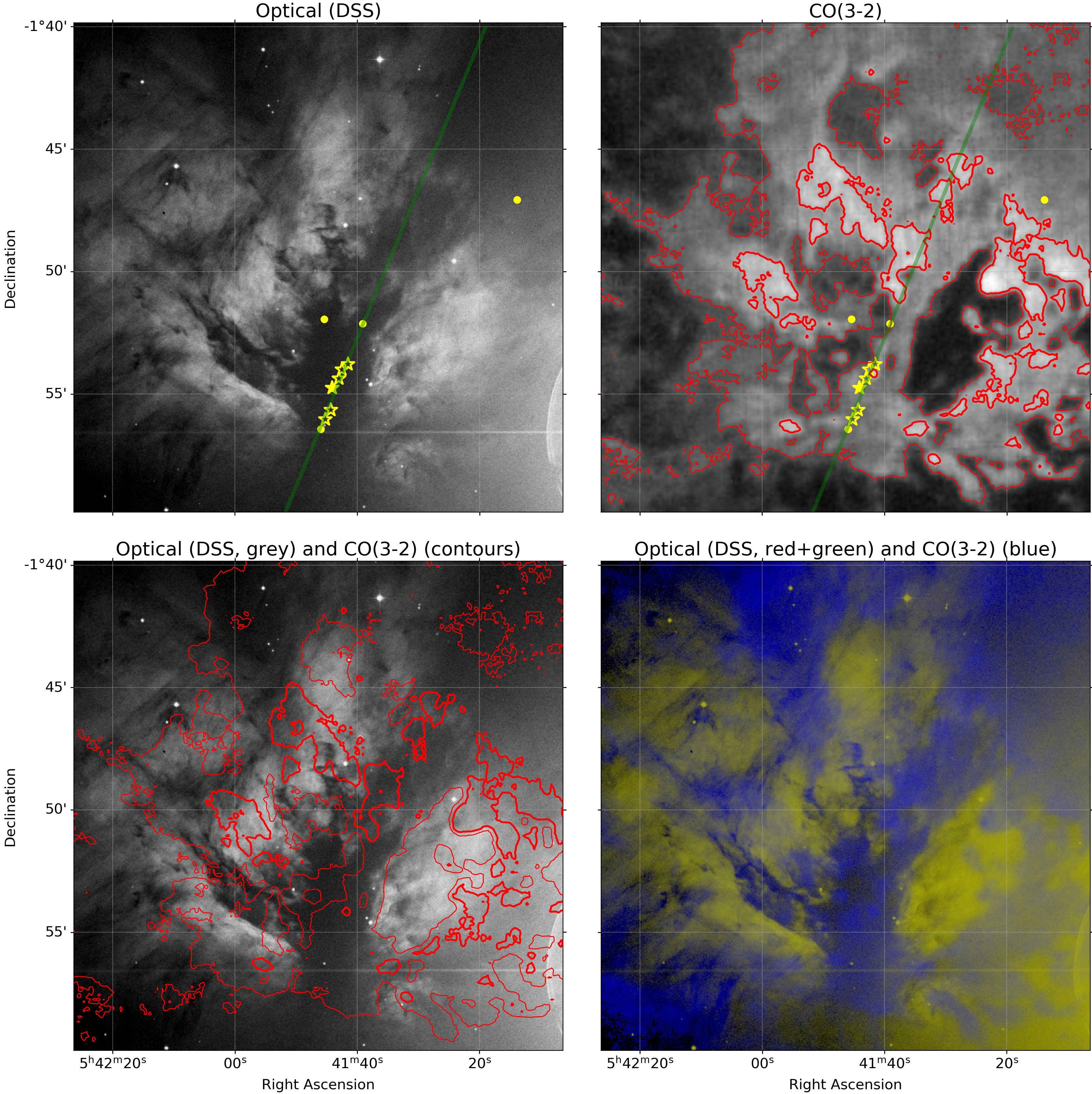}
   \caption{Comparison of the NGC~2024 \ion{H}{II} region at optical wavelengths and CO(3-2) emission at 8~km/s.
     The panel to the top left shows a DSS2-blue image; yellow stars mark the location of the sub-millimetre cores FIR1 to FIR6
     \citep[counting from north to south;][]{mezgeretal1988}, yellow dots mark HOPS protostars in the field, the green line marks the
     location of the position-velocity cut shown in Fig.~\ref{fig:N2024_pv_darklane}. The panel to the top right shows the CO(3-2)
     emission in a single velocity channel at 8~km/s (greyscale and contours). The bottom left panel shows an overlay of the CO(3-2)
     8~km/s velocity channel (using the same contour levels as in the upper right panel) on the DSS2-blue image, the bottom right
     panel shows a colour-composite image, where the optical DSS2-blue image is shown in the red and green channels, and CO at 8~km/s
     in the blue channel.
   }
              \label{fig:N2024_opt_8kms}%
\end{figure*}

Figure~\ref{fig:N2024_pv_darklane} shows a position-velocity cut through the NGC~2024 \ion{H}{II} region along the optical dark lane,
as indicated by the green line in Figs.~\ref{fig:N2023N2024_overview}, \ref{fig:N2024_PDRs}, and \ref{fig:N2024_opt_8kms}.
Our data reproduce the complex CO(3-2) line shapes seen there before. For example, a double peaked line noted by \cite{grafetal1993} is seen immediately
south of FIR6, while three peaks noted by \cite{emprechtingeretal2009} are seen between FIR6 and FIR4. We also see  substantial
variations in the line properties. For example, the number of emission and absorption components and their respective velocities and
brightnesses can change dramatically even on sub-arcminute spatial scales.

The absorption dip at $\sim$12~km/s drifts by a few velocity channels (i.e. by more than 0.5~km/s) over just 60\arcsec{}
around the position of FIR5, while only 2\arcmin{} further north bright emission is seen at the same velocity. The bright emission
component at intermediate velocities (around $\sim$11.0~km/s) is seen as a small, but clearly resolved patch in the channel maps (green
contours in Fig.~\ref{fig:N2024_PDRs}). It coincides with the brightest portion of the rim seen at infrared wavelengths marking the PDR
at the southernmost boundary of the \ion{H}{II} region \citep[e.g.][]{gaumeetal1992,watanabemitchell2008,renli2016}. This location marks both the smallest separation between the infrared rim and the
main ionizing source IRS2b, and the location of IRS3, which was once proposed as a possible ionizing source and which is actually
a small group of stars \citep{biketal2003}. Either IRS2b or IRS3 seem to heat the molecular gas around this
location. The bright emission at $\sim$13~km/s (red contours in
Fig.~\ref{fig:N2024_PDRs}) is seen south of the bright infrared rim, with the rim precisely curving around its northern edge.
The shape of the infrared rim suggests that the \ion{H}{II} region at the southern end is bounded by a dense clump, around which the PDR
is wrapping, and to which the most red-shifted CO emission at 13~km/s is associated. FIR5 and FIR6 are seen (in projection) towards this clump, but
might still be located further in the background. The radial velocities of the gas in the FIR1-6 dense ridge itself ranges between $<$10~km/s
to 11~km/s \citep[e.g.][]{gaumeetal1992,watanabemitchell2008,renli2016}, that is, slightly less redshifted than the rear cloud wall
bounding the \ion{H}{II} region, indicating that only the surface layer of the cloud has been accelerated by the pressure from the
\ion{H}{II} region and is moving into the densest portion of the cloud.

The basic picture of a more red-shifted background molecular component and a more blue-shifted foreground component provides
a plausible explanation of many features seen in previous multi-line, multi-isotopologue CO line studies. In particular, it explains the double-peaked
profiles of low-J, low opacity isotopologue lines. The true three dimensional structure of the CO emitting gas at the southern
end of the \ion{H}{II} region, however, is clearly more complex. For example, \cite{emprechtingeretal2009} assume layers with constant velocities
and no spatial substructure over the 80\arcsec{} beam used in their study towards the position of FIR5. A fuller understanding will
require multiline observations at substantially better spatial resolution and adequate spatial sampling of the region.

Going further north along the dark lane, we observe a gradual increase in the maximum red-shifted extent of the line, from $\sim$12~km/s
at an offset of $\sim$650\arcsec{} up to $\sim$15~km/s at an offset of $\sim$1200\arcsec{}. The blue-shifted maximum velocity remains
roughly constant (apart from some wiggles) at $\sim$7-7.5~km/s. The brightest emission is consistently seen at the most blue-shifted edge
of the line in the position-velocity diagram. In Fig.~\ref{fig:N2024_opt_8kms} we compare the spatial distribution of the blue-shifted
CO(3-2) emission (using a single channel of the CO(3-2) cube at a velocity of 8~km/s) with an optical image (Digitized Sky Survey:
DSS2-blue), which shows the distribution of obscuring dust in the foreground of the \ion{H}{II} region. A number of structures 
correspond well between the two maps. In particular, the western rim of the optical dark lane coincides with a sharp drop in CO intensity.
This drop in CO emission is also clearly seen in maps of velocity-integrated $^{13}$CO emission
\citep[e.g.][]{subrahmanyanetal1997,buckleetal2010}. It is not clearly apparent, however, in the velocity integrated $^{13}$CO maps shown in
\cite{petyetal2017} as their velocity interval (9-12~km/s) excludes the majority of the blue-shifted component. The $^{13}$CO channel
maps shown in \cite{enokiyaetal2021} also suggest that in the northern part of the cloud this blue-shifted component dominates the
total emission. Overall, this indicates that the blue-shifted component contributes very significantly
to the total CO column density at least towards the northern part of the \ion{H}{II} region. The good correspondance between the
blue-shifted CO distribution and the optical obscuration clearly places the blue-shifted molecular gas in the foreground of the
\ion{H}{II} region.

Taken together, our data suggest a revised model for the NGC~2024 \ion{H}{II} region, particularly for its northern. The more
blue-shifted CO(3-2) emission clearly traces a layer in the foreground of the \ion{H}{II} region. As it is significantly brighter
in our $^{12}$CO data, and as it also seems to dominate $^{13}$CO maps, we argue that the foreground,
more blue-shifted CO actually traces the bulk of the cloud column density in the area. We see the brightest line emission at the most
blue-shifted velocities, tracing a warm layer interfacing the molecular foreground cloud and the \ion{H}{II} region, and being
pushed towards us (to slightly higher velocities than the bulk foreground cloud) by the \ion{H}{II} region. Overall
the bulk velocity of the blue-shifted component does not change systematically with increasing distance from the central cluster,
indicating that the gas there is too massive to be accelerated as a whole by the impact of the \ion{H}{II} region.
The red-shifted
CO then traces the molecular gas behind the \ion{H}{II} region. The increasing red-shift with distance from the ionizing cluster
could indicate that we see the (southern) edge of an inflating bubble, or that the amount of gas decreases with distance from the
cluster and is thus more easily accelerated to larger velocities. This situation would be comparable to that found in the \object{Orion Nebula}
(albeit seen here 'from behind'),
where blue-shifted [\ion{C}{II}] emission traces a thin foreground layer that is accelerated by the expanding \ion{H}{II} region, while the
background layer remains at a roughly constant velocity of the bulk background cloud \citep{pabstetal2020}.
Finally, observations of hydrogen radio recombination lines
show increasingly blue-shifted velocities towards the north \citep{kruegeletal1982,subrahmanyanetal1997}, which could be explained as
an accelerating flow directed north and inclined towards the observer. Consequently, the foreground molecular layer must also
be inclined towards the observer.

\begin{figure}
   \centering
   \includegraphics[width=\columnwidth]{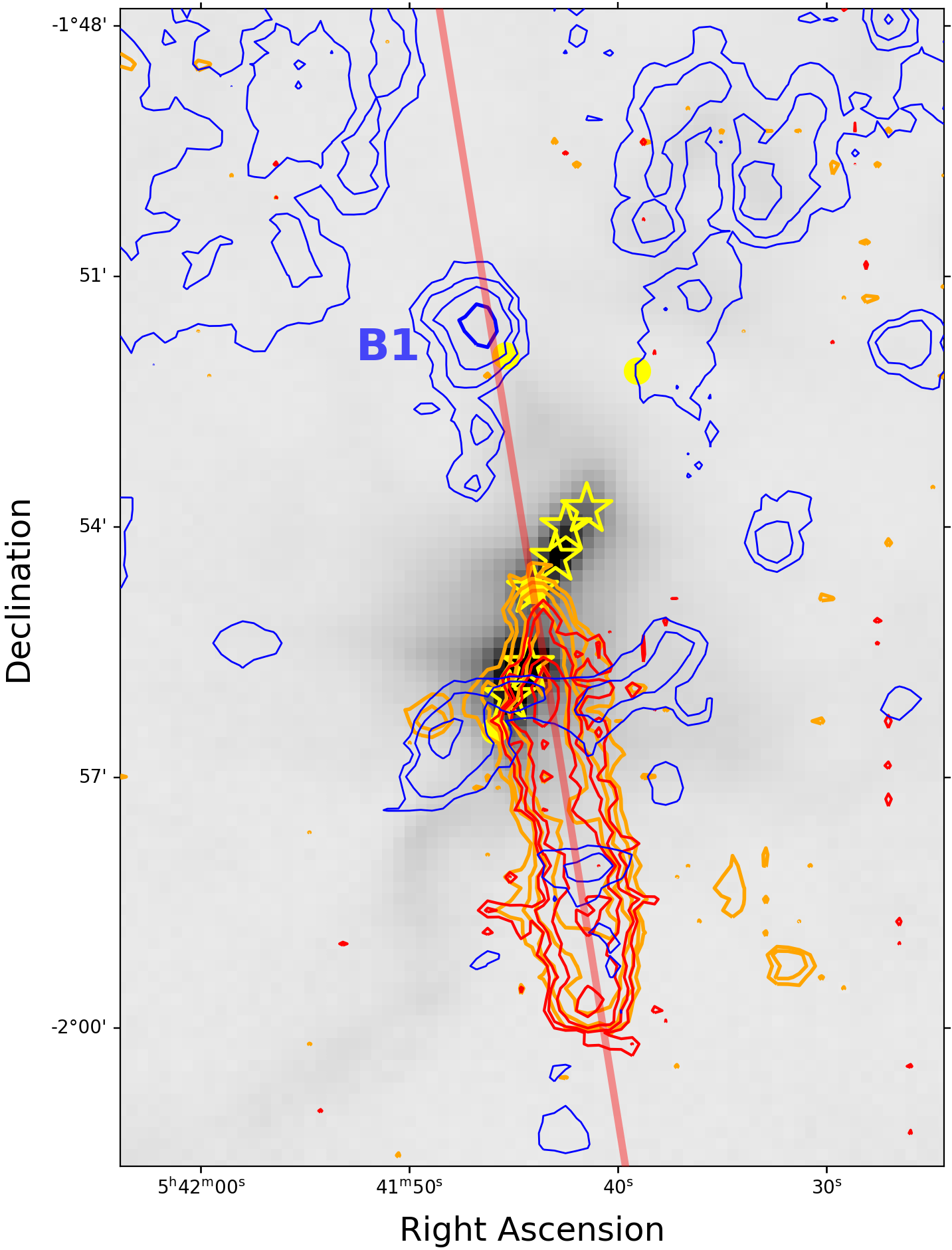}
   \caption{Maps of high-velocity CO emission in the central NGC~2024 region. Greyscale: Laboca 850~$\mu$m continuum emission.
     Orange contours: CO emission integrated from 16.0~km/s to 19.75~km/s. Red contours: CO integrated from 20.0~km/s to 36.0~km/s.
     Blue contours: CO integrated from 5.5~km/s to 7.0~km/s. Yellow stars mark the positions of the 1.3~mm dust continuum cores FIR1 to
     FIR6 (counting from north to south, from \cite{mezgeretal1988}), while yellow dots mark HOPS protostars (where FIR4 coincides
     with \object{HOPS~384}). The red line marks the position of the velocity cut shown in Fig.~\ref{fig:N2024_flow_pv}.
   }
              \label{fig:N2024_FIR4_flow}%
\end{figure}

\begin{figure}
   \centering
   \includegraphics[width=\columnwidth]{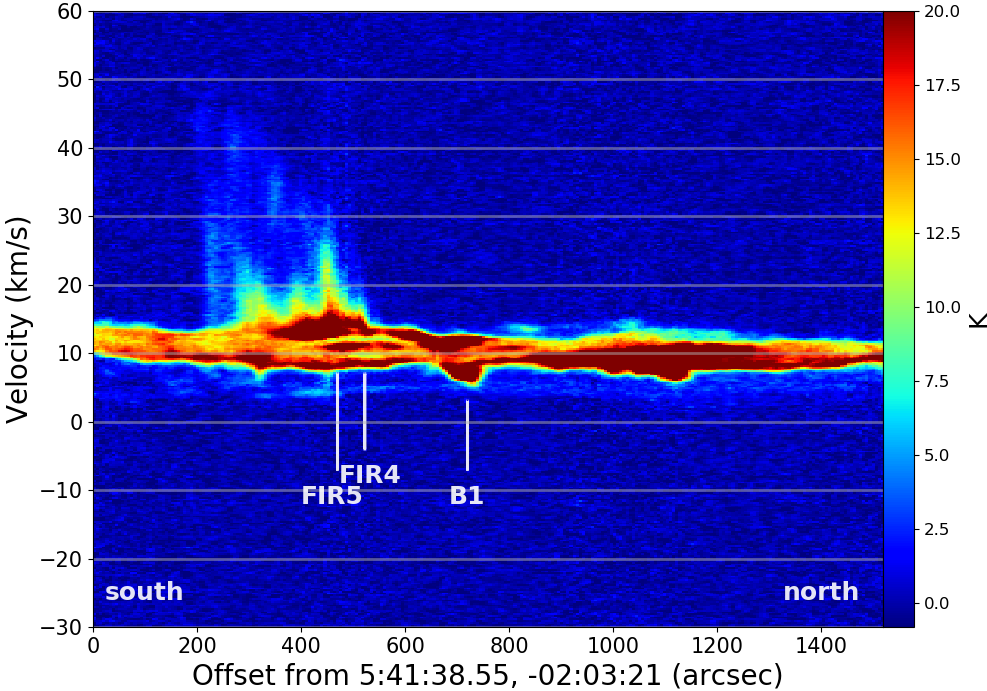}
   \caption{Position velocity cut along the NGC~2024 outflow axis, starting from 5$^h$41$^m$38.$\!\!^s$55,
   $-$2$^\circ$03\arcmin21\arcsec{} (south), ending at 5$^h$41$^m$54.$\!\!^s$65, $-$01$^\circ$38\arcmin20\arcsec{} (north; marked
   with a red line in Fig.~\ref{fig:N2023N2024_overview} and Fig.~\ref{fig:N2024_channel}).
   }
              \label{fig:N2024_flow_pv}%
\end{figure}

\begin{figure}
   \centering
   \includegraphics[width=\columnwidth]{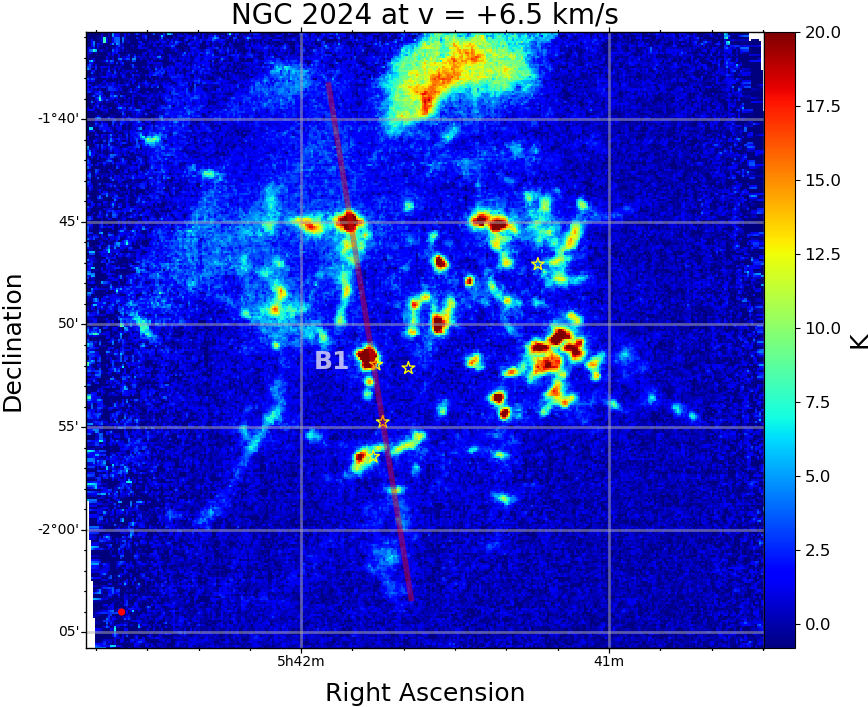}
   \caption{Map of the CO emission in the NGC~2024 region in a single velocity channel at $+$6.5~km/s, showing the highly fragmented
     distribution of emission features at this moderately blue-shifted velocity. The red line shows the location of the position-velocity
     cut along the NGC~2024 outflow axis shown in Fig.~\ref{fig:N2024_flow_pv}. The red dot in the bottom left corner indicates the beam
     size.}
     \label{fig:N2024_channel}%
\end{figure}

\subsubsection{The NGC~2024~collimated outflow}
\label{section:N2024flow}

The presence of high velocity CO emission in the NGC~2024 area was first noticed by \cite{ballylada1983}. First mapping
observations by \cite{sanderswillner1985} indicated a bipolar outflow with its highly collimated red-shifted lobe extending south,
and blue-shifted emission to the north. Subsequent observations \citep{richeretal1989, richeretal1992} did not confirm the
northern, blueshifted lobe, making the \object{NGC~2024} outflow a rare example of a monopolar protostellar outflow. The proposed
driving source of the outflow is the sub-millimetre condensation \object{NGC~2024~FIR5}, the brightest condensation in a chain
of 7 running along a north-south high column density ridge \citep{mezgeretal1988, mezgeretal1992}. The reason for the absence of
the blue lobe is not entirely clear, but has been attributed to \object{NGC~2024~FIR5} being located immediately south of a PDR
interfacing the dense core and the NGC~2024 \ion{H}{II} region.

Figure~\ref{fig:N2024_FIR4_flow} shows the distribution of high velocity CO emission in the central NGC~2024 region, and
Fig.~\ref{fig:N2024_flow_pv} shows a position-velocity cut along the outflow axis, marked by the red line in
Fig.~\ref{fig:N2024_FIR4_flow} and Fig.~\ref{fig:N2023N2024_overview}. We clearly detect red-shifted high-velocity
emission with a spatial distribution and kinematic structure very similar to previous observations (see e.g. Fig.~4 of
\cite{richeretal1989} for a comparison in position-velocity space, and \cite{buckleetal2010} for a CO(3-2) high velocity
emission map).

The redshifted lobe extends north-south, starting from the position of FIR4, passing just west of the position of FIR5.
In position-velocity, space the upper envelope in velocity shows the maximum velocity in the lobe increasing steadily
from north to south in a 'Hubble-type' flow. This increase in maximum velocity starts at FIR4 and continues
towards the southern end of the lobe without any noticable change or feature visible at the position of FIR5. We hence
propose that FIR4 is the actual driving source of the outflow. In fact, this identification is consistent with a number of observations
reported in the literature. \cite{richeretal1992} note the existence of red-shifted emission north of FIR5, but discard
a driving source further north. The CO(3-2) map of \cite{buckleetal2010} also show the red-shifted lobe to extend up to FIR4.
Meanwhile, \cite{chandlercarlstrom1996} detect red-shifted outflow emission to the south of FIR4. They see the brightest emission
extending towards the south-east from FIR4, but faint and more diffuse emission is also seen to its south and south-west, possibly
indicating a conical base of a south-oriented outflow. Certainly these early interferometric observations are
compromised by poor sensitivity and uv coverage. Perhaps even more interestingly, the CS(2-1) maps in the same paper
show two filaments first extending to the south-east and south-west of FIR4, respectively, then both turning
towards a more southerly orientation. These filaments could indicate the walls of an outflow cavity extending south of
FIR4. \cite{alvesetal2011} present SMA interferometric CO(3-2) observations of the FIR5/6 region and find two nearly
parallel, north-south oriented filaments of red-shifted emission at intermediate velocities (with an additional north-south
oriented filament at higher velocities located between the lower-velocity filaments). They suggest that the eastern filament
is driven by FIR5, and the western filament is either driven by a separate, fainter continuum peak along the filament
or is part of a precessing outflow from FIR5, along with the high-velocity filament. We instead propose that the two filaments
at intermediate velocity trace the eastern and western edge of a single outflow lobe originating further north. The lack of
CO detection at similar velocities between the filaments may either indicate a hollow outflow structure, or (more likely) is due
to filtering of extended emission by absent short spacings in their observations. \cite{chernin1996} argue that FIR5 may not
be the driving source, but propose that its origin is in a still undiscovered low-mass young stellar object rather than any of the
FIR sources of \cite{mezgeretal1988} and \cite{mezgeretal1992}.

\cite{choietal2015} describe a radio continuum jet and water maser features which line up in a north-north-west to
south-south-east orientation (position angle $-$25\degr, broadly consistent with the orientation of the brightest part
of the red-shifted CO outflow found by \cite{chandlercarlstrom1996}. This might either indicate a second flow, or
a variation of outflow direction with time. In fact, at large scales the red lobe shows evidence for some wiggling,
as seen in the maps of \cite{sanderswillner1985}, \cite{richeretal1992}, \cite{buckleetal2010}, and in the present data.
Finally, we note that recent ALMA observations taken with the 7m ACA array also show a redshifted, conical outflow lobe
opening towards the south of FIR4 (Megeath, pers.\ comm.), broadly consistent with \cite{chandlercarlstrom1996}, and
perfectly coinciding with the northern end of the large-scale red-shifted lobe.

FIR5 (the brightest of the sub-millimetre cores identified by \cite{mezgeretal1988} and \cite{mezgeretal1992}) was originally proposed
as driving source, as the large mass/momentum/energy of the flow requires at least an intermediate-mass
object as driving source. The newly proposed driving source FIR4 corresponds to \object{HOPS~384}, a Class~0
protostar with a bolometric luminosity $L_{\rm bol} = 1478 L_\odot$, that is, also an intermediate-mass object, and thus consistent with
the original postulation.

We also note that the emission from the red-shifted lobe seems to curve around the continuum emission of FIR5, as seen more cleary in
the slightly higher angular resolution JCMT map of \cite{buckleetal2010}). We propose that the outflow lobe is located behind
the FIR5 core and that the dust in the centre of that core is optically thick enough to absorb the CO emission in its background.
This scenario is in line with the notion that FIR5 and FIR6 are seen immediately south of the ionization front at the southern boundary
of the NGC~2024  \ion{H}{II} region, and FIR1-4 are located just behind the southernmost part of the  \ion{H}{II} region.

The blue-shifted counterlobe remains elusive. While we detect the east-west oriented, mostly blue-shifted flow attributed to
FIR6 (see Fig.~\ref{fig:N2024_FIR4_flow}), we do not detect any emission at
blue-shifted velocities north of FIR4 that could obviously form a counterlobe to the prominent red-shifted outflow lobe
south of FIR4. Figure~\ref{fig:N2024_FIR4_flow} shows a roundish patch of blue-shifted emission about 3\farcm{}3 north of
FIR4, which corresponds to a triangular feature in the position-velocity diagram of Fig.~\ref{fig:N2024_flow_pv} extending towards
velocities lower than the region's systemic velocity. We refer to this feature as B1 in the following. It is
accompanied by two smaller blobs to its south, which also show clear blue outflow line wings in their spectra.  
Further north (outside the field shown in Fig.~\ref{fig:N2024_FIR4_flow}) the position-velocity diagram in Fig.~\ref{fig:N2024_flow_pv}
shows another slow increase towards blue-shifted velocities, which is however much less clearly separated from the
region's systemic velocity in the north. Both features are also visible in the channel map in Fig.~\ref{fig:N2024_channel}, which
shows a map of the CO emission in a single, 0.25~km/s wide spectral channel, along with a population of
compact, bright, slightly blue-shifted blobs seen scattered over the entire area of the NGC~2024 \ion{H}{II} region.

It remains to be seen whether this population of blue-shifted blobs are due to outflows driven by the embedded young stellar object population
in the area, or whether it is just a signpost of the complex kinematics of the molecular gas in front of the NGC~2024 \ion{H}{II}
region. The location of the blue-shifted feature B1 on the axis defined by the red-shifted outflow lobe, and its
size, which is comparable to the east-west diameter of the red-shifted lobe at similar separations from FIR4, however, leads us to
speculate that it marks the location where the blue counterlobe of the FIR4 flow punches through the layer of foreground
molecular gas. In this picture, FIR4 is embedded in a dense ridge behind the \ion{H}{II} region (but very close to the surface
of the molecular material). The northern flow immediately enters the \ion{H}{II} region, and first remains invisible in molecular lines,
as there is no molecular gas to entrain, and any molecules in the flow very quickly are dissociated. It then enters the
foreground molecular layer, where it can entrain molecular gas over some distance and accelerate it to modest velocities.
Then, it leaves the molecular layer after a short distance, where molecules are dissociated again as they are exposed to the
ambient UV field.

\subsection{L~1641-S}
\label{section:L1641S}

\begin{figure}
   \centering
   \includegraphics[width=0.98\columnwidth]{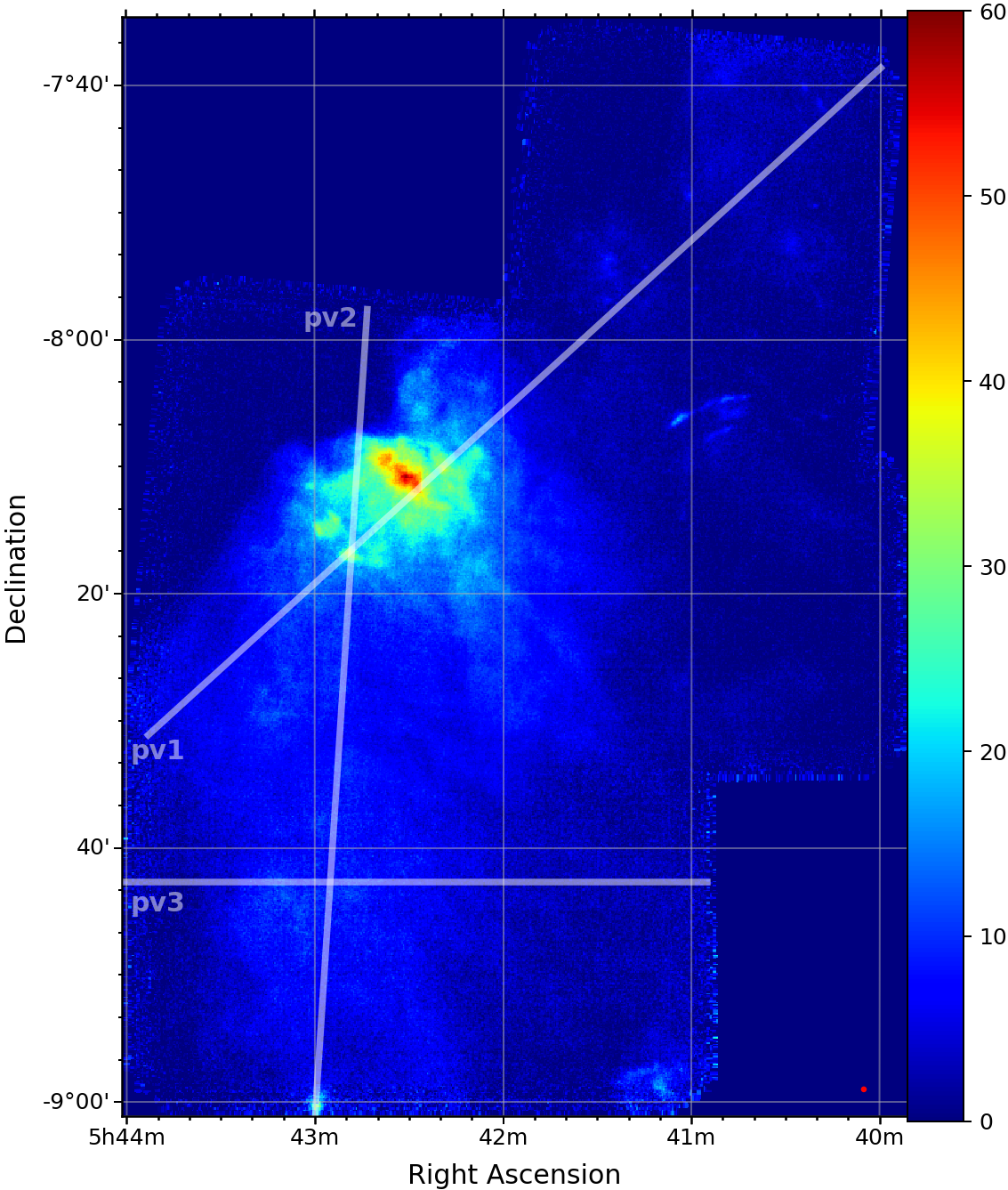}
   \includegraphics[width=0.98\columnwidth]{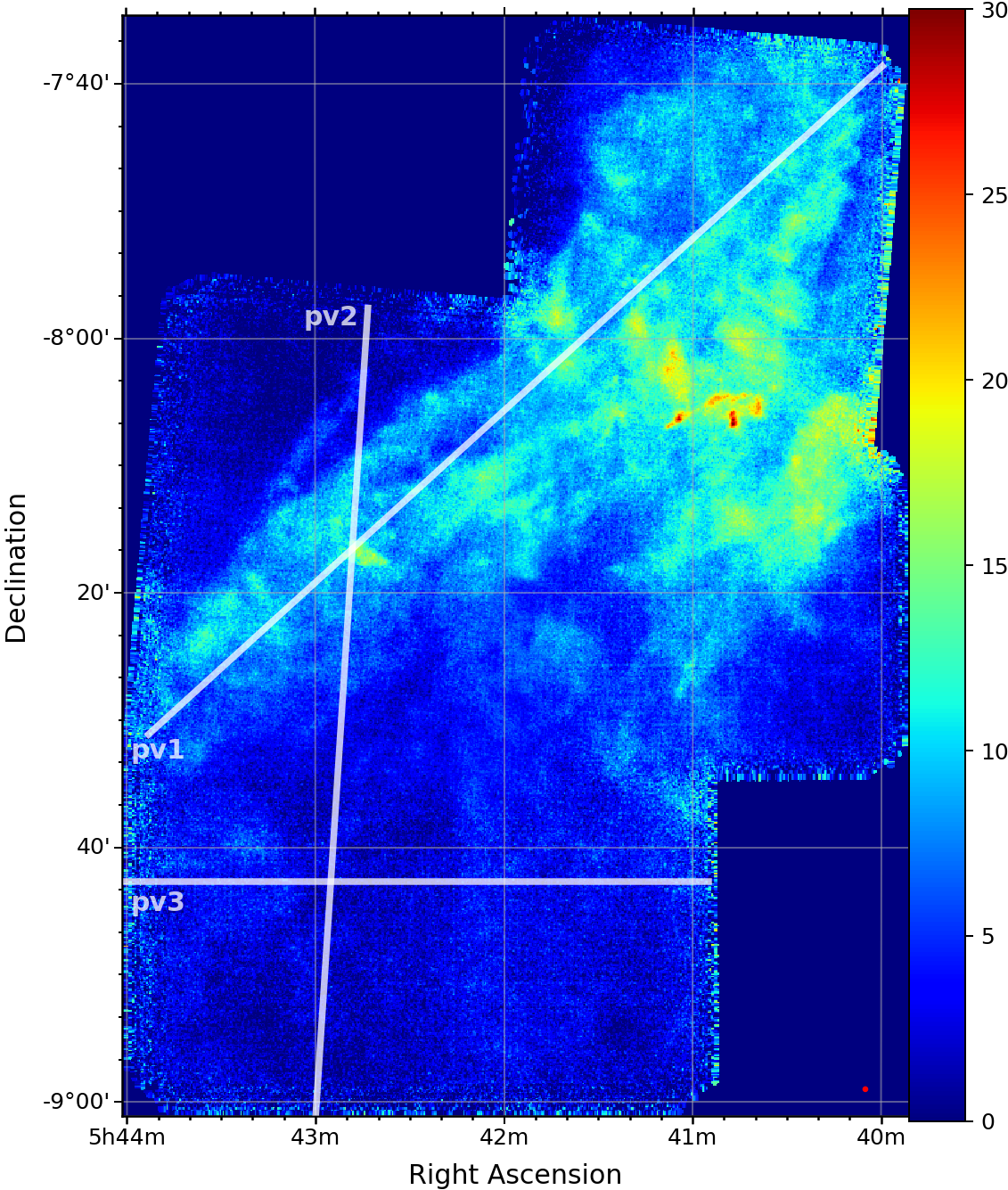}
   \caption{CO emission in the L~1641-S field, integrated from 0.5~km/s to 3.5~km/s (top, in K~km/s) and
            from 4.5~km/s to 7.25~km/s (bottom)
        The red dot in the bottom right corner indicates the beam size.}
              \label{fig:intmap_L1641S}%
\end{figure}

\begin{figure}
   \centering
   \includegraphics[width=\columnwidth]{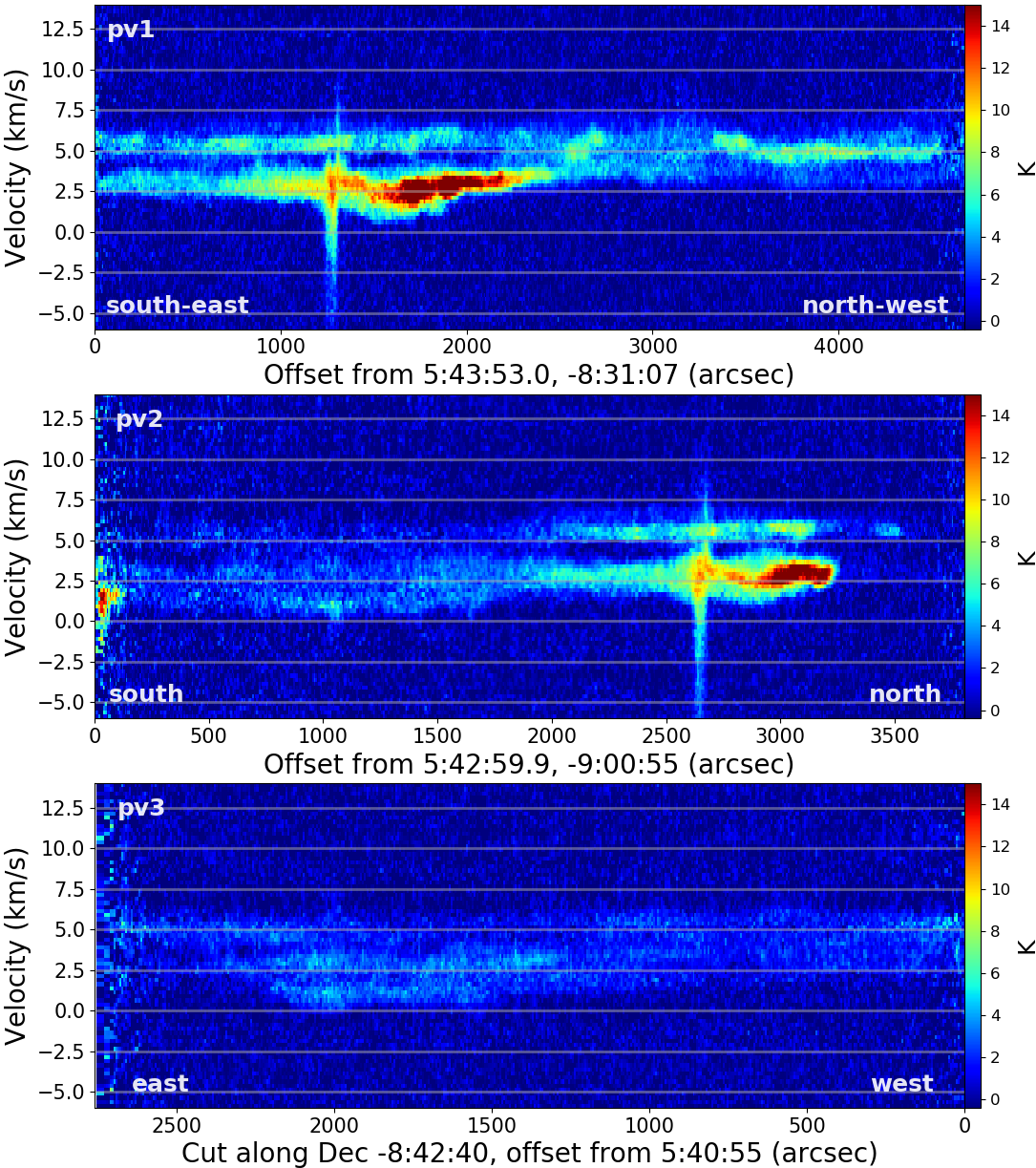}
   \caption{Position-velocity cuts through the L~1641-S field. See Fig.~\ref{fig:intmap_L1641S} for definition.}
              \label{fig:L1641S_pv}%
\end{figure}

The spectrum integrated over the entire L~1641-S survey field shows two velocity components (Fig.~\ref{fig:meanspectra}). Figure~\ref{fig:intmap_L1641S} shows
maps of the CO emission integrated over the velocity ranges corresponding to each component. We see the cloud component with velocities
around 6~km/s mainly in the north-western part of the mapped field, getting fainter towards the south. The 3~km/s component is brightest
around the location of the \object{L~1641-S cluster} and its associated reflection nebula 
\citep{strometal1993, carpenter2000, allendavis2008, megeathetal2016, pillitterietal2013}
and extends mostly to its south. 

Figure~\ref{fig:L1641S_pv} shows position-velocity cuts through the L~1641-S field. The top panel shows a cut going south-east to north-west through
the northern part of the survey field, intersecting the L~1641-S cluster area (bright emission in the 3~km/s component around offsets of
2000\arcsec{}). Further to the north-west, the two components become entangled, and in the north-westernmost part, the 6~km/s component
becomes the brightest (with a velocity centred around 5~km/s). The panel in the middle shows a south-to-north cut through the eastern
part of the survey field, again intersecting the L~1641-S cluster area at its northern end. The 3~km/s CO emission has a sharp edge
in the north. Overall its shape, particularly in individual channel maps at velocities around 2~km/s, is that of a large cometary cloud,
with its head pointing north and the tail opening to the south. In the southern part of the position-velocity cut the 3~km/s component
seems to split up again, showing an additional velocity component at around 1~km/s. This third component is seen more clearly in the
east-west position-velocity cut shown in the bottom panel.

A large-scale north-to-south velocity gradient over the Orion~A cloud has long been known \citep[e.g.][]{ballyetal1987, wilsonetal2005},
with the gas exhibiting increasingly more blue-shifted velocities further south. \cite{hacaretal2016} showed that the stars associated with the cloud
follow the same trend in their radial velocities. There is also mounting evidence that the L~1641 cloud has a
significant extent perpendicular to the plane of the sky, with the northernmost part being at around 400~pc, and the southernmost
areas at up to 500~pc, as determined from Gaia distance measurements of young stellar objects in the cloud \citep[][]{grossschedletal2018}.
This inclination suggests that overall the radial velocities of the cloud and stars are correlated with their distance. We suggest that the
presence of multiple, separated CO velocity components over much of the L~1641-S survey field is due to the cloud actually being
composed of a chain of sub-clouds extending north-west to south-east and through the plane of the sky, rather than being one long,
contiguous filamentary cloud.


\section{Summary and future work}
\label{section:conclusions}

We present extensive mapping observations in the $^{12}$CO(3-2) line of the Orion~A and B GMCs. The survey covers a total
area of $\sim$2.7~square degrees, with an angular resolution of $\sim$19\arcsec{} ($\sim$7500~AU, $\sim$0.04~pc), and a sensitivity
of 0.7--0.8~K (at a velocity resolution of 0.25~km/s). The survey covers the major star formation sites in the Orion~B GMC
(L~1622, NGC~2068 and NGC~2071, the Ori~B9 area, and NGC~2023 and NGC~2024, including the Horsehead Nebula)
and a $\sim$1.1~square degree area in the southern part of the L~1641 cloud in Orion~A, complementing existing (at the time of proposal
writing) JCMT HARP observations.

In the present paper, we give an overview of the survey regions and discuss the overall velocity fields of the CO emission
(using moment maps of the line emission and position-velocity cuts). We find that the velocity fields in L~1622, Ori~B9, and L~1641-S
are characterized by narrow lines (of the order of 2--3~km/s; 'quiet' regions), while the more 'active' regions in NGC~2023, NGC~2024,
NGC~2068, and NGC~2071 (which are all associated with optical reflection nebulae, \ion{H}{II} regions, and large embedded clusters) show
larger line-widths ($\sim$4~km/s). In several regions, we see multiple velocity components (L~1622, Ori~B9, NGC~2024) indicating the
presence of more than one cloud or cloud component along the line of sight. In L~1641-S, in particular, we see up to three distinct
velocity components. Together with previous measurements showing that there is a gradient in the distance to the cloud 
\citep[more distant when going further south-east,][]{grossschedletal2018} along with a gradient in the radial velocities of the
associated stars \citep{hacaretal2016}, we interpret this finding as an indication that the southern L~1641-S cloud consists of a
sequence of clouds along the line of sight rather than a contiguous filament.

In the more 'active' regions, we see several instances of lines splitting up into two components, coinciding with regions of
optical nebulosities (most prominently in the NGC~2024 \ion{H}{II} region). We interpret this behaviour as being due to expanding bubbles
driven by early-type stars into the molecular clouds.
In particular, we rediscuss the structure of the molecular gas surrounding
the NGC~2024 \ion{H}{II} region, concluding that, contrary to previous work, the bulk of that cloud's material (particularly in the
northern part of the \ion{H}{II} region) lies in the foreground.

We identify a trend for CO emission to be slightly more red-shifted
and have narrower lines in regions that are directly exposed to incident irradiation, for example in the PDRs at the west-facing
edges of NGC~2023 and NGC~2024 (including the Horsehead Nebula) and in the NGC~2068 area. In NGC~2068, around the dense dust
ridge harbouring (among others) the protostars driving the HH~24 system of outflows and the HH~26 outflow, we see
a transition from narrow (west) to broad (east) lines right at the location of the dense ridge, which stands out as having
particularly strong ongoing star formation activity given the presence of an exceptionally high number of very red
(hence young) protostars.
Further evidence for the impact of the PDR along the western edge of the cloud in the NGC~2023/NGC2024 region is indicated by
elevated brightness ratios between our CO(3-2) data and the CO(1-0) maps of \cite{petyetal2017}, indicative of high ($> 40$~K)
  molecular gas temperatures as constrained through exploratory RADEX radiative transfer modelling. In contrast, a fainter cloud component
  seen at velocities offset by a few km/s (around 4-5~km/s) from the main cloud is found to be cooler and more tenuous, indicating that it
  is not subject to significant UV irradiation and might be located somewhere in the fore- or background of the Orion region.

We also report the discovery of a small, round CO cloud north of NGC~2071, which is seen as a cometary cloud in optical images.
The cloud apparently is not associated with any protostar or dense cloud core. It is illuminated and heated from the west, as indicated
by a bright, west-facing rim seen in Spitzer images. While further observations will be needed to determine the true column density
structure of the cloud, its circular appearance in CO suggests an underlying near spherical geometry in three dimensional space.
The simple geometry makes it a prime target for comparisons with theoretical work on cloud structure and dynamics, particularly that under
the influence of external irradiation.
Due to its simple, apparently spherical geometry, we name this cloud the "Cow~Nebula" globule.

One of the main scientific drivers for this project is to obtain an unbiased account of protostellar CO outflows over a significant
part of a GMC, covering the full extent of the outflows to measure their entire energy and momentum content.
While we defer a detailed identification and analysis of the outflow population to a subsequent paper, we here presented a comparison
of the CO(3-2) line emission for the two most prominent protostellar outflows in the Orion~B GMC (the
NGC~2071-IR outflow and the monopolar outflow in NGC~2024). In both cases we find good agreement with previous, targeted observations.
In particular we reproduce the wide velocity extent of the NGC~2071-IR outflow, including the pronounced 'bump' seen in the redshifted
lobe at extremely high velocities, confirming the validity of our data reduction approach. We confirm the monopolar nature of the
outflow in NGC~2024 in our wide-field maps, but revise its driving source to be the source FIR4 of \cite{mezgeretal1988} rather than
FIR5 as thought previously. No conclusive evidence for a blue-shifted counterlobe to the well-collimated, jet-like red-shifted outflow
lobe is seen, particularly in the immediate vicinity of the driving source, but we speculate that a patch of blue-shifted emission (B1)
further north along the outflow axis marks the point where the blue-shifted outflow lobe punches through the foreground cloud bounding
the \ion{H}{II} region.

The line moment maps (moment~1 and moment~2, i.e. mean CO velocity and CO line-widths) and numerous high-velocity spikes in 
the position-velocity cuts indicate the presence of a multitude of additional outflows in the survey fields. We will provide a full
account of these outflows in a follow-up publication, including an identification of their driving sources (making use of the
catalogue of protostars identified in the course of the Herschel Orion Protostar Survey HOPS), a characterization of the outflows
in terms of mass, momentum, and energy and their respective flow rates.
With our discussions of the structure of the NGC~2024 \ion{H}{II} region, the elevated CO(3-2)/CO(1-0) line ratios at the western
edge of the cloud in the NGC~2023/NGC~2024 region, the multi-component structure of the cloud in the L~1641-S region,
and the comparison of the ALCOHOLS CO(3-2) maps with the SOFIA [\ion{C}{II}] maps of the Horsehead Nebula and the \object{IC~434}
ionization front by \cite{ballyetal2018} \citep[see also][]{pabstetal2017} we only scratch the
surface of the potential applications of our wide-field CO(3-2) survey and future, even larger, wide-band molecular line surveys.

\begin{acknowledgements}
      Our warmest thanks go to the entire APEX crew for their continued support during the preparation of the SuperCAM
      visiting run, during installation of the instrument, commissioning, and operation, and for general hospitality on site. 
      We thank G\"{o}ran Sandell for providing us fits cubes of the NGC~2023 FLASH+ CO data.
      Part of this research was conducted at the Jet Propulsion Laboratory, California Institute of Technology under contract with the
      National Aeronautics and Space Administration. The Digitized Sky Surveys were produced at the Space Telescope Science Institute
      under U.S. Government grant NAG W-2166. The images of these surveys are based on photographic data obtained using the Oschin 
      Schmidt Telescope on Palomar Mountain and the UK Schmidt Telescope. The plates were processed into the present compressed 
      digital form with the permission of these institutions. This research made use of Astropy,\footnote{http://www.astropy.org} a community-developed core Python package for Astronomy \citep{astropy:2013, astropy:2018}.
\end{acknowledgements}

%
\bibliographystyle{aa} 
\bibliography{ALCOHOLS_survey.bib} 

\begin{thebibliography}{112}
\expandafter\ifx\csname natexlab\endcsname\relax\def\natexlab#1{#1}\fi

\bibitem[{{Allen} \& {Davis}(2008)}]{allendavis2008}
{Allen}, L.~E. \& {Davis}, C.~J. 2008, {Low Mass Star Formation in the Lynds
  1641 Molecular Cloud}, ed. B.~{Reipurth}, Vol.~4, 621

\bibitem[{{Alves} {et~al.}(2011){Alves}, {Girart}, {Lai}, {Rao}, \&
  {Zhang}}]{alvesetal2011}
{Alves}, F.~O., {Girart}, J.~M., {Lai}, S.-P., {Rao}, R., \& {Zhang}, Q. 2011,
  \apj, 726, 63

\bibitem[{{Aoyama} {et~al.}(2001){Aoyama}, {Mizuno}, {Yamamoto}, {Onishi},
  {Mizuno}, \& {Fukui}}]{aoyamaetal2001}
{Aoyama}, H., {Mizuno}, N., {Yamamoto}, H., {et~al.} 2001, \pasj, 53, 1053

\bibitem[{{Astropy Collaboration} {et~al.}(2018){Astropy Collaboration},
  {Price-Whelan}, {Sip{\H o}cz}, {G{\"u}nther}, {Lim}, {Crawford}, {Conseil},
  {Shupe}, {Craig}, {Dencheva}, {Ginsburg}, {VanderPlas}, {Bradley},
  {P{\'e}rez-Su{\'a}rez}, {de Val-Borro}, {Aldcroft}, {Cruz}, {Robitaille},
  {Tollerud}, {Ardelean}, {Babej}, {Bach}, {Bachetti}, {Bakanov}, {Bamford},
  {Barentsen}, {Barmby}, {Baumbach}, {Berry}, {Biscani}, {Boquien}, {Bostroem},
  {Bouma}, {Brammer}, {Bray}, {Breytenbach}, {Buddelmeijer}, {Burke},
  {Calderone}, {Cano Rodr{\'{\i}}guez}, {Cara}, {Cardoso}, {Cheedella},
  {Copin}, {Corrales}, {Crichton}, {D'Avella}, {Deil}, {Depagne}, {Dietrich},
  {Donath}, {Droettboom}, {Earl}, {Erben}, {Fabbro}, {Ferreira}, {Finethy},
  {Fox}, {Garrison}, {Gibbons}, {Goldstein}, {Gommers}, {Greco}, {Greenfield},
  {Groener}, {Grollier}, {Hagen}, {Hirst}, {Homeier}, {Horton}, {Hosseinzadeh},
  {Hu}, {Hunkeler}, {Ivezi{\'c}}, {Jain}, {Jenness}, {Kanarek}, {Kendrew},
  {Kern}, {Kerzendorf}, {Khvalko}, {King}, {Kirkby}, {Kulkarni}, {Kumar},
  {Lee}, {Lenz}, {Littlefair}, {Ma}, {Macleod}, {Mastropietro}, {McCully},
  {Montagnac}, {Morris}, {Mueller}, {Mumford}, {Muna}, {Murphy}, {Nelson},
  {Nguyen}, {Ninan}, {N{\"o}the}, {Ogaz}, {Oh}, {Parejko}, {Parley}, {Pascual},
  {Patil}, {Patil}, {Plunkett}, {Prochaska}, {Rastogi}, {Reddy Janga},
  {Sabater}, {Sakurikar}, {Seifert}, {Sherbert}, {Sherwood-Taylor}, {Shih},
  {Sick}, {Silbiger}, {Singanamalla}, {Singer}, {Sladen}, {Sooley},
  {Sornarajah}, {Streicher}, {Teuben}, {Thomas}, {Tremblay}, {Turner},
  {Terr{\'o}n}, {van Kerkwijk}, {de la Vega}, {Watkins}, {Weaver}, {Whitmore},
  {Woillez}, {Zabalza}, \& {Astropy Contributors}}]{astropy:2018}
{Astropy Collaboration}, {Price-Whelan}, A.~M., {Sip{\H o}cz}, B.~M., {et~al.}
  2018, \aj, 156, 123

\bibitem[{{Astropy Collaboration} {et~al.}(2013){Astropy Collaboration},
  {Robitaille}, {Tollerud}, {Greenfield}, {Droettboom}, {Bray}, {Aldcroft},
  {Davis}, {Ginsburg}, {Price-Whelan}, {Kerzendorf}, {Conley}, {Crighton},
  {Barbary}, {Muna}, {Ferguson}, {Grollier}, {Parikh}, {Nair}, {Unther},
  {Deil}, {Woillez}, {Conseil}, {Kramer}, {Turner}, {Singer}, {Fox}, {Weaver},
  {Zabalza}, {Edwards}, {Azalee Bostroem}, {Burke}, {Casey}, {Crawford},
  {Dencheva}, {Ely}, {Jenness}, {Labrie}, {Lim}, {Pierfederici}, {Pontzen},
  {Ptak}, {Refsdal}, {Servillat}, \& {Streicher}}]{astropy:2013}
{Astropy Collaboration}, {Robitaille}, T.~P., {Tollerud}, E.~J., {et~al.} 2013,
  \aap, 558, A33

\bibitem[{{Bally}(1981)}]{bally1981}
{Bally}, J. 1981, in \baas, Vol.~13, 540

\bibitem[{{Bally}(2008)}]{bally2008}
{Bally}, J. 2008, {Overview of the Orion Complex}, ed. B.~{Reipurth}, Vol.~4,
  459

\bibitem[{{Bally} {et~al.}(2018){Bally}, {Chambers}, {Guzman}, {Keto},
  {Mookerjea}, {Sandell}, {Stanke}, \& {Zinnecker}}]{ballyetal2018}
{Bally}, J., {Chambers}, E., {Guzman}, V., {et~al.} 2018, \aj, 155, 80

\bibitem[{{Bally} \& {Lada}(1983)}]{ballylada1983}
{Bally}, J. \& {Lada}, C.~J. 1983, \apj, 265, 824

\bibitem[{{Bally} {et~al.}(1987){Bally}, {Langer}, {Stark}, \&
  {Wilson}}]{ballyetal1987}
{Bally}, J., {Langer}, W.~D., {Stark}, A.~A., \& {Wilson}, R.~W. 1987, \apjl,
  312, L45

\bibitem[{{Bally} {et~al.}(2009){Bally}, {Walawender}, {Reipurth}, \&
  {Megeath}}]{ballyetal2009}
{Bally}, J., {Walawender}, J., {Reipurth}, B., \& {Megeath}, S.~T. 2009, \aj,
  137, 3843

\bibitem[{{Barnes} {et~al.}(1989){Barnes}, {Crutcher}, {Bieging}, {Storey}, \&
  {Willner}}]{barnesetal1989}
{Barnes}, P.~J., {Crutcher}, R.~M., {Bieging}, J.~H., {Storey}, J.~W.~V., \&
  {Willner}, S.~P. 1989, \apj, 342, 883

\bibitem[{{Belloche} {et~al.}(2011){Belloche}, {Schuller}, {Parise},
  {Andr{\'e}}, {Hatchell}, {J{\o}rgensen}, {Bontemps}, {Wei{\ss}}, {Menten}, \&
  {Muders}}]{bellocheetal2011}
{Belloche}, A., {Schuller}, F., {Parise}, B., {et~al.} 2011, \aap, 527, A145

\bibitem[{{Bern{\'e}} {et~al.}(2014){Bern{\'e}}, {Marcelino}, \&
  {Cernicharo}}]{berneetal2014}
{Bern{\'e}}, O., {Marcelino}, N., \& {Cernicharo}, J. 2014, \apj, 795, 13

\bibitem[{{Bernes}(1977)}]{bernes1977}
{Bernes}, C. 1977, \aaps, 29, 65

\bibitem[{{Bik} {et~al.}(2003){Bik}, {Lenorzer}, {Kaper}, {Comer{\'o}n},
  {Waters}, {de Koter}, \& {Hanson}}]{biketal2003}
{Bik}, A., {Lenorzer}, A., {Kaper}, L., {et~al.} 2003, \aap, 404, 249

\bibitem[{{Buckle} {et~al.}(2010){Buckle}, {Curtis}, {Roberts}, {White},
  {Hatchell}, {Brunt}, {Butner}, {Cavanagh}, {Chrysostomou}, {Davis},
  {Duarte-Cabral}, {Etxaluze}, {di Francesco}, {Friberg}, {Friesen}, {Fuller},
  {Graves}, {Greaves}, {Hogerheijde}, {Johnstone}, {Matthews}, {Matthews},
  {Nutter}, {Rawlings}, {Richer}, {Sadavoy}, {Simpson}, {Tothill}, {Tsamis},
  {Viti}, {Ward-Thompson}, {Wouterloot}, \& {Yates}}]{buckleetal2010}
{Buckle}, J.~V., {Curtis}, E.~I., {Roberts}, J.~F., {et~al.} 2010, \mnras, 401,
  204

\bibitem[{{Buckle} {et~al.}(2012){Buckle}, {Davis}, {di Francesco}, {Graves},
  {Nutter}, {Richer}, {Roberts}, {Ward-Thompson}, {White}, {Brunt}, {Butner},
  {Cavanagh}, {Chrysostomou}, {Curtis}, {Duarte-Cabral}, {Etxaluze}, {Fich},
  {Friberg}, {Friesen}, {Fuller}, {Greaves}, {Hatchell}, {Hogerheijde},
  {Johnstone}, {Matthews}, {Matthews}, {Rawlings}, {Sadavoy}, {Simpson},
  {Tothill}, {Tsamis}, {Viti}, {Wouterloot}, \& {Yates}}]{buckleetal2012}
{Buckle}, J.~V., {Davis}, C.~J., {di Francesco}, J., {et~al.} 2012, \mnras,
  422, 521

\bibitem[{{Carpenter}(2000)}]{carpenter2000}
{Carpenter}, J.~M. 2000, \aj, 120, 3139

\bibitem[{{Caselli} \& {Myers}(1995)}]{casellimyers1995}
{Caselli}, P. \& {Myers}, P.~C. 1995, \apj, 446, 665

\bibitem[{{Castets} {et~al.}(1990){Castets}, {Duvert}, {Dutrey}, {Bally},
  {Langer}, \& {Wilson}}]{castetsetal1990}
{Castets}, A., {Duvert}, G., {Dutrey}, A., {et~al.} 1990, \aap, 234, 469

\bibitem[{{Chandler} \& {Carlstrom}(1996)}]{chandlercarlstrom1996}
{Chandler}, C.~J. \& {Carlstrom}, J.~E. 1996, \apj, 466, 338

\bibitem[{{Chernin}(1996)}]{chernin1996}
{Chernin}, L.~M. 1996, \apj, 460, 711

\bibitem[{{Chernin} \& {Masson}(1992)}]{cherninmasson1992}
{Chernin}, L.~M. \& {Masson}, C.~R. 1992, \apjl, 396, L35

\bibitem[{{Chernin} \& {Welch}(1995)}]{cherninwelch1995}
{Chernin}, L.~M. \& {Welch}, W.~J. 1995, \apjl, 440, L21

\bibitem[{{Choi} {et~al.}(1993){Choi}, {Evans}, \& {Jaffe}}]{choietal1993}
{Choi}, M., {Evans}, Neal~J., I., \& {Jaffe}, D.~T. 1993, \apj, 417, 624

\bibitem[{{Choi} {et~al.}(2015){Choi}, {Kang}, \& {Lee}}]{choietal2015}
{Choi}, M., {Kang}, M., \& {Lee}, J.-E. 2015, \aj, 150, 29

\bibitem[{{Curtis} {et~al.}(2010){Curtis}, {Richer}, \&
  {Buckle}}]{curtisetal2010}
{Curtis}, E.~I., {Richer}, J.~S., \& {Buckle}, J.~V. 2010, \mnras, 401, 455

\bibitem[{{Davis} {et~al.}(2010){Davis}, {Chrysostomou}, {Hatchell},
  {Wouterloot}, {Buckle}, {Nutter}, {Fich}, {Brunt}, {Butner}, {Cavanagh},
  {Curtis}, {Duarte-Cabral}, {di Francesco}, {Etxaluze}, {Friberg}, {Friesen},
  {Fuller}, {Graves}, {Greaves}, {Hogerheijde}, {Johnstone}, {Matthews},
  {Matthews}, {Rawlings}, {Richer}, {Roberts}, {Sadavoy}, {Simpson}, {Tothill},
  {Tsamis}, {Viti}, {Ward-Thompson}, {White}, \& {Yates}}]{davisetal2010}
{Davis}, C.~J., {Chrysostomou}, A., {Hatchell}, J., {et~al.} 2010, \mnras, 405,
  759

\bibitem[{{Dorschner} \& {G{\"u}rtler}(1963)}]{dorschnerguertler1963}
{Dorschner}, J. \& {G{\"u}rtler}, J. 1963, Astronomische Nachrichten, 287, 257

\bibitem[{{Emprechtinger} {et~al.}(2009){Emprechtinger}, {Wiedner}, {Simon},
  {Wieching}, {Volgenau}, {Bielau}, {Graf}, {G{\"u}sten}, {Honingh}, {Jacobs},
  {Rabanus}, {Stutzki}, \& {Wyrowski}}]{emprechtingeretal2009}
{Emprechtinger}, M., {Wiedner}, M.~C., {Simon}, R., {et~al.} 2009, \aap, 496,
  731

\bibitem[{{Enokiya} {et~al.}(2021){Enokiya}, {Ohama}, {Yamada}, {Sano},
  {Fujita}, {Hayashi}, {Tsutsumi}, {Torii}, {Nishimura}, {Konishi}, {Yamamoto},
  {Tachihara}, {Hasegawa}, {Kimura}, {Ogawa}, \& {Fukui}}]{enokiyaetal2021}
{Enokiya}, R., {Ohama}, A., {Yamada}, R., {et~al.} 2021, \pasj, 73, S256

\bibitem[{{Furlan} {et~al.}(2016){Furlan}, {Fischer}, {Ali}, {Stutz}, {Stanke},
  {Tobin}, {Megeath}, {Osorio}, {Hartmann}, {Calvet}, {Poteet}, {Booker},
  {Manoj}, {Watson}, \& {Allen}}]{furlanetal2016}
{Furlan}, E., {Fischer}, W.~J., {Ali}, B., {et~al.} 2016, \apjs, 224, 5

\bibitem[{{Gaume} {et~al.}(1992){Gaume}, {Johnston}, \&
  {Wilson}}]{gaumeetal1992}
{Gaume}, R.~A., {Johnston}, K.~J., \& {Wilson}, T.~L. 1992, \apj, 388, 489

\bibitem[{{Genzel} \& {Stutzki}(1989)}]{genzelstutzki1989}
{Genzel}, R. \& {Stutzki}, J. 1989, \araa, 27, 41

\bibitem[{{Gibb}(2008)}]{gibb2008}
{Gibb}, A.~G. 2008, {Star Formation in NGC 2068, NGC 2071, and Northern L1630},
  ed. B.~{Reipurth}, Vol.~4, 693

\bibitem[{{Graf} {et~al.}(1993){Graf}, {Eckart}, {Genzel}, {Harris},
  {Poglitsch}, {Russell}, \& {Stutzki}}]{grafetal1993}
{Graf}, U.~U., {Eckart}, A., {Genzel}, R., {et~al.} 1993, \apj, 405, 249

\bibitem[{{Graves} {et~al.}(2010){Graves}, {Richer}, {Buckle}, {Duarte-Cabral},
  {Fuller}, {Hogerheijde}, {Owen}, {Brunt}, {Butner}, {Cavanagh},
  {Chrysostomou}, {Curtis}, {Davis}, {Etxaluze}, {di Francesco}, {Friberg},
  {Friesen}, {Greaves}, {Hatchell}, {Johnstone}, {Matthews}, {Matthews},
  {Matzner}, {Nutter}, {Rawlings}, {Roberts}, {Sadavoy}, {Simpson}, {Tothill},
  {Tsamis}, {Viti}, {Ward-Thompson}, {White}, {Wouterloot}, \&
  {Yates}}]{gravesetal2010}
{Graves}, S.~F., {Richer}, J.~S., {Buckle}, J.~V., {et~al.} 2010, \mnras, 409,
  1412

\bibitem[{{Gro{\ss}schedl} {et~al.}(2018){Gro{\ss}schedl}, {Alves}, {Meingast},
  {Ackerl}, {Ascenso}, {Bouy}, {Burkert}, {Forbrich}, {F{\"u}rnkranz},
  {Goodman}, {Hacar}, {Herbst-Kiss}, {Lada}, {Larreina}, {Leschinski},
  {Lombardi}, {Moitinho}, {Mortimer}, \& {Zari}}]{grossschedletal2018}
{Gro{\ss}schedl}, J.~E., {Alves}, J., {Meingast}, S., {et~al.} 2018, \aap, 619,
  A106

\bibitem[{{G{\"u}sten} {et~al.}(2006){G{\"u}sten}, {Nyman}, {Schilke},
  {Menten}, {Cesarsky}, \& {Booth}}]{guestenetal2006}
{G{\"u}sten}, R., {Nyman}, L.~{\r{A}}., {Schilke}, P., {et~al.} 2006, \aap,
  454, L13

\bibitem[{{Hacar} {et~al.}(2016){Hacar}, {Alves}, {Forbrich}, {Meingast},
  {Kubiak}, \& {Gro{\ss}schedl}}]{hacaretal2016}
{Hacar}, A., {Alves}, J., {Forbrich}, J., {et~al.} 2016, \aap, 589, A80

\bibitem[{{Harju} {et~al.}(1993){Harju}, {Walmsley}, \&
  {Wouterloot}}]{harjuetal1993}
{Harju}, J., {Walmsley}, C.~M., \& {Wouterloot}, J.~G.~A. 1993, \aaps, 98, 51

\bibitem[{{Ishii} {et~al.}(2019){Ishii}, {Nakamura}, {Shimajiri}, {Kawabe},
  {Tsukagoshi}, {Dobashi}, \& {Shimoikura}}]{ishiietal2019}
{Ishii}, S., {Nakamura}, F., {Shimajiri}, Y., {et~al.} 2019, \pasj, 87

\bibitem[{{Ishii} {et~al.}(2016){Ishii}, {Seta}, {Nagai}, {Miyamoto}, {Nakai},
  {Nagasaki}, {Arai}, {Imada}, {Miyagawa}, {Maezawa}, {Maehashi}, {Bronfman},
  \& {Finger}}]{ishiietal2016}
{Ishii}, S., {Seta}, M., {Nagai}, M., {et~al.} 2016, \pasj, 68, 10

\bibitem[{{Kirk} {et~al.}(2016){Kirk}, {Di Francesco}, {Johnstone},
  {Duarte-Cabral}, {Sadavoy}, {Hatchell}, {Mottram}, {Buckle}, {Berry},
  {Broekhoven-Fiene}, {Currie}, {Fich}, {Jenness}, {Nutter}, {Pattle},
  {Pineda}, {Quinn}, {Salji}, {Tisi}, {Hogerheijde}, {Ward-Thompson},
  {Bastien}, {Bresnahan}, {Butner}, {Chen}, {Chrysostomou}, {Coude}, {Davis},
  {Drabek-Maunder}, {Fiege}, {Friberg}, {Friesen}, {Fuller}, {Graves},
  {Greaves}, {Gregson}, {Holland}, {Joncas}, {Kirk}, {Knee}, {Mairs}, {Marsh},
  {Matthews}, {Moriarty-Schieven}, {Mowat}, {Rawlings}, {Richer}, {Robertson},
  {Rosolowsky}, {Rumble}, {Thomas}, {Tothill}, {Viti}, {White}, {Wouterloot},
  {Yates}, \& {Zhu}}]{kirketal2016}
{Kirk}, H., {Di Francesco}, J., {Johnstone}, D., {et~al.} 2016, \apj, 817, 167

\bibitem[{{Klein} {et~al.}(2014){Klein}, {Ciechanowicz}, {Leinz}, {Heyminck},
  {Gusten}, {Kasemann}, {Wunsch}, {Maier}, \& {Sekimoto}}]{kleinetal2014}
{Klein}, T., {Ciechanowicz}, M., {Leinz}, C., {et~al.} 2014, IEEE Transactions
  on Terahertz Science and Technology, 4, 588

\bibitem[{{Kloosterman} {et~al.}(2012){Kloosterman}, {Cottam}, {Swift},
  {Lesser}, {Schickling}, {Groppi}, {Borden}, {Towner}, {Schmidt}, {Kulesa},
  {d'Aubigny}, {Walker}, {Golish}, {Weinreb}, {Jones}, {Mani}, {Kooi},
  {Lichtenberger}, {Puetz}, \& {Narayanan}}]{kloostermanetal2012}
{Kloosterman}, J., {Cottam}, T., {Swift}, B., {et~al.} 2012, in Society of
  Photo-Optical Instrumentation Engineers (SPIE) Conference Series, Vol. 8452,
  \procspie, 845204

\bibitem[{{Kloosterman}(2014)}]{kloosterman2014}
{Kloosterman}, J.~L. 2014, PhD thesis, The University of Arizona

\bibitem[{{Kong} {et~al.}(2018){Kong}, {Arce}, {Feddersen}, {Carpenter},
  {Nakamura}, {Shimajiri}, {Isella}, {Ossenkopf-Okada}, {Sargent},
  {S{\'a}nchez-Monge}, {Suri}, {Kauffmann}, {Pillai}, {Pineda}, {Koda},
  {Bally}, {Lis}, {Padoan}, {Klessen}, {Mairs}, {Goodman}, {Goldsmith},
  {McGehee}, {Schilke}, {Teuben}, {Maureira}, {Hara}, {Ginsburg}, {Burkhart},
  {Smith}, {Schmiedeke}, {Pineda}, {Ishii}, {Sasaki}, {Kawabe}, {Urasawa},
  {Oyamada}, \& {Tanabe}}]{kongetal2018}
{Kong}, S., {Arce}, H.~G., {Feddersen}, J.~R., {et~al.} 2018, \apjs, 236, 25

\bibitem[{{Kounkel} {et~al.}(2017){Kounkel}, {Hartmann}, {Loinard},
  {Ortiz-Le{\'o}n}, {Mioduszewski}, {Rodr{\'\i}guez}, {Dzib}, {Torres}, {Pech},
  {Galli}, {Rivera}, {Boden}, {Evans}, {Brice{\~n}o}, \&
  {Tobin}}]{kounkeletal2017}
{Kounkel}, M., {Hartmann}, L., {Loinard}, L., {et~al.} 2017, \apj, 834, 142

\bibitem[{{Kramer} {et~al.}(1996){Kramer}, {Stutzki}, \&
  {Winnewisser}}]{krameretal1996}
{Kramer}, C., {Stutzki}, J., \& {Winnewisser}, G. 1996, \aap, 307, 915

\bibitem[{{Kr\"{u}gel} {et~al.}(1982){Kr\"{u}gel}, {Thum}, {Pankonin}, \&
  {Martin-Pintado}}]{kruegeletal1982}
{Kr\"{u}gel}, E., {Thum}, C., {Pankonin}, V., \& {Martin-Pintado}, J. 1982,
  \aaps, 48, 345

\bibitem[{{Kun} {et~al.}(2008){Kun}, {Balog}, {Mizuno}, {Kawamura},
  {G{\'a}sp{\'a}r}, {Kenyon}, \& {Fukui}}]{kunetal2008}
{Kun}, M., {Balog}, Z., {Mizuno}, N., {et~al.} 2008, \mnras, 391, 84

\bibitem[{{Kutner} {et~al.}(1977){Kutner}, {Tucker}, {Chin}, \&
  {Thaddeus}}]{kutneretal1977}
{Kutner}, M.~L., {Tucker}, K.~D., {Chin}, G., \& {Thaddeus}, P. 1977, \apj,
  215, 521

\bibitem[{{Lada} {et~al.}(1991){Lada}, {Bally}, \& {Stark}}]{ladaetal1991}
{Lada}, E.~A., {Bally}, J., \& {Stark}, A.~A. 1991, \apj, 368, 432

\bibitem[{{Lichten}(1982)}]{lichten1982}
{Lichten}, S.~M. 1982, \apj, 253, 593

\bibitem[{{Lombardi} {et~al.}(2014){Lombardi}, {Bouy}, {Alves}, \&
  {Lada}}]{lombardietal2014}
{Lombardi}, M., {Bouy}, H., {Alves}, J., \& {Lada}, C.~J. 2014, \aap, 566, A45

\bibitem[{{Maddalena} {et~al.}(1986){Maddalena}, {Morris}, {Moscowitz}, \&
  {Thaddeus}}]{maddalenaetal1986}
{Maddalena}, R.~J., {Morris}, M., {Moscowitz}, J., \& {Thaddeus}, P. 1986,
  \apj, 303, 375

\bibitem[{{Magakian}(2003)}]{magakian2003}
{Magakian}, T.~Y. 2003, \aap, 399, 141

\bibitem[{{Margulis} \& {Snell}(1989)}]{margulissnell1989}
{Margulis}, M. \& {Snell}, R.~L. 1989, \apj, 343, 779

\bibitem[{{Matthews} {et~al.}(2002){Matthews}, {Fiege}, \&
  {Moriarty-Schieven}}]{matthewsetal2002}
{Matthews}, B.~C., {Fiege}, J.~D., \& {Moriarty-Schieven}, G. 2002, \apj, 569,
  304

\bibitem[{{Megeath} {et~al.}(2012){Megeath}, {Gutermuth}, {Muzerolle},
  {Kryukova}, {Flaherty}, {Hora}, {Allen}, {Hartmann}, {Myers}, {Pipher},
  {Stauffer}, {Young}, \& {Fazio}}]{megeathetal2012}
{Megeath}, S.~T., {Gutermuth}, R., {Muzerolle}, J., {et~al.} 2012, \aj, 144,
  192

\bibitem[{{Megeath} {et~al.}(2016){Megeath}, {Gutermuth}, {Muzerolle},
  {Kryukova}, {Hora}, {Allen}, {Flaherty}, {Hartmann}, {Myers}, {Pipher},
  {Stauffer}, {Young}, \& {Fazio}}]{megeathetal2016}
{Megeath}, S.~T., {Gutermuth}, R., {Muzerolle}, J., {et~al.} 2016, \aj, 151, 5

\bibitem[{{Menten} {et~al.}(2007){Menten}, {Reid}, {Forbrich}, \&
  {Brunthaler}}]{mentenetal2007}
{Menten}, K.~M., {Reid}, M.~J., {Forbrich}, J., \& {Brunthaler}, A. 2007, \aap,
  474, 515

\bibitem[{{Meyer} {et~al.}(2008){Meyer}, {Flaherty}, {Levine}, {Lada},
  {Bowler}, \& {Kandori}}]{meyeretal2008}
{Meyer}, M.~R., {Flaherty}, K., {Levine}, J.~L., {et~al.} 2008, {Star Formation
  in NGC 2023, NGC 2024, and Southern L1630}, ed. B.~{Reipurth}, Vol.~4, 662

\bibitem[{{Mezger} {et~al.}(1988){Mezger}, {Chini}, {Kreysa}, {Wink}, \&
  {Salter}}]{mezgeretal1988}
{Mezger}, P.~G., {Chini}, R., {Kreysa}, E., {Wink}, J.~E., \& {Salter}, C.~J.
  1988, \aap, 191, 44

\bibitem[{{Mezger} {et~al.}(1992){Mezger}, {Sievers}, {Haslam}, {Kreysa},
  {Lemke}, {Mauersberger}, \& {Wilson}}]{mezgeretal1992}
{Mezger}, P.~G., {Sievers}, A.~W., {Haslam}, C.~G.~T., {et~al.} 1992, \aap,
  256, 631

\bibitem[{{Miettinen} {et~al.}(2009){Miettinen}, {Harju}, {Haikala},
  {Kainulainen}, \& {Johansson}}]{miettinenetal2009}
{Miettinen}, O., {Harju}, J., {Haikala}, L.~K., {Kainulainen}, J., \&
  {Johansson}, L.~E.~B. 2009, \aap, 500, 845

\bibitem[{{Moriarty-Schieven} {et~al.}(1989){Moriarty-Schieven}, {Snell}, \&
  {Hughes}}]{moriartyschievenetal1989}
{Moriarty-Schieven}, G.~H., {Snell}, R.~L., \& {Hughes}, V.~A. 1989, \apj, 347,
  358

\bibitem[{{Motte} {et~al.}(2001){Motte}, {Andr{\'e}}, {Ward-Thompson}, \&
  {Bontemps}}]{motteetal2001}
{Motte}, F., {Andr{\'e}}, P., {Ward-Thompson}, D., \& {Bontemps}, S. 2001,
  \aap, 372, L41

\bibitem[{{Muders} \& {Hafok}(2019)}]{mudershafok2019}
{Muders}, D. \& {Hafok}, H. 2019, APEX Calibration and Data Reduction Manual

\bibitem[{{Nagahama} {et~al.}(1998){Nagahama}, {Mizuno}, {Ogawa}, \&
  {Fukui}}]{nagahamaetal1998}
{Nagahama}, T., {Mizuno}, A., {Ogawa}, H., \& {Fukui}, Y. 1998, \aj, 116, 336

\bibitem[{{Nakamura} {et~al.}(2019){Nakamura}, {Ishii}, {Dobashi},
  {Shimoikura}, {Shimajiri}, {Kawabe}, {Tanabe}, {Hirose}, {Oyamada},
  {Urasawa}, {Takemura}, {Tsukagoshi}, {Momose}, {Sugitani}, {Nishi},
  {Okumura}, {Sanhueza}, {Nguyen-Luong}, \& {Kusune}}]{nakamuraetal2019}
{Nakamura}, F., {Ishii}, S., {Dobashi}, K., {et~al.} 2019, \pasj, 71, S3

\bibitem[{{Nakamura} {et~al.}(2012){Nakamura}, {Miura}, {Kitamura},
  {Shimajiri}, {Kawabe}, {Akashi}, {Ikeda}, {Tsukagoshi}, {Momose}, {Nishi}, \&
  {Li}}]{nakamuraetal2012}
{Nakamura}, F., {Miura}, T., {Kitamura}, Y., {et~al.} 2012, \apj, 746, 25

\bibitem[{{Nishimura} {et~al.}(2015){Nishimura}, {Tokuda}, {Kimura}, {Muraoka},
  {Maezawa}, {Ogawa}, {Dobashi}, {Shimoikura}, {Mizuno}, {Fukui}, \&
  {Onishi}}]{nishimuraetal2015}
{Nishimura}, A., {Tokuda}, K., {Kimura}, K., {et~al.} 2015, \apjs, 216, 18

\bibitem[{{Pabst} {et~al.}(2020){Pabst}, {Goicoechea}, {Teyssier}, {Bern{\'e}},
  {Higgins}, {Chambers}, {Kabanovic}, {G{\"u}sten}, {Stutzki}, \&
  {Tielens}}]{pabstetal2020}
{Pabst}, C.~H.~M., {Goicoechea}, J.~R., {Teyssier}, D., {et~al.} 2020, \aap,
  639, A2

\bibitem[{{Pabst} {et~al.}(2017){Pabst}, {Goicoechea}, {Teyssier}, {Bern{\'e}},
  {Ochsendorf}, {Wolfire}, {Higgins}, {Riquelme}, {Risacher}, {Pety}, {Le
  Petit}, {Roueff}, {Bron}, \& {Tielens}}]{pabstetal2017}
{Pabst}, C.~H.~M., {Goicoechea}, J.~R., {Teyssier}, D., {et~al.} 2017, \aap,
  606, A29

\bibitem[{{Pety} {et~al.}(2017){Pety}, {Guzm{\'a}n}, {Orkisz}, {Liszt},
  {Gerin}, {Bron}, {Bardeau}, {Goicoechea}, {Gratier}, {Le Petit}, {Levrier},
  {{\"O}berg}, {Roueff}, \& {Sievers}}]{petyetal2017}
{Pety}, J., {Guzm{\'a}n}, V.~V., {Orkisz}, J.~H., {et~al.} 2017, \aap, 599, A98

\bibitem[{{Pillitteri} {et~al.}(2013){Pillitteri}, {Wolk}, {Megeath}, {Allen},
  {Bally}, {Gagn{\'e}}, {Gutermuth}, {Hartmann}, {Micela}, {Myers}, {Oliveira},
  {Sciortino}, {Walter}, {Rebull}, \& {Stauffer}}]{pillitterietal2013}
{Pillitteri}, I., {Wolk}, S.~J., {Megeath}, S.~T., {et~al.} 2013, \apj, 768, 99

\bibitem[{{Reipurth} {et~al.}(2010){Reipurth}, {Herbig}, \&
  {Aspin}}]{reipurthetal2010}
{Reipurth}, B., {Herbig}, G., \& {Aspin}, C. 2010, \aj, 139, 1668

\bibitem[{{Reipurth} \& {Madsen}(1989)}]{reipurthmadsen1989}
{Reipurth}, B. \& {Madsen}, C. 1989, The Messenger, 55, 32

\bibitem[{{Reipurth} {et~al.}(2008){Reipurth}, {Megeath}, {Bally}, \&
  {Walawender}}]{reipurthetal2008}
{Reipurth}, B., {Megeath}, S.~T., {Bally}, J., \& {Walawender}, J. 2008, {The
  L1617 and L1622 Cometary Clouds in Orion}, ed. B.~{Reipurth}, Vol.~4, 782

\bibitem[{{Ren} \& {Li}(2016)}]{renli2016}
{Ren}, Z. \& {Li}, D. 2016, \apj, 824, 52

\bibitem[{{Richer} {et~al.}(1992){Richer}, {Hills}, \&
  {Padman}}]{richeretal1992}
{Richer}, J.~S., {Hills}, R.~E., \& {Padman}, R. 1992, \mnras, 254, 525

\bibitem[{{Richer} {et~al.}(1989){Richer}, {Hills}, {Padman}, \&
  {Russell}}]{richeretal1989}
{Richer}, J.~S., {Hills}, R.~E., {Padman}, R., \& {Russell}, A. P.~G. 1989,
  \mnras, 241, 231

\bibitem[{{Ripple} {et~al.}(2013){Ripple}, {Heyer}, {Gutermuth}, {Snell}, \&
  {Brunt}}]{rippleetal2013}
{Ripple}, F., {Heyer}, M.~H., {Gutermuth}, R., {Snell}, R.~L., \& {Brunt},
  C.~M. 2013, \mnras, 431, 1296

\bibitem[{{Sakamoto} {et~al.}(1997){Sakamoto}, {Hasegawa}, {Hayashi}, {Morino},
  \& {Sato}}]{sakamotoetal1997}
{Sakamoto}, S., {Hasegawa}, T., {Hayashi}, M., {Morino}, J.-I., \& {Sato}, K.
  1997, \apj, 481, 302

\bibitem[{{Sakamoto} {et~al.}(1994){Sakamoto}, {Hayashi}, {Hasegawa}, {Handa},
  \& {Oka}}]{sakamotoetal1994}
{Sakamoto}, S., {Hayashi}, M., {Hasegawa}, T., {Handa}, T., \& {Oka}, T. 1994,
  \apj, 425, 641

\bibitem[{{Sandell} {et~al.}(1999){Sandell}, {Avery}, {Baas}, {Coulson},
  {Dent}, {Friberg}, {Gear}, {Greaves}, {Holland}, {Jenness}, {Jewell},
  {Lightfoot}, {Matthews}, {Moriarty-Schieven}, {Prestage}, {Robson},
  {Stevens}, {Tilanus}, \& {Watt}}]{sandelletal1999}
{Sandell}, G., {Avery}, L.~W., {Baas}, F., {et~al.} 1999, \apj, 519, 236

\bibitem[{{Sandell} {et~al.}(2015){Sandell}, {Mookerjea}, {G{\"u}sten},
  {Requena-Torres}, {Riquelme}, \& {Okada}}]{sandelletal2015}
{Sandell}, G., {Mookerjea}, B., {G{\"u}sten}, R., {et~al.} 2015, \aap, 578, A41

\bibitem[{{Sanders} \& {Willner}(1985)}]{sanderswillner1985}
{Sanders}, D.~B. \& {Willner}, S.~P. 1985, \apjl, 293, L39

\bibitem[{{Shimajiri} {et~al.}(2011){Shimajiri}, {Kawabe}, {Takakuwa}, {Saito},
  {Tsukagoshi}, {Momose}, {Ikeda}, {Akiyama}, {Austermann}, {Ezawa}, {Fukue},
  {Hiramatsu}, {Hughes}, {Kitamura}, {Kohno}, {Kurono}, {Scott}, {Wilson},
  {Yoshida}, \& {Yun}}]{shimajirietal2011}
{Shimajiri}, Y., {Kawabe}, R., {Takakuwa}, S., {et~al.} 2011, \pasj, 63, 105

\bibitem[{{Siringo} {et~al.}(2009){Siringo}, {Kreysa}, {Kov{\'a}cs},
  {Schuller}, {Wei{\ss}}, {Esch}, {Gem{\"u}nd}, {Jethava}, {Lundershausen},
  {Colin}, {G{\"u}sten}, {Menten}, {Beelen}, {Bertoldi}, {Beeman}, \&
  {Haller}}]{siringoetal2009}
{Siringo}, G., {Kreysa}, E., {Kov{\'a}cs}, A., {et~al.} 2009, \aap, 497, 945

\bibitem[{{Snell} {et~al.}(1984){Snell}, {Scoville}, {Sanders}, \&
  {Erickson}}]{snelletal1984}
{Snell}, R.~L., {Scoville}, N.~Z., {Sanders}, D.~B., \& {Erickson}, N.~R. 1984,
  \apj, 284, 176

\bibitem[{{Strom} {et~al.}(1993){Strom}, {Strom}, \& {Merrill}}]{strometal1993}
{Strom}, K.~M., {Strom}, S.~E., \& {Merrill}, K.~M. 1993, \apj, 412, 233

\bibitem[{{Stutz}(2018)}]{stutz2018}
{Stutz}, A.~M. 2018, \mnras, 473, 4890

\bibitem[{{Stutz} \& {Gould}(2016)}]{stutzgould2016}
{Stutz}, A.~M. \& {Gould}, A. 2016, \aap, 590, A2

\bibitem[{{Stutz} \& {Kainulainen}(2015)}]{stutzkainulainen2015}
{Stutz}, A.~M. \& {Kainulainen}, J. 2015, \aap, 577, L6

\bibitem[{{Stutz} {et~al.}(2013){Stutz}, {Tobin}, {Stanke}, {Megeath},
  {Fischer}, {Robitaille}, {Henning}, {Ali}, {di Francesco}, {Furlan},
  {Hartmann}, {Osorio}, {Wilson}, {Allen}, {Krause}, \&
  {Manoj}}]{stutzetal2013}
{Stutz}, A.~M., {Tobin}, J.~J., {Stanke}, T., {et~al.} 2013, \apj, 767, 36

\bibitem[{{Subrahmanyan} {et~al.}(1997){Subrahmanyan}, {Goss}, {Megeath}, \&
  {Barnes}}]{subrahmanyanetal1997}
{Subrahmanyan}, R., {Goss}, W.~M., {Megeath}, S.~T., \& {Barnes}, P.~J. 1997,
  \mnras, 290, 431

\bibitem[{{Takahashi} {et~al.}(2008){Takahashi}, {Saito}, {Ohashi}, {Kusakabe},
  {Takakuwa}, {Shimajiri}, {Tamura}, \& {Kawabe}}]{takahashietal2008}
{Takahashi}, S., {Saito}, M., {Ohashi}, N., {et~al.} 2008, \apj, 688, 344

\bibitem[{{Ulich} \& {Haas}(1976)}]{ulichhaas1976}
{Ulich}, B.~L. \& {Haas}, R.~W. 1976, \apjs, 30, 247

\bibitem[{{van der Tak} {et~al.}(2007){van der Tak}, {Black}, {Sch{\"o}ier},
  {Jansen}, \& {van Dishoeck}}]{vandertaketal2007}
{van der Tak}, F.~F.~S., {Black}, J.~H., {Sch{\"o}ier}, F.~L., {Jansen}, D.~J.,
  \& {van Dishoeck}, E.~F. 2007, \aap, 468, 627

\bibitem[{{Walther} \& {Geballe}(2019)}]{walthergeballe2019}
{Walther}, D.~M. \& {Geballe}, T.~R. 2019, \apj, 875, 153

\bibitem[{{Ward-Thompson} {et~al.}(2007){Ward-Thompson}, {Di Francesco},
  {Hatchell}, {Hogerheijde}, {Nutter}, {Bastien}, {Basu}, {Bonnell}, {Bowey},
  {Brunt}, {Buckle}, {Butner}, {Cavanagh}, {Chrysostomou}, {Curtis}, {Davis},
  {Dent}, {van Dishoeck}, {Edmunds}, {Fich}, {Fiege}, {Fissel}, {Friberg},
  {Friesen}, {Frieswijk}, {Fuller}, {Gosling}, {Graves}, {Greaves}, {Helmich},
  {Hills}, {Holland}, {Houde}, {Jayawardhana}, {Johnstone}, {Joncas}, {Kirk},
  {Kirk}, {Knee}, {Matthews}, {Matthews}, {Matzner}, {Moriarty-Schieven},
  {Naylor}, {Padman}, {Plume}, {Rawlings}, {Redman}, {Reid}, {Richer},
  {Shipman}, {Simpson}, {Spaans}, {Stamatellos}, {Tsamis}, {Viti}, {Weferling},
  {White}, {Whitworth}, {Wouterloot}, {Yates}, \& {Zhu}}]{wardthompsonetal2007}
{Ward-Thompson}, D., {Di Francesco}, J., {Hatchell}, J., {et~al.} 2007, \pasp,
  119, 855

\bibitem[{{Watanabe} \& {Mitchell}(2008)}]{watanabemitchell2008}
{Watanabe}, T. \& {Mitchell}, G.~F. 2008, \aj, 136, 1947

\bibitem[{{Weiland} {et~al.}(2011){Weiland}, {Odegard}, {Hill}, {Wollack},
  {Hinshaw}, {Greason}, {Jarosik}, {Page}, {Bennett}, {Dunkley}, {Gold},
  {Halpern}, {Kogut}, {Komatsu}, {Larson}, {Limon}, {Meyer}, {Nolta}, {Smith},
  {Spergel}, {Tucker}, \& {Wright}}]{weilandetal2011}
{Weiland}, J.~L., {Odegard}, N., {Hill}, R.~S., {et~al.} 2011, \apjs, 192, 19

\bibitem[{{White} {et~al.}(2015){White}, {Drabek-Maunder}, {Rosolowsky},
  {Ward-Thompson}, {Davis}, {Gregson}, {Hatchell}, {Etxaluze}, {Stickler},
  {Buckle}, {Johnstone}, {Friesen}, {Sadavoy}, {Natt}, {Currie}, {Richer},
  {Pattle}, {Spaans}, {di Francesco}, \& {Hogerheijde}}]{whiteetal2015}
{White}, G.~J., {Drabek-Maunder}, E., {Rosolowsky}, E., {et~al.} 2015, \mnras,
  447, 1996

\bibitem[{{Wilson} {et~al.}(2005){Wilson}, {Dame}, {Masheder}, \&
  {Thaddeus}}]{wilsonetal2005}
{Wilson}, B.~A., {Dame}, T.~M., {Masheder}, M.~R.~W., \& {Thaddeus}, P. 2005,
  \aap, 430, 523

\bibitem[{{Wilson} {et~al.}(1970){Wilson}, {Jefferts}, \&
  {Penzias}}]{wilsonetal1970}
{Wilson}, R.~W., {Jefferts}, K.~B., \& {Penzias}, A.~A. 1970, \apjl, 161, L43

\bibitem[{{Zhang} {et~al.}(2019){Zhang}, {Li}, {Zhou}, {Yuan}, \&
  {Zhu}}]{zhangetal2019}
{Zhang}, C.-P., {Li}, G.-X., {Zhou}, C., {Yuan}, L., \& {Zhu}, M. 2019, \aap,
  631, A110

\bibitem[{{Zucker} {et~al.}(2019){Zucker}, {Speagle}, {Schlafly}, {Green},
  {Finkbeiner}, {Goodman}, \& {Alves}}]{zuckeretal2019}
{Zucker}, C., {Speagle}, J.~S., {Schlafly}, E.~F., {et~al.} 2019, \apj, 879,
  125

\end{thebibliography}
%

\begin{appendix} 
\section{Noise maps and additional rms statistics}
\label{app:noisemaps}

\begin{figure}
   \centering
   \includegraphics[width=\hsize]{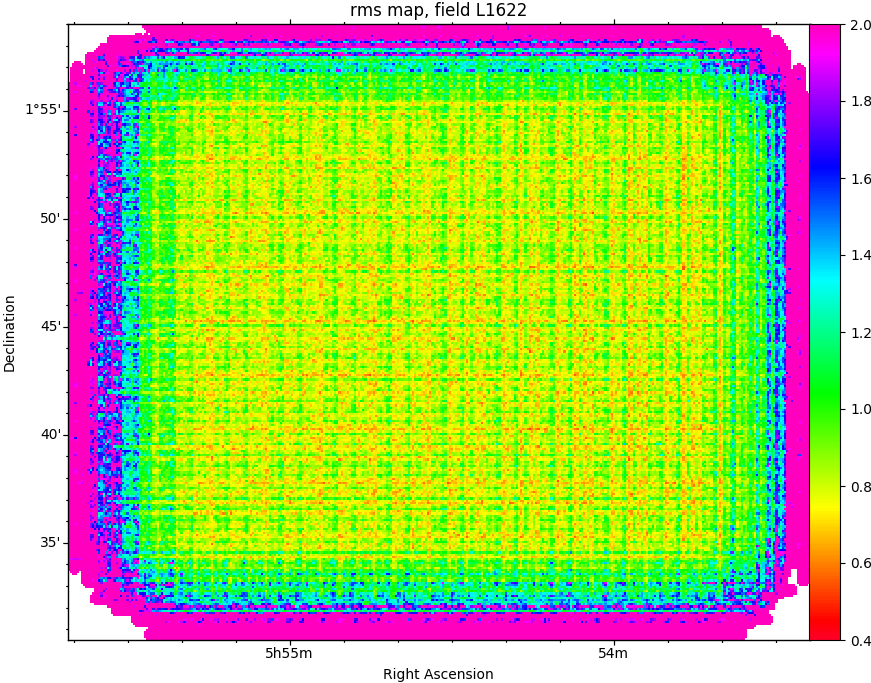}
      \caption{Noise map for the L1622 field.
              }
         \label{fig:rmsmap_L1622}
\end{figure}

\begin{figure}
   \centering
   \includegraphics[width=\hsize]{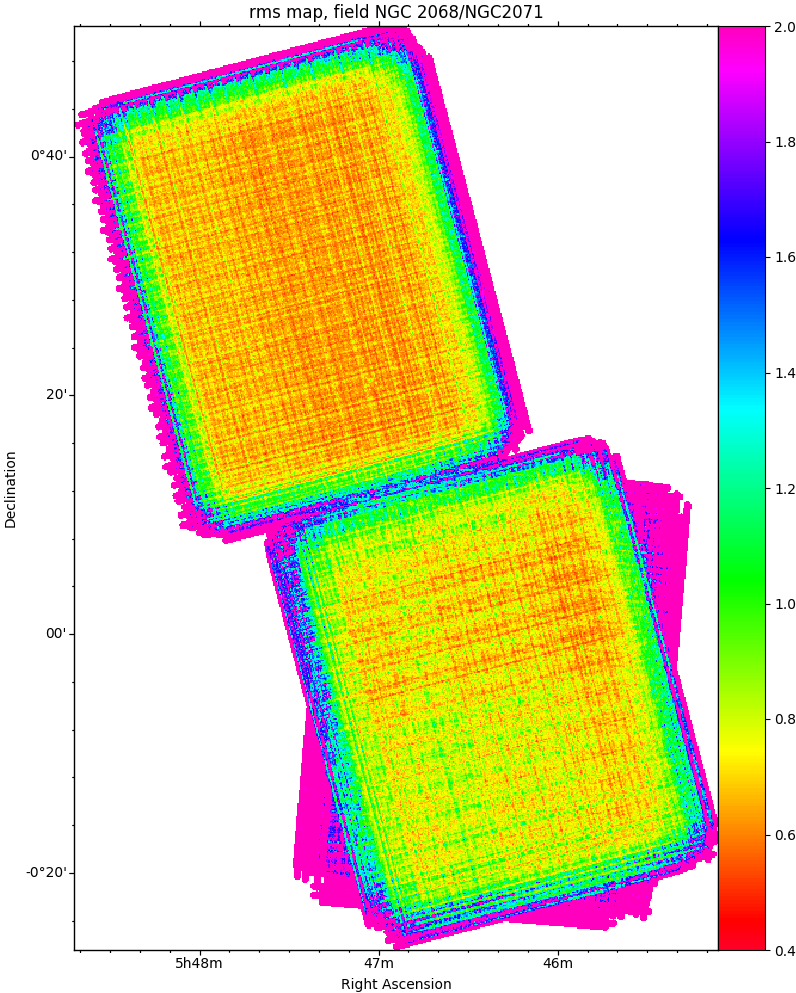}
      \caption{Noise map for the NGC~2068 and NGC~2071 field.
              }
         \label{fig:rmsmap_N2068N2071}
\end{figure}

\begin{figure}
   \centering
   \includegraphics[width=\hsize]{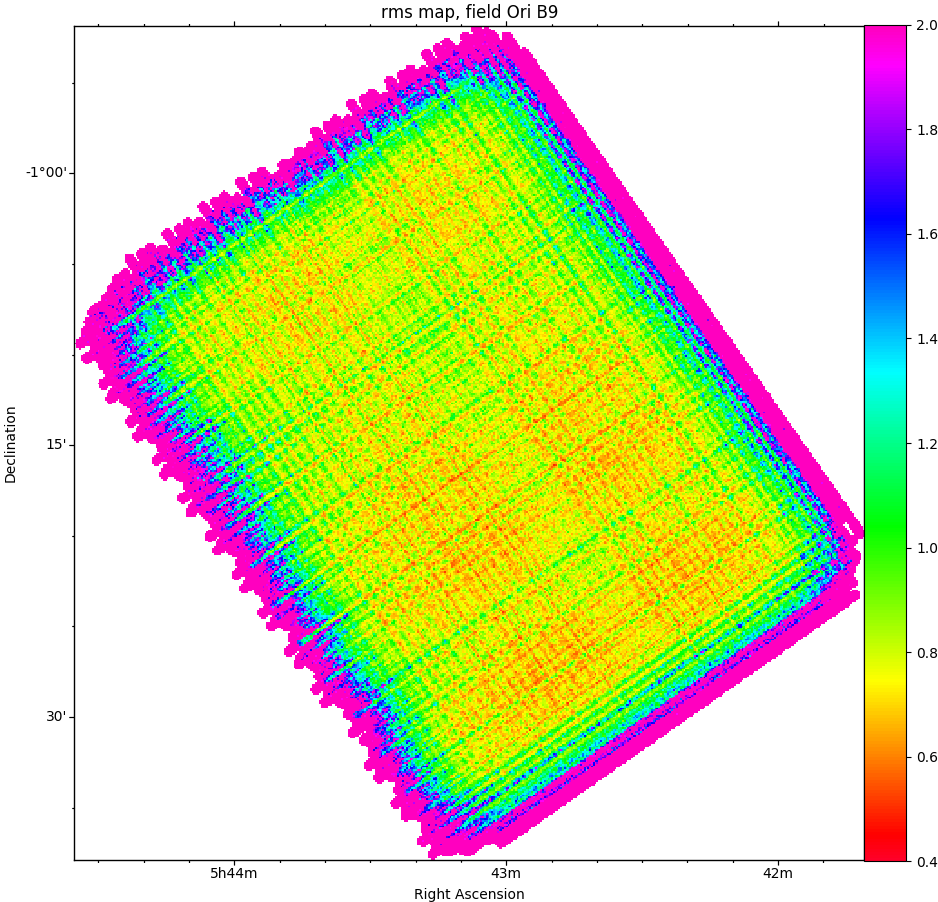}
      \caption{Noise map for the Ori~B9 field.
              }
         \label{fig:rmsmap_ORIB9}
\end{figure}

\begin{figure}
   \centering
   \includegraphics[width=\hsize]{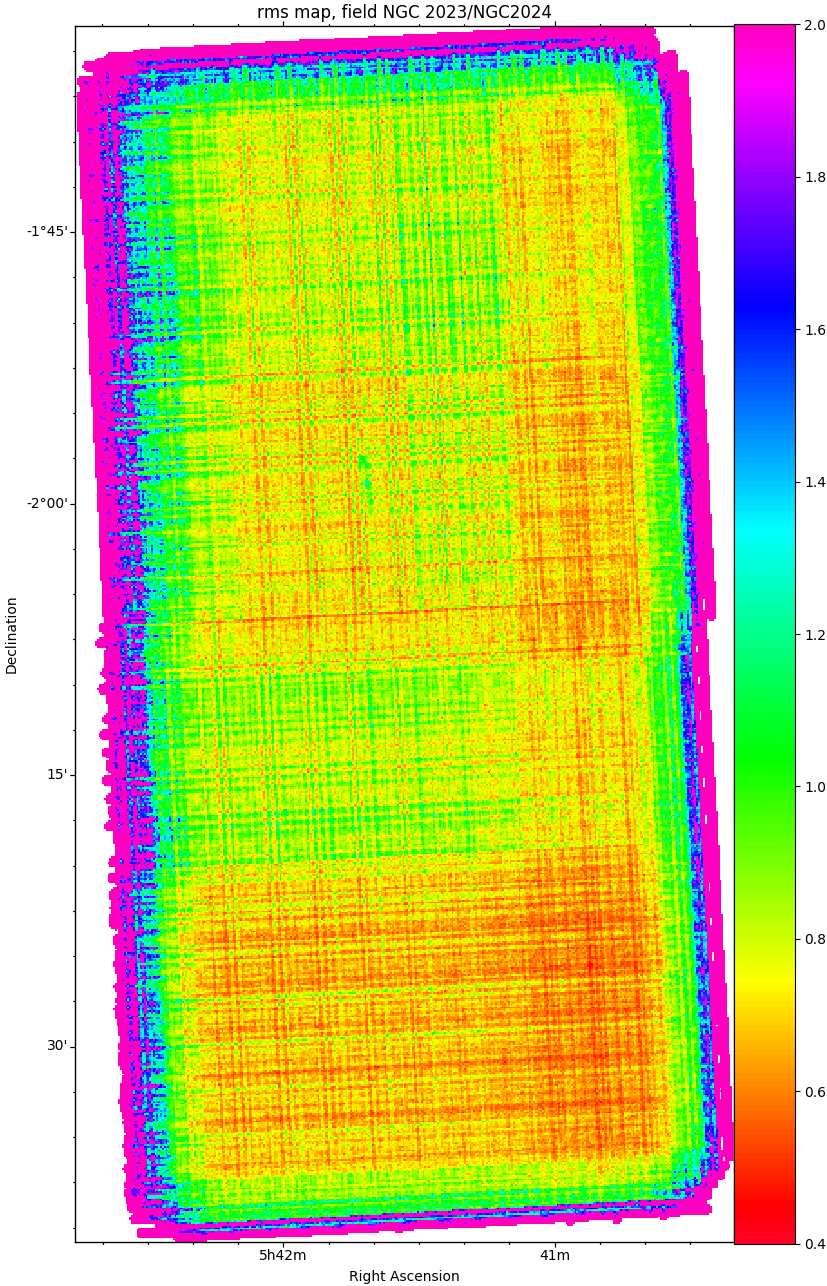}
      \caption{Noise map for the NGC~2023 and NGC~2024 field.
              }
         \label{fig:rmsmap_N2023N2024}
\end{figure}

\begin{figure}
   \centering
   \includegraphics[width=\hsize]{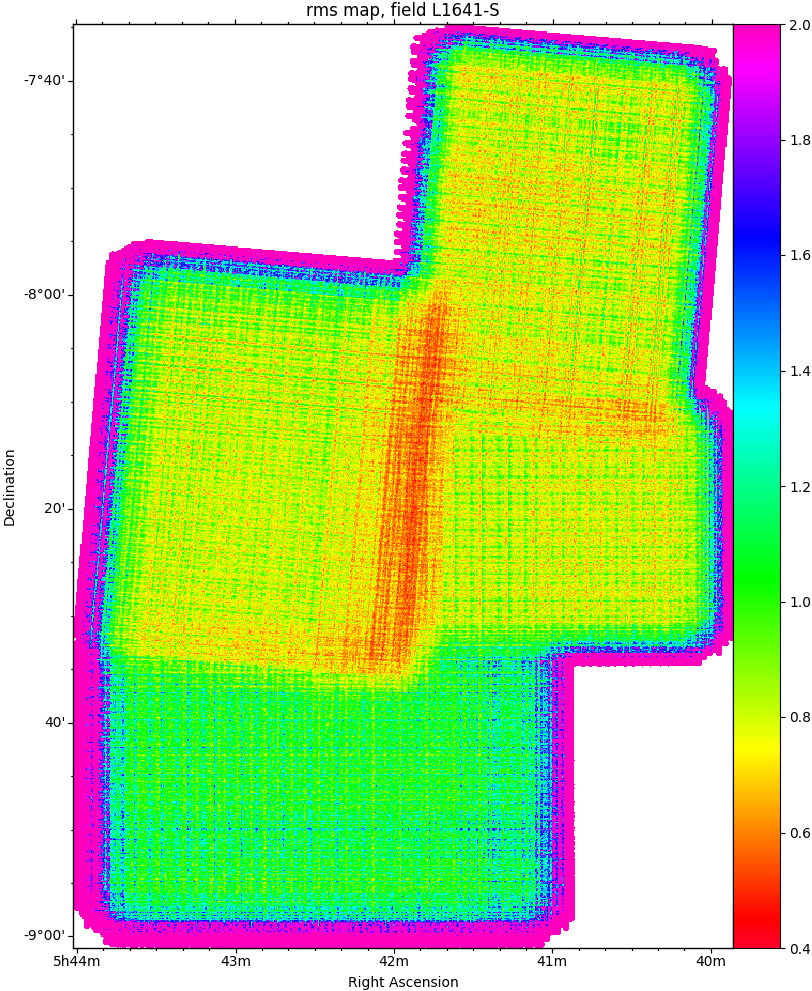}
      \caption{Noise map for the L1641-S field.
              }
         \label{fig:rmsmap_L1641S}
\end{figure}

\begin{figure*}
   \centering
   \includegraphics[width=\hsize]{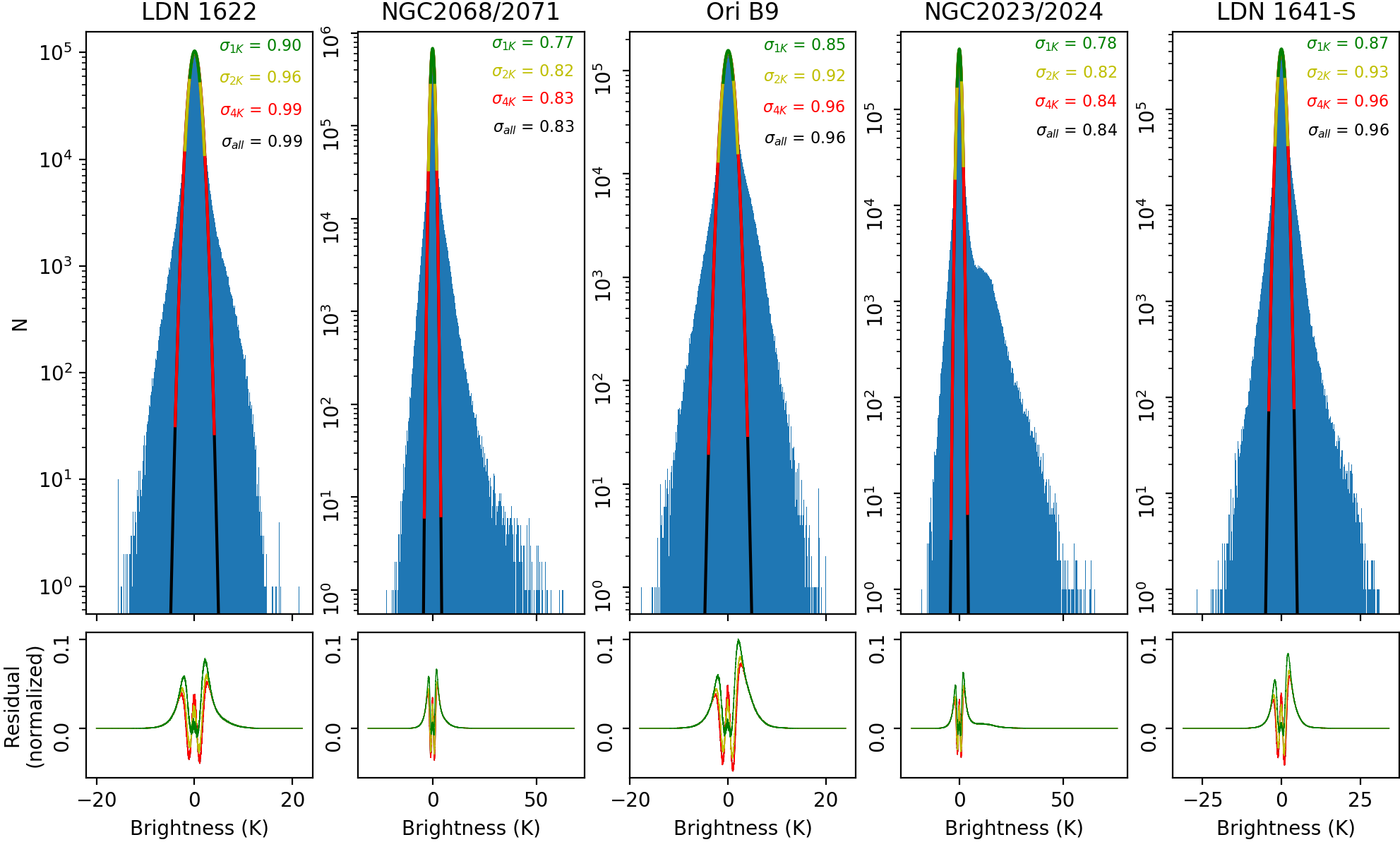}
   \caption{Histograms of the brightness values for the five survey fields, along with the results of Gaussian fits to
     the central peak of the distribution (green: fitting range $0 \pm 1$~K; yellow: fitting range $0 \pm 2$~K; red: fitting range
     $0 \pm 4$~K; black: fitting range entire brightness range). The top row shows the histograms, while the bottom row show the
     residuals of the Gaussian fits (normalized to the maximum pixel count).
              }
         \label{fig:rms_hist}
\end{figure*}

\begin{figure}
   \centering
   \includegraphics[width=\hsize]{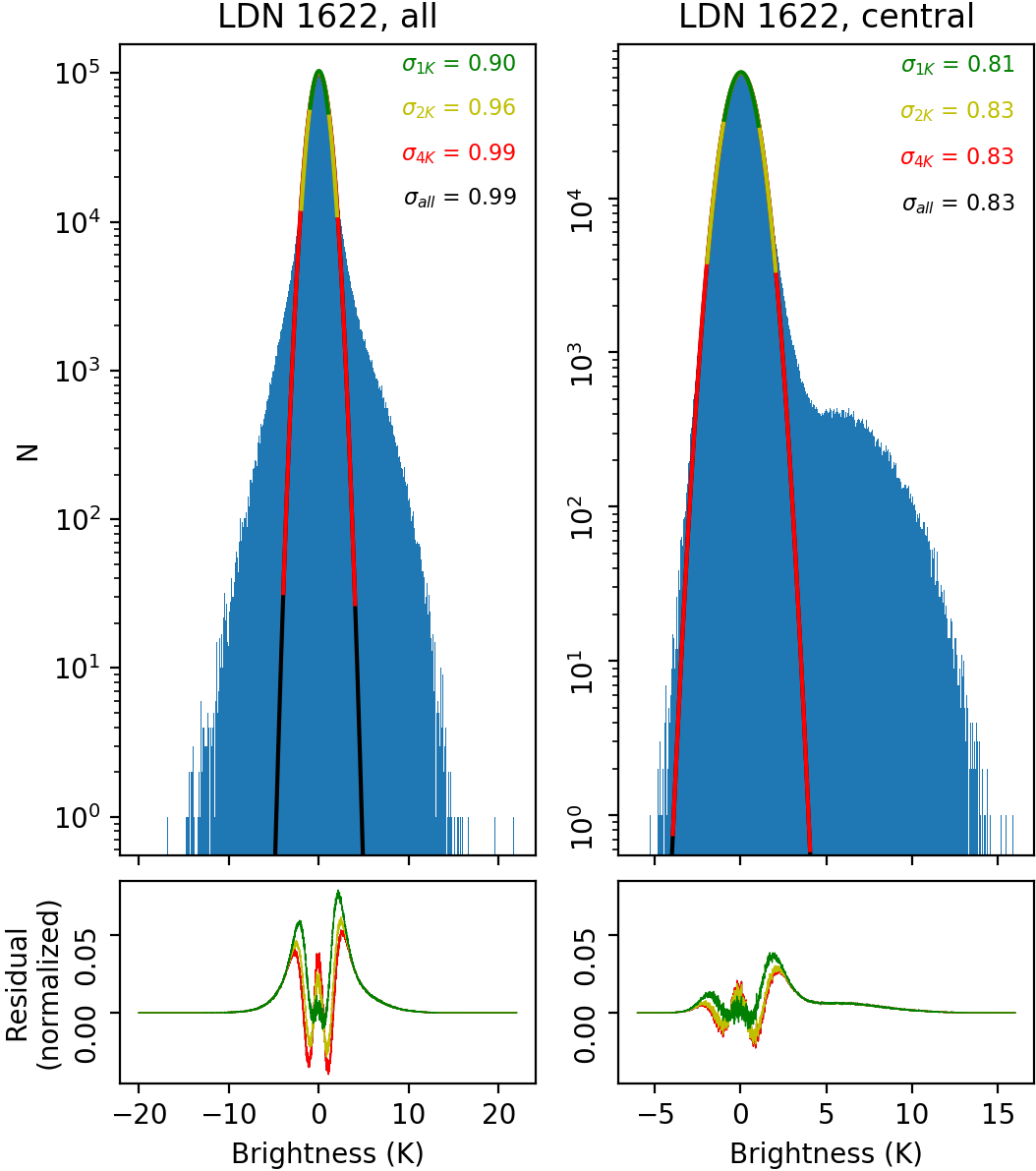}
   \caption{Histograms of the brightness values for the LDN~1622 survey field, along with the results of Gaussian fits to
     the central peak of the distribution (green: fitting range $0 \pm 1$~K; yellow: fitting range $0 \pm 2$~K; red: fitting range
     $0 \pm 4$~K; black: fitting range entire brightness range). The top row shows the histograms, while the bottom row show the
     residuals of the Gaussian fits (normalized to the maximum pixel count). Left: distribution for the full field. Right: distribution
     for the central part of the field, excluding the noisy map edges.
              }
         \label{fig:rms_hist_L1622}
\end{figure}

We here present noise maps for the five survey fields (Figs.~\ref{fig:rmsmap_ORIB9} to \ref{fig:rmsmap_L1641S}). 
The rms for each spatial pixel is measured directly in the cube, deriving statistics for the spectrum over
the velocity range beyond the maximum velocity extent of CO emission.
The colour mapping is the
same for all fields to allow for a comparison between them and an easy assessment of the
(non)uniformity of the rms within each.

We also show the full frequency distribution of brightness values in the cubes in Fig.~\ref{fig:rms_hist}.
To characterize the noise in the cubes further, we have approximated the frequency distributions by
Gaussians, varying the fitting ranges to check at which extent the non-Gaussian wings of the distributions
(stemming from noisier sections at the cube edges, where the exposure time per spatial pixel is lower, and from
source emission) influences the noise estimate. We provide the derived rms values for various fitting ranges around
the maximum of the distribution ($\pm 1$~K: green; $\pm 2$~K: yellow; $\pm 4$~K: red; full intensity range: black)
in Fig.~\ref{fig:rms_hist}, along with the fits plotted over the histograms. We also plot the normalized residuals
after subtraction of the Gaussian fits.

As expected, increasing the fitting range results in broader Gaussian fits, as the contribution of the noisy
spatial pixels at the map edges and astronomical signal contribute increasingly to the wider range of brightness values.
The residuals indicate that overall a Gaussian is not a very good representation of the brightness distribution, which can
be understood as the overall distribution is the sum over the distributions of the individual spatial pixels, each of which
should be a Gaussian (in the emission-free parts of the spectrum) with its own standard deviation. Adding Gaussians with
differing standard deviations will result in an overall distribution that is more triangular than a Gaussian. In particular, a
Gaussian fit will underestimate the values in the very centre of the distribution. Using a narrow fitting range reduces
the residual in the centre, but implies larger residuals in the wings of the distributions. Larger fitting ranges somewhat
reduce the residuals in the wings, but also produce larger residuals on the central peak.

The derived standard deviations are systematically larger than the typical rms values derived in Sect.~\ref{section:sensitivity}. This can be attributed
to the significant contribution of the noisy spatial pixels at the map edges (and the inclusion of the full spectral range, while in
Sect.~\ref{section:sensitivity} we excluded the spectral range with bright line emission. To illustrate this further, Fig.~\ref{fig:rms_hist_L1622} shows the frequency
distributions of brightness values in the LDN~1622 survey field over the full cube (left) and of the central part excluding the noisy
spatial pixels at the map edges (right). Excluding the map edges results in a more Gaussian shape of the distribution
(the residuals are significantly reduced, the width of the fitting range has a much smaller impact on the resulting standard deviations
of the fits). Moreover, the derived rms of 0.82~K agrees well with the typical rms derived in Sect.~\ref{section:sensitivity} of 0.81~K, indicating that our
approach for deriving typical rms values for the well-observed central parts of the maps gives sensible values.

\section{Line velocities}
\label{app:momentmaps}

\begin{figure*}
   \centering
   \includegraphics[width=\hsize]{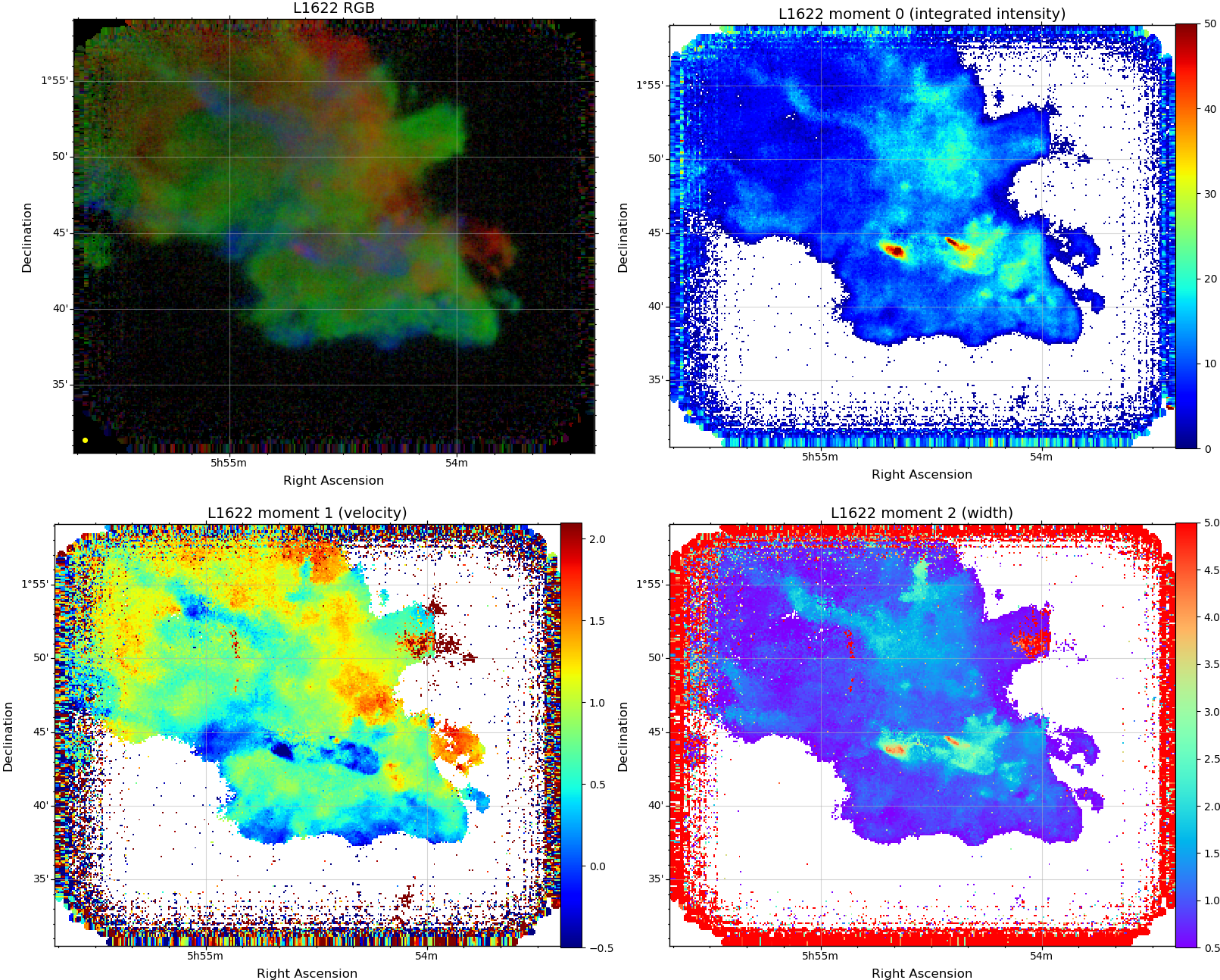}
      \caption{Properties of the velocity field in the L~1622 area. Top, left panel: red/green/blue representation of the velocity
          field around the clouds ambient velocity; red: $v = +1.25$~km/s to $+2.25$~km/s; green: $v = +0.25$~km/s to $+1.0$~km/s; 
          blue: $v = -1.0$~km/s to $0.0$~km/s. The yellow dot in the lower left corner indicates the beam size. The remaining three panels show maps of the first three moments of the line emission, computed
          over the velocity range from $-$6~km/s to $+$8~km/s, and limited to emission brighter than 3.5~K.
              }
         \label{fig:momentmaps_L1622}
\end{figure*}

\begin{figure*}
   \centering
   \includegraphics[width=\hsize]{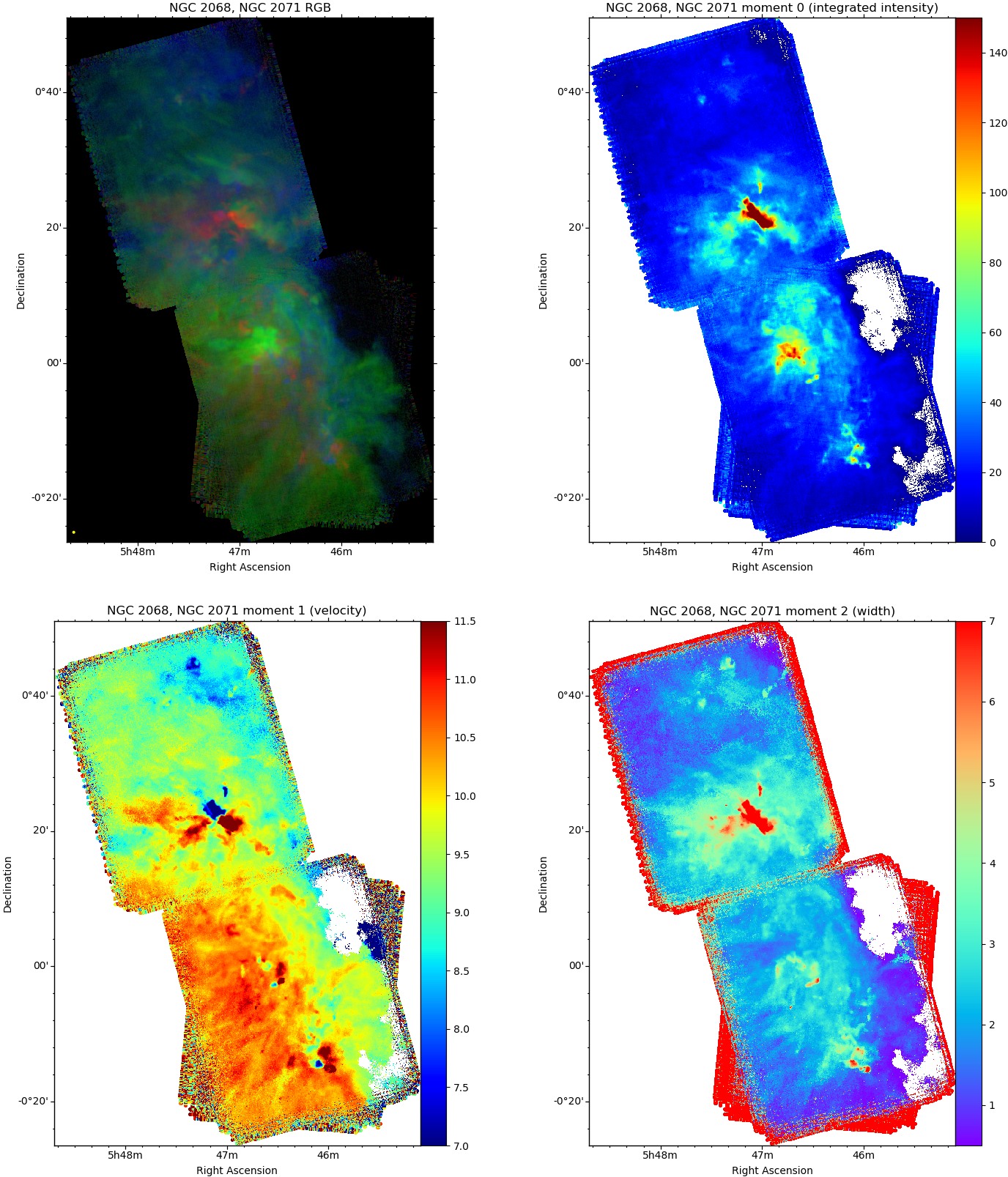}
      \caption{Properties of the velocity field in the NGC~2068 and NGC~2071 area. Top, left panel: red/green/blue representation of
          the velocity
          field around the clouds ambient velocity; red: $v = +11.25$~km/s to $+13.0$~km/s; green: $v = +9.5$~km/s to $+11.0$~km/s; 
          blue: $v = +7.5$~km/s to $+9.25$~km/s. The yellow dot in the lower left corner indicates the beam size. The remaining three panels show maps of the first three moments of the line emission, computed
          over the velocity range from 0~km/s to $+$20~km/s, and limited to emission brighter than 3.5~K.
              }
         \label{fig:momentmaps_N2068N2071}
\end{figure*}

\begin{figure*}
   \centering
   \includegraphics[width=\hsize]{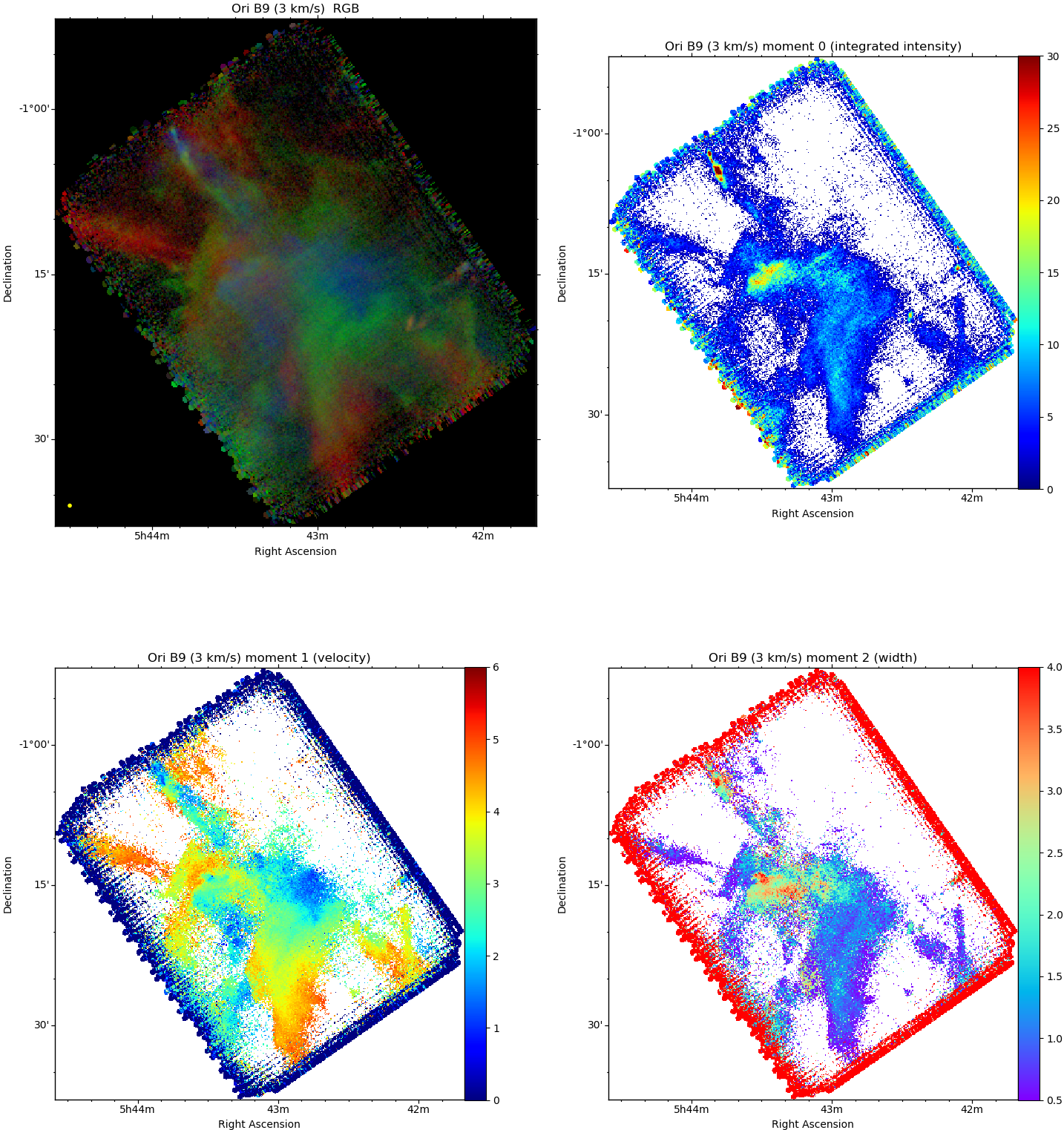}
      \caption{Properties of the velocity field in the Ori~B9 area, for velocities around $+3$~km/s. Top, left panel: red/green/blue
          representation of the velocity
          field around the central velocity of $+3$~km/s; red: $v = +4.0$~km/s to $+5.5$~km/s; green: $v = +2.25$~km/s to $+3.75$~km/s; 
          blue: $v = +0.5$~km/s to $+2.0$~km/s. The yellow dot in the lower left corner indicates the beam size. The remaining three panels show maps of the first three moments of the line emission, computed
          over the velocity range from $-$6~km/s to $+$8~km/s, and limited to emission brighter than 3.5~K.
              }
         \label{fig:momentmaps_ORIB9_2kms}
\end{figure*}

\begin{figure*}
   \centering
   \includegraphics[width=\hsize]{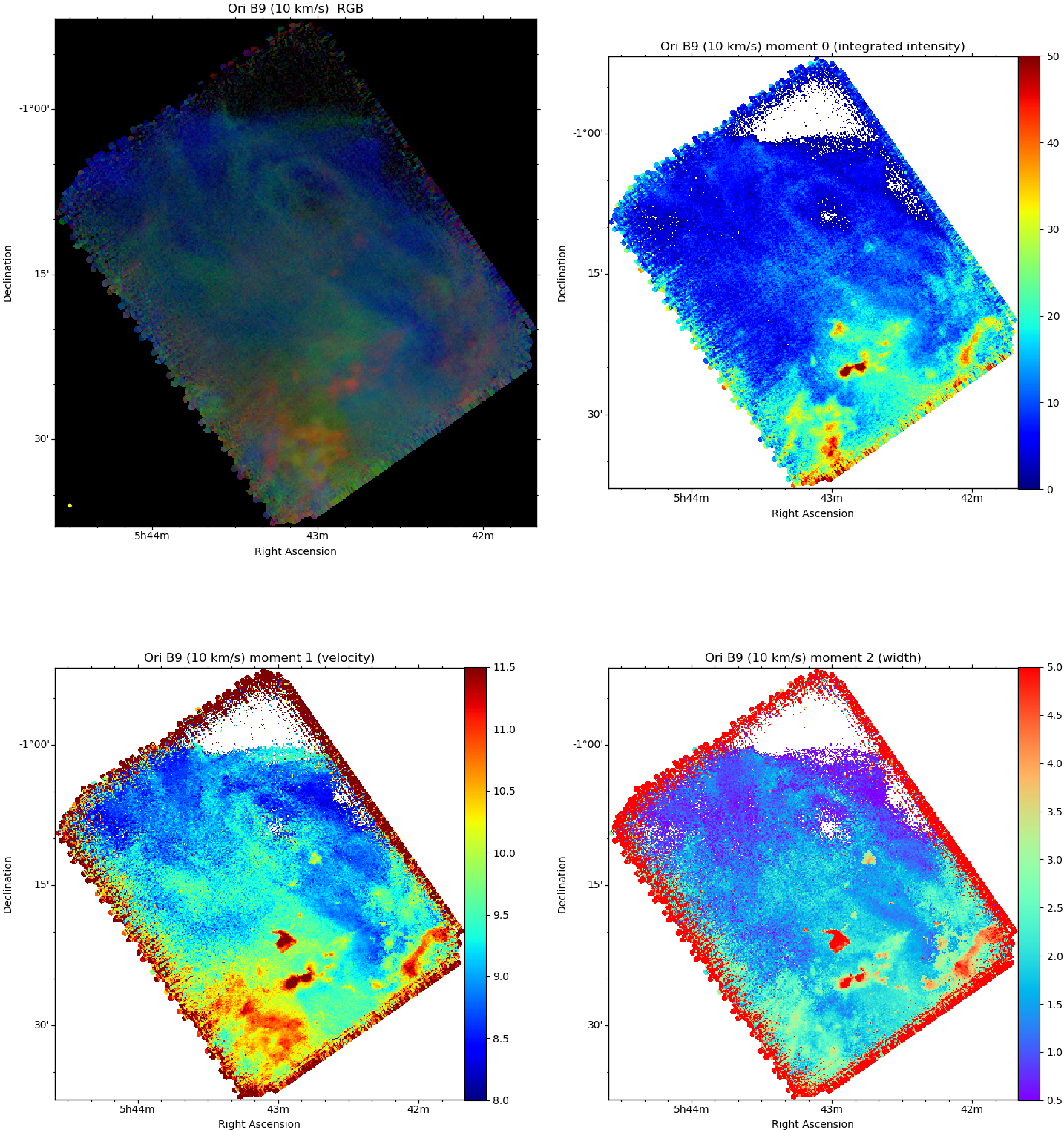}
      \caption{Properties of the velocity field in the Ori~B9 area, for velocities around $+10$~km/s.  Top, left panel: red/green/blue
          representation of the velocity
          field around the central velocity of $+10$~km/s; red: $v = +10.75$~km/s to $+12.0$~km/s; green: $v = +9.5$~km/s to $+10.5$~km/s; 
          blue: $v = +8.0$~km/s to $+9.25$~km/s. The yellow dot in the lower left corner indicates the beam size. The remaining three panels show maps of the first three moments of the line emission, computed
          over the velocity range from $+$8~km/s to $+$20~km/s, and limited to emission brighter than 3.5~K.
              }
         \label{fig:momentmaps_ORIB9_10kms}
\end{figure*}

\begin{figure*}
   \centering
   \includegraphics[width=\hsize]{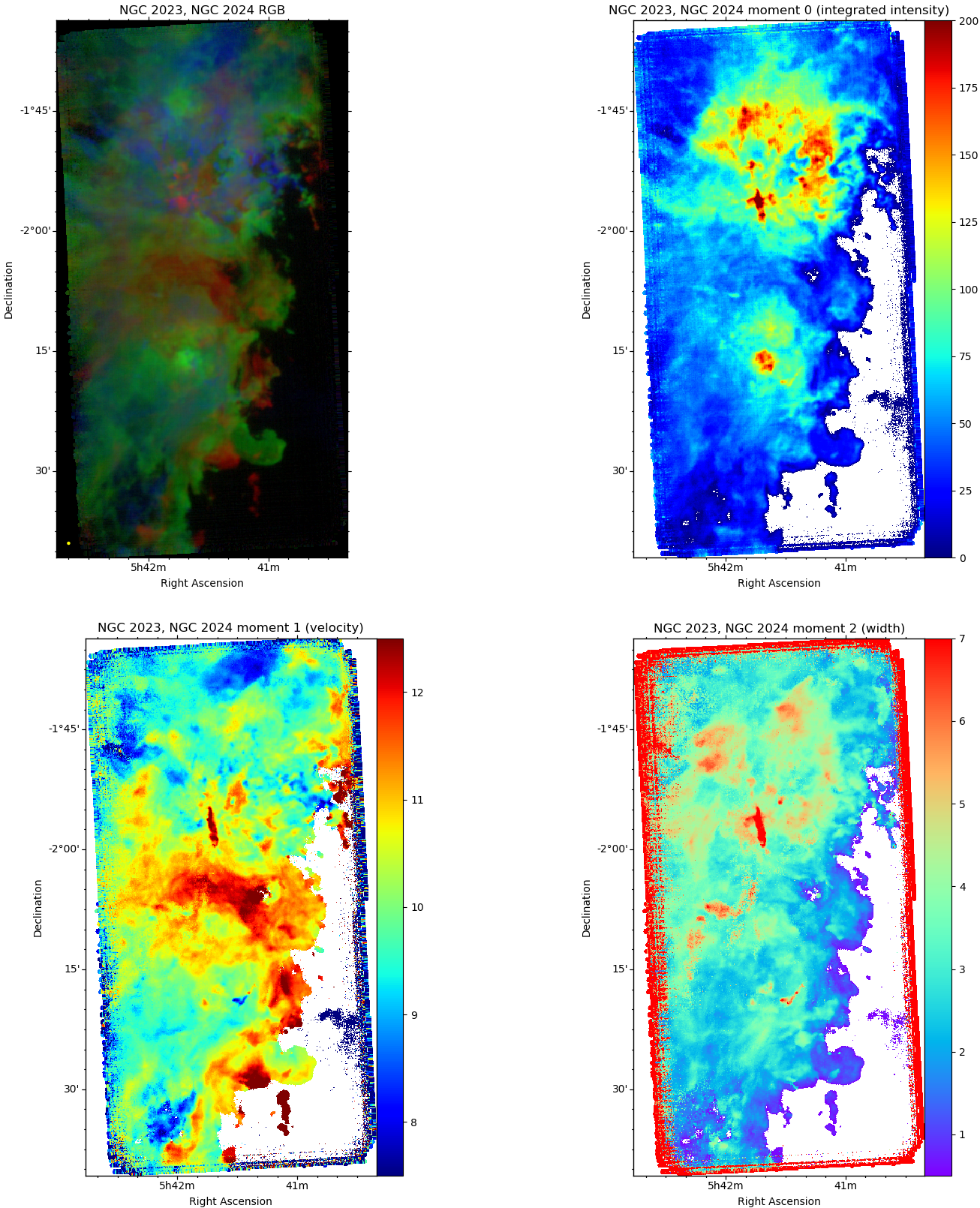}
      \caption{Properties of the velocity field in the NGC~2023 and NGC~2024 area.  Top, left panel: red/green/blue representation of
          the velocity
          field around the clouds ambient velocity; red: $v = +11.5$~km/s to $+14.0$~km/s; green: $v = +9.25$~km/s to $+11.25$~km/s; 
          blue: $v = +6.5$~km/s to $+9.0$~km/s. The yellow dot in the lower left corner indicates the beam size. The remaining three panels show maps of the first three moments of the line emission, computed
          over the velocity range from $-$5~km/s to $+$20~km/s, and limited to emission brighter than 3.5~K.
              }
         \label{fig:momentmaps_N2023N2024}
\end{figure*}

\begin{figure*}
   \centering
   \includegraphics[width=\hsize]{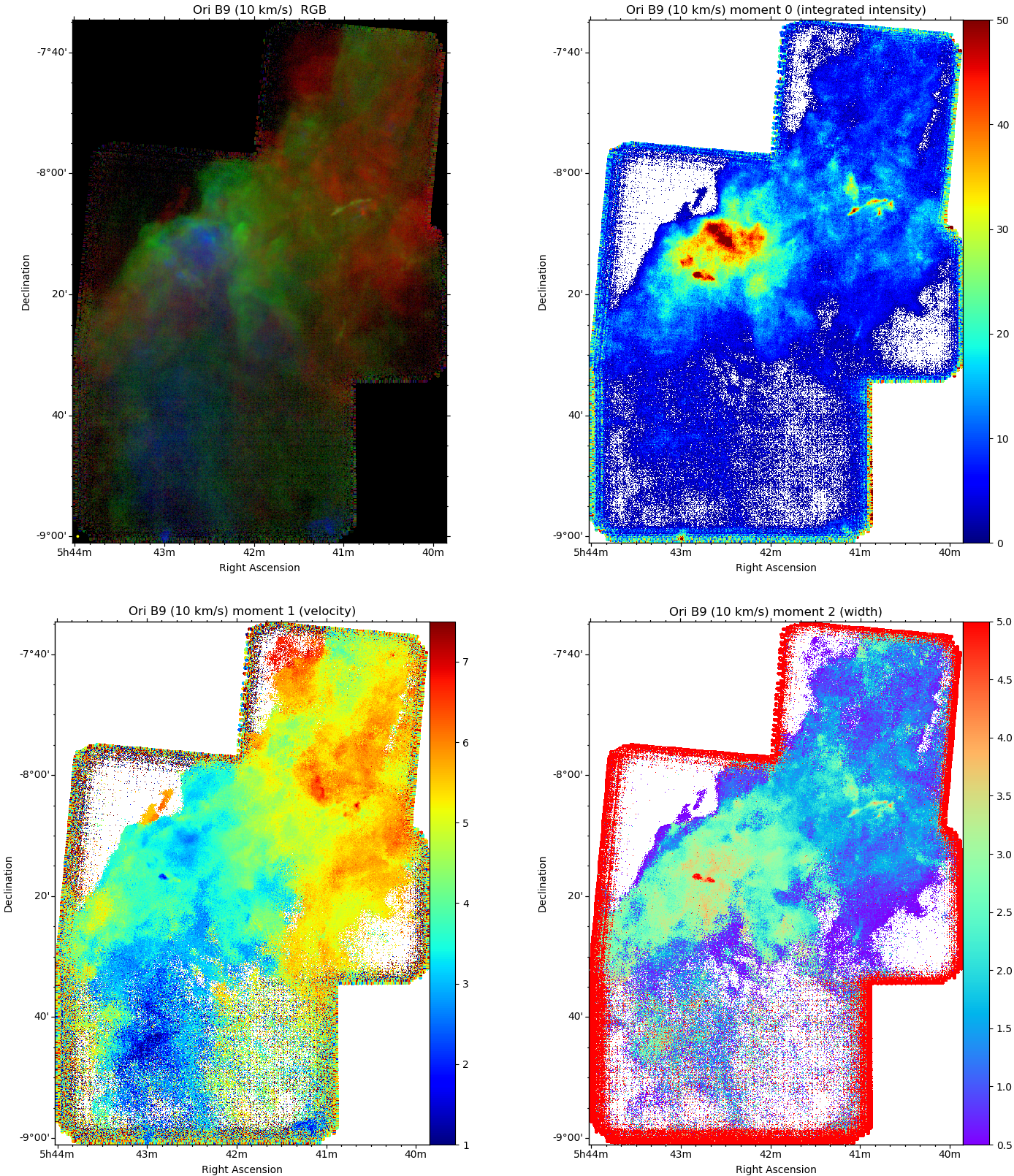}
      \caption{Properties of the velocity field in the L~1641-S area.  Top, left panel: red/green/blue representation of
          the velocity
          field around the clouds ambient velocity; red: $v = +5.0$~km/s to $+7.25$~km/s; green: $v = +3.0$~km/s to $+4.75$~km/s; 
          blue: $v = +0.5$~km/s to $+2.75$~km/s. The yellow dot in the lower left corner indicates the beam size. The remaining three panels show maps of the first three moments of the line emission, computed
          over the velocity range from $-$5~km/s to $+$15~km/s, and limited to emission brighter than 3.5~K.
              }
         \label{fig:momentmaps_L1641S}
\end{figure*}

We here give an overview of the CO line properties within the survey fields. For each field, we show a RGB coded map of the
red-shifted/central/blue-shifted velocity range of the main part of the line (i.e. excluding high velocity line-wings).
We furthermore show maps of the first three line moments, computed using the GILDAS MOMENTS task: moment~0: integrated intensity,
moment~1: mean velocity, and moment~2:
line width, where we limit the calculation to emission brighter than 3.5~K (while the integrated maps presented in
Figs.~\ref{fig:intmap_L1622}, \ref{fig:intmap_N2068N2071}, \ref{fig:intmap_ORIB9}, \ref{fig:intmap_N2023N2024}, and 
\ref{fig:intmap_L1641S} add up individual velocity channels regardless of line detection and brightness) and to a certain velocity range,
which varies from field to field. For the L~1622 field we only include the main cloud component around 1~km/s. For the
Ori~B9, field we show RGB and moment maps for both velocity components (around 3~km/s and around 10~km/s).

\section{CO(3-2)/CO(1-0) RADEX models}
\label{app:radex}

\begin{figure}
  \centering
  \includegraphics[width=\columnwidth]{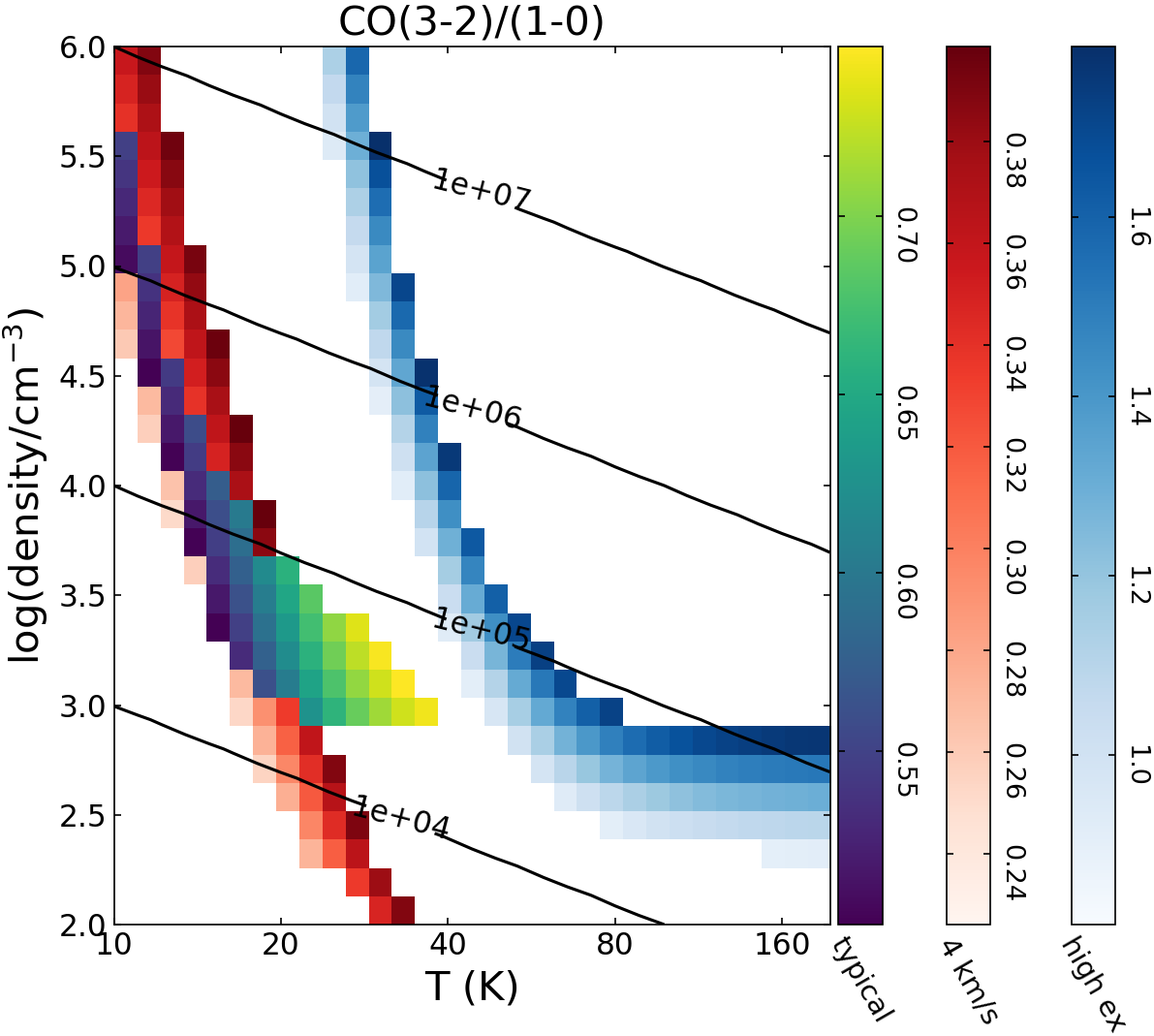}
  \caption{Parameter space of RADEX models reproducing the line intensities and line ratios
    found in the NGC~2023 and NGC~2024 region, for three diverse emission components. Line ratios of models reproducing
    the 'typical' emission are shown on a multi-colour scale, models corresponding to the emission component seen
    at $\sim$4\,km/s are shown on a red colour scale, and models reproducing the high-excitation component at the western
    cloud edge are shown on a blue colour scale. For each component, models for a column density in the (logarithmic) middle
    of the column density range reproducing the respective
    characteristic line ratios and intensities are shown: $N = 10^{18}$\,cm$^{-2}$ (typical emission), $N = 2 \cdot 10^{17}$\,cm$^{-2}$
    (4\,km/s component),
    and $N = 10^{16}$\,cm$^{-2}$ (high-excitation component). Straight lines mark locations of constant pressure $p = n \cdot T$.
     Top: RADEX models for the 'typical' emission. 
   }
              \label{fig:N2024_Radex}%
\end{figure}

Here, we explore the use of the CO(3-2) and CO(1-0) line intensities and intensity ratios to constrain the
physical conditions in the emitting gas, using the RADEX line radiative transfer code \citep{vandertaketal2007}.
We calculated grids of line main beam brightnesses and their ratios, covering a temperature range from 10\,K to 200\,K and
H$_2$ densities ranging from $n = 10^2$\,cm$^{-3}$ to 10$^6$\,cm$^{-3}$, and for CO column densities ranging from $N = 10^{15}$\,cm$^{-2}$
to 10$^{19}$\,cm$^{-2}$.
In Fig.~\ref{fig:N2024_Radex} we plot RADEX CO(3-2)/CO(1-0) line ratios,
restricted to models reproducing three diverse emission regimes, as follows:

The multi-colour scale in Fig.~\ref{fig:N2024_Radex} corresponds to models resulting in 'typical' line brightnesses and line ratios
as judged from Fig.~\ref{fig:N2024_ratioCO_TCO}.
As 'typical' we take CO(1-0) brightness values between 15\,K and 23\,K (corresponding to 45\,K\,km/s to 70\,K\,km/s integrated over
3\,km/s), CO(3-2) between 8\,K and 16\,K (24\,K\,km/s to 48\,K\,km/s integrated over 3\,km/s), and ratios between 0.5 and 0.75.
We find that there are valid models
for column densities ranging from $10^{17}$\,cm$^{-2}$ to $10^{19}$\,cm$^{-2}$ (the multi-colour scale models in Fig.~\ref{fig:N2024_Radex}
are for an intermediate column density of $N = 10^{18}$\,cm$^{-2}$).
Depending on the CO column density a large range of parameters can reproduce the observed emission. Low column densities require
moderately high temperatures (20\,K-40\,K) to reach the measured brightness values, along with high densities. At higher column densities,
temperatures of a 10\,K to $\sim$25\,K are sufficient along with lower densities (few $\times 10^3$cm$^{-3}$).

The red colour scale in Fig.~\ref{fig:N2024_Radex} marks conditions that might yield the line ratios that characterize
the faint cloud component seen at velocities around 4-5\,km/s towards NGC~2023/NGC~2024. While we do not have a map of the CO(1-0)
emission for this component available, the comparison of the integrated spectra (Fig.~\ref{fig:N2024_avespectrum}) indicates that
the CO(3-2)/CO(1-0) ratio is $\sim 2$~times smaller than for the 'typical' cloud emission, we thus plot only models that result in
line ratios between 0.25 and 0.4; further we only show models with CO(3-2) brightness less than 10\,K (the maximum we see in our cubes
for that velocity component). Again a wide range of parameters can reproduce the observed emission, with column densities being
lower by a factor of a few than for the 'typical' molecular gas. Valid models are obtained for column densities ranging from
2$\times 10^{16}$~cm$^{-2}$ to 5$\times 10^{18}$~cm$^{-2}$ (where the red colour scale indicates models with an intermediate column
density of $N = 2 \times 10^{17}$~cm$^{-2}$). Lower column densities generally result in a shift to higher temperatures, while there
is a degeneracy between temperature and density (higher temperatures need lower densities to keep the CO(3-2)/(1-0) line ratio low through
subthermal excitation at low densities).

Finally, the blue colour scale in Fig.~\ref{fig:N2024_Radex} shows models reproducing the high line ratios and comparably low intensities
found along the western edge of the clouds interfacing molecular gas and the \object{IC~434} ionization front. We only show models that
result in CO(3-2) brightness between 2\,K and 13\,K (6\,K\,km/s to $\sim$40\,K\,km/s integrated over 3\,km/s), CO(1-0) brightness between
0\,K and 10\,K (0\,K\,km/s to $\sim$30~K\,km/s integrated over 3\,km/s), and line ratios between 0.9 and 1.8. Valid models are found
for column densities between 5$\times 10^{15}$~cm$^{-2}$ and 5$\times 10^{16}$~cm$^{-2}$ (where the blue scale in Fig.~\ref{fig:N2024_Radex}
shows models with an intermediate column density of $N = 10^{16}$~cm$^{-2}$). Compared to the 'typical' molecular gas
discussed above, significantly lower CO column densities along with a higher temperatures gas temperature are required to reproduce the line
emission in this high-excitation component, as could be expected from gas being exposed to and heated by intense UV irradiation.

\section{APEX/Laboca 870~$\mu$m dust continuum maps}
\label{app:laboca}

\begin{figure}
   \centering
   \includegraphics[width=\hsize]{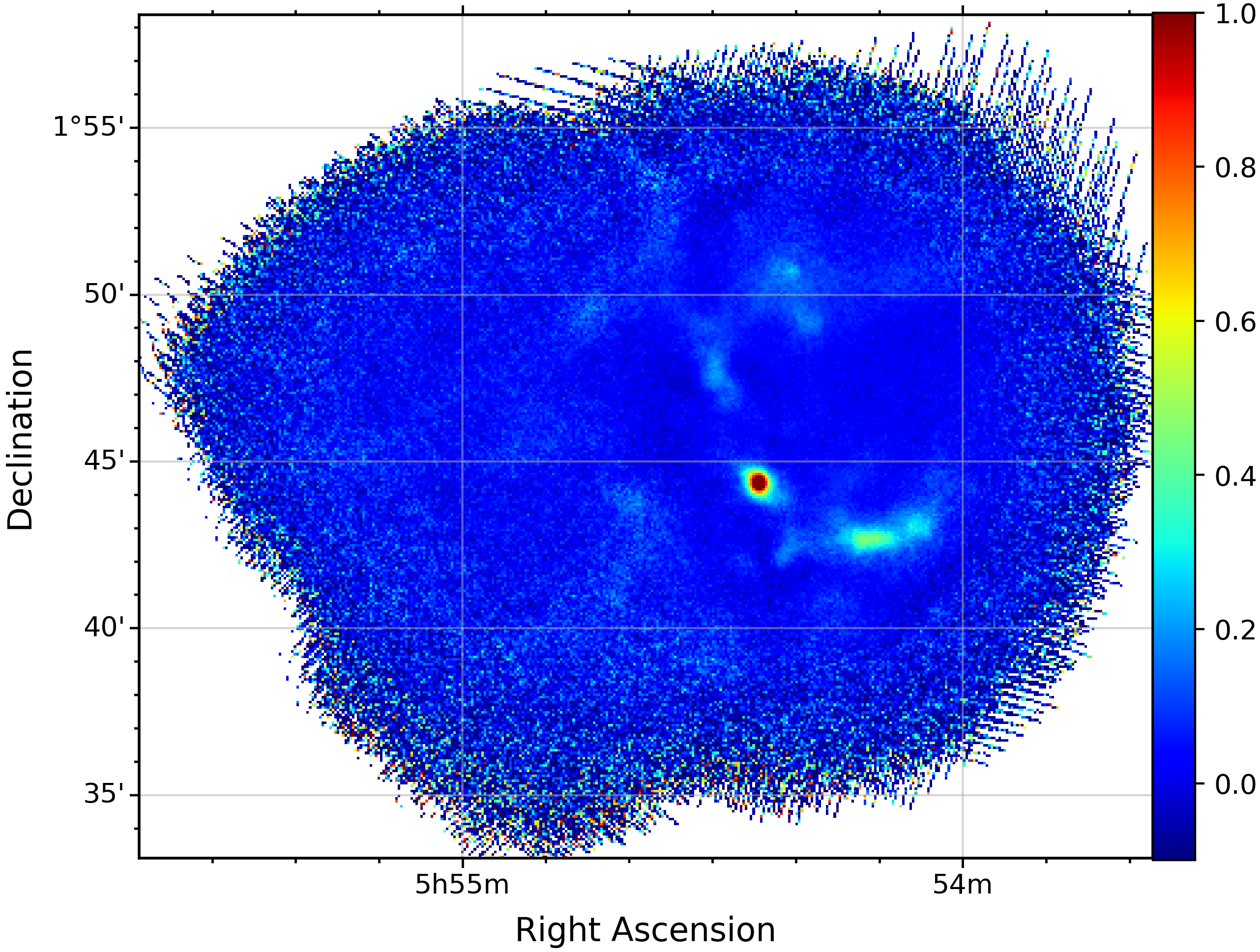}
      \caption{APEX Laboca 870~$\mu$m dust continuum emission maps in L~1622. Brightness is given in units of Jy/beam.
              Contours (at 2, 4, 8, 16, and 32~Jy/beam)
              are used to represent the brightest features in the maps.
              }
         \label{fig:laboca_L1622}
\end{figure}

\begin{figure}
   \centering
   \includegraphics[width=\hsize]{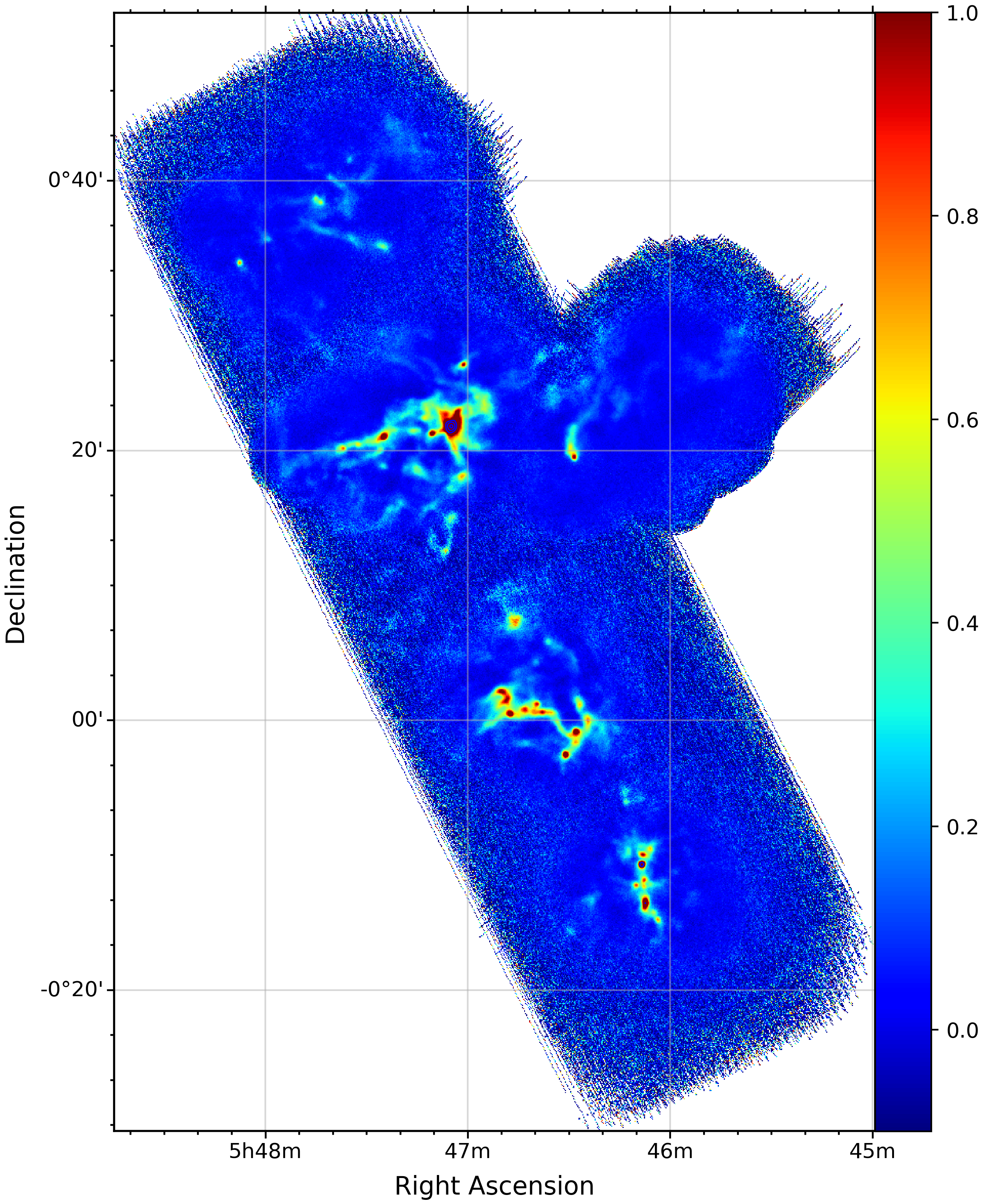}
      \caption{As Fig.~\ref{fig:laboca_L1622}, but for the NGC~2068/NGC~2071 field.
              }
         \label{fig:laboca_N2068N2071}
\end{figure}

\begin{figure}
   \centering
   \includegraphics[width=\hsize]{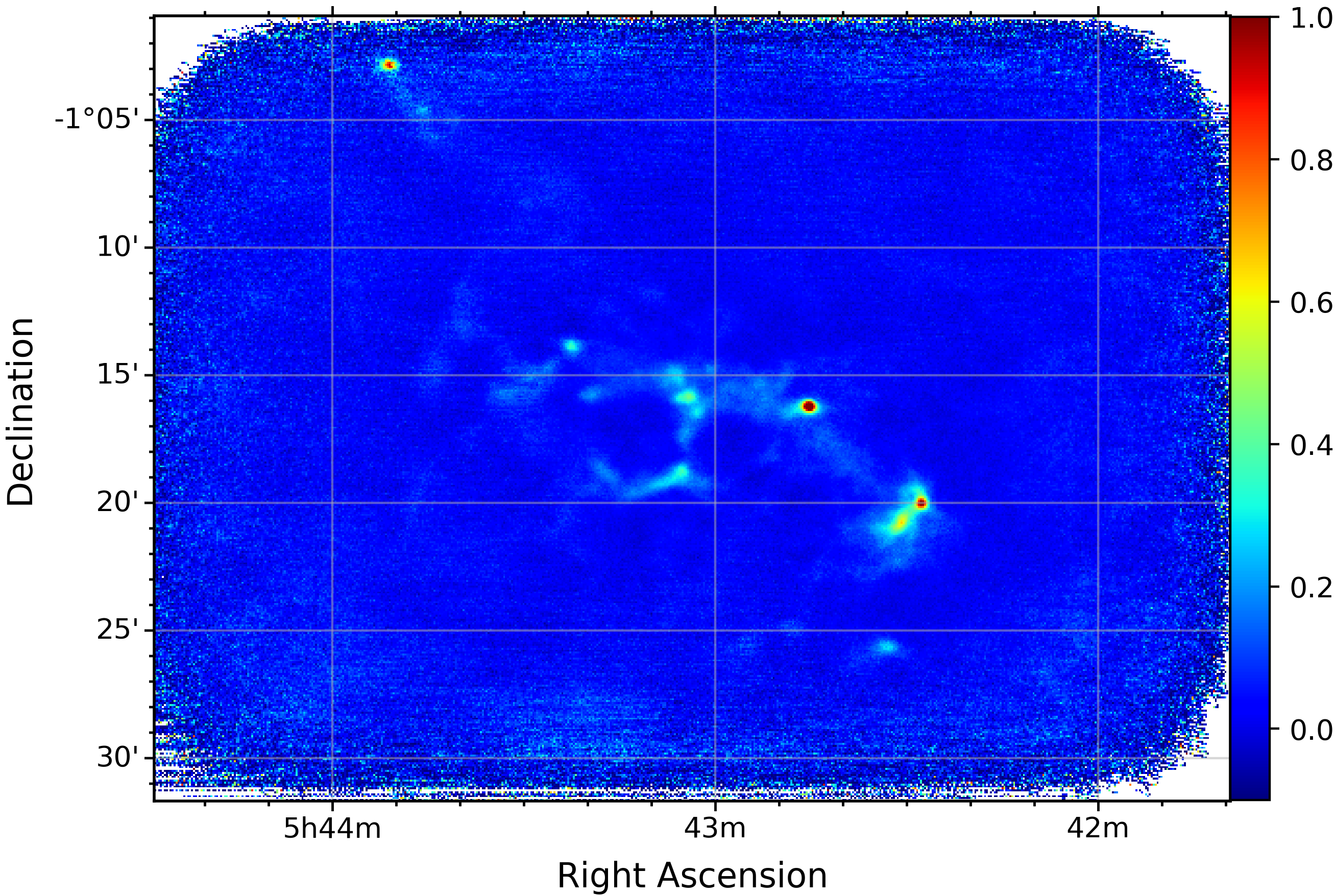}
      \caption{As Fig.~\ref{fig:laboca_L1622}, but for the Ori~B9 area (combining own observations with the data presented by \cite{miettinenetal2009}).
              }
         \label{fig:laboca_ORIB9}
\end{figure}

\begin{figure}
   \centering
   \includegraphics[width=\hsize]{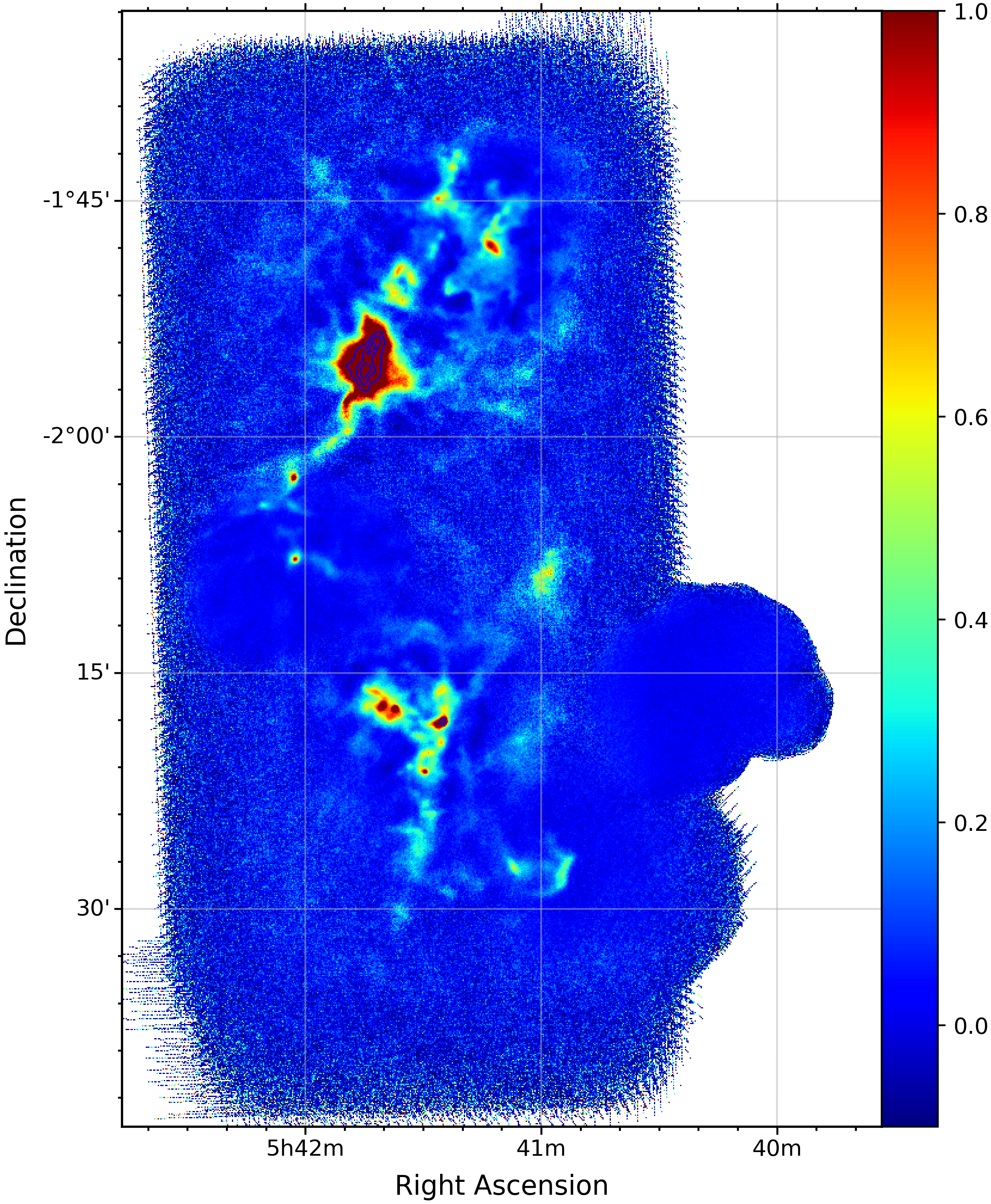}
      \caption{As Fig.~\ref{fig:laboca_L1622}, but for the NGC~2023/NGC~2024 area.
              }
         \label{fig:laboca_N2023N2024}
\end{figure}

\begin{figure*}
   \centering
   \includegraphics[width=0.95\hsize]{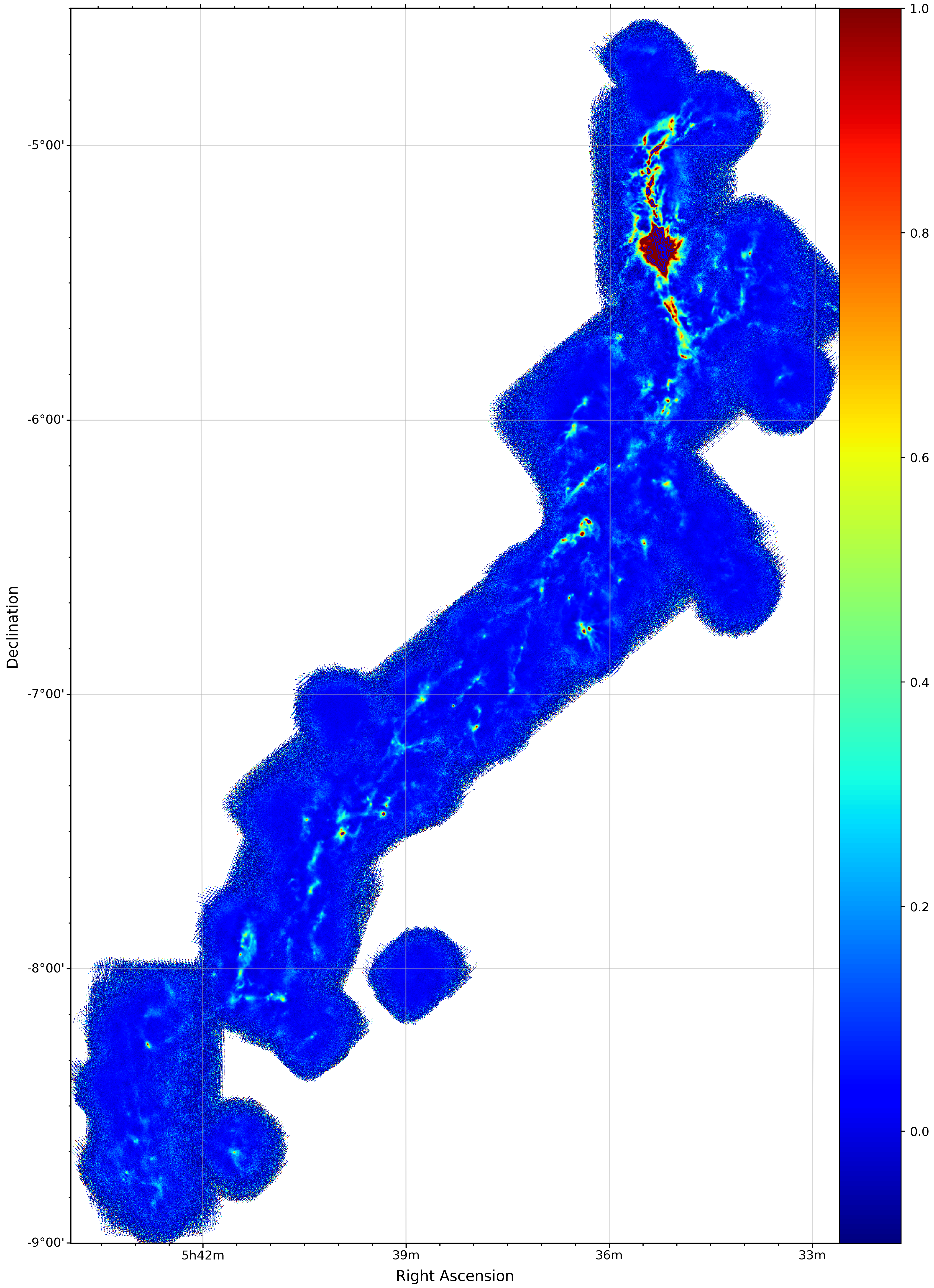}
      \caption{As Fig.~\ref{fig:laboca_L1622}, but for Orion~A (including the L~1641-S CO survey field in the south).
              }
         \label{fig:laboca_OriA}
\end{figure*}

We here include 870~$\mu$m dust continuum maps, taken with the Laboca bolometer array \citep{siringoetal2009} on APEX.
The main purpose of the observations
was to provide sub-millimetre photometry of the protostars observed in the course of the Herschel Orion Protostar Survey (HOPS) and so
determine the amount of circumstellar material left for further accretion in the envelope and disk \citep{stutzetal2013, furlanetal2016}. The maps are used at several instances in the present paper to demonstrate the relation between dense gas and the
bulk cloud emission as traced by the CO emission.

The observing strategy was optimized to cover efficiently the sample of Spitzer identified protostars observed with Herschel
in the HOPS project. This goal was achieved by implementing a number of well placed spiral raster maps (each covering a roughly circular patch
with a diameter of the order of 15\arcmin{}). To these maps, we added linear On-The-Fly scans to achieve more complete areal
coverage and better sensitivity to larger scale emission. We emphasize that the noise is not uniform over the maps as a
consequence of this observing strategy. In the Ori~B9 field, we also extracted the data obtained by \cite{miettinenetal2009}
from the archive and included them in the reduction and analysis. 

Data calibration and reduction followed standard methods as outlined in \cite{siringoetal2009}, with an iterative procedure
designed to filter out rapid atmospheric brightness variations (sky-noise) while retaining as much as possible of the extended
astronomical emission, similar to the method outlined by \cite{bellocheetal2011}. The iterative procedure uses the resulting map
of the previous iteration as input for a sky-brightness model. This model emission is subtracted from the bolometer time-stream
data before the sky-noise filtering, allowing for more stringent sky-noise filtering in later iterations, as the sky-brightness
model gets increasingly better. The sky-brightness model uses the resulting map of the previous iteration and assumes that
negative emission is an artefact due to filtering of extended emission components. This behaviour is accounted for by replacing negative
values in the map, below a certain threshold, by 'zero' values. For our reduction we used a slightly different approach than
\cite{bellocheetal2011} to construct the sky emission model for each iteration, smoothing the map from the previous iteration
to a larger beam before replacing negative values (below a certain threshold) by 'zero' values (and excluding areas with higher
rms, mostly at map edges, from replacing negative values, to avoid injecting too much 'artificial' flux in these areas).
The amount of smoothing was gradually reduced in later iterations, while at the same time the radius over which correlated 
emission variations (sky-noise) is determined was reduced (allowing for an increasingly more efficient removal of sky-noise). 
Overall, the earlier iterations result in increasingly better recovery of extended emission, while the later iterations give
increasingly better noise removal, while preserving the extended emission components. In the end we use 35 iterations for our
final maps \citep[as compared to 21 iterations in][]{bellocheetal2011}.

\end{appendix}
\end{document}